\documentclass[fleqn,usenatbib]{mnras}

\usepackage{newtxtext,newtxmath}

\usepackage[T1]{fontenc}

\DeclareRobustCommand{\VAN}[3]{#2}
\let\VANthebibliography\thebibliography
\def\thebibliography{\DeclareRobustCommand{\VAN}[3]{##3}\VANthebibliography}

\usepackage{wrapfig,graphicx,amsmath,multirow,array,bm,bbm,esint,slashed,hyperref,caption,threeparttable,gensymb}
\usepackage[dvipsnames]{xcolor}
\def\orcid#1{\href{https://orcid.org/#1}{\includegraphics[scale=0.3]{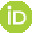}}}

\title[]{Protoclusters as Drivers of Stellar Mass Growth in the Early Universe, a Case Study: Taralay - a Massive Protocluster at z $\sim$ 4.57}

\author[Priti Staab et al.]{
Priti Staab\orcid{0000-0002-8877-4320},$^{1}$\thanks{E-mail: ppatil@ucdavis.edu}
Brian C. Lemaux\orcid{0000-0002-1428-7036},$^{2,1}$
Ben Forrest\orcid{0000-0001-6003-0541},$^{1}$
Ekta Shah\orcid{0000-0001-7811-9042},$^{1}$
Olga Cucciati\orcid{0000-0002-9336-7551},$^{4}$
Lori Lubin\orcid{0000-0003-2119-8151},$^{1}$
\newauthor
Roy R. Gal\orcid{0000-0001-8255-6560},$^{5}$
Denise Hung\orcid{0000-0001-7523-140X}$^{2,5}$
Lu Shen\orcid{0000-0001-9495-7759}$^{6,7}$
Finn Giddings\orcid{0009-0003-2158-1246},$^{5}$
Yana Khusanova\orcid{0000-0002-7220-397X},$^{8,9}$
\newauthor
Giovanni Zamorani\orcid{0000-0002-2318-301X},$^{4}$
Sandro Bardelli\orcid{0000-0002-8900-0298},$^{4}$
Letizia Pasqua Cassara\orcid{0000-0001-5760-089X},$^{10}$
Paolo Cassata\orcid{0000-0002-6716-4400},$^{11}$
\newauthor
Yi-Kuan Chiang\orcid{0000-0001-6320-261X},$^{12}$
Yoshinobu Fudamoto\orcid{0000-0001-7440-8832},$^{13}$
Shuma Fukushima\orcid{0009-0007-9580-6291},$^{14}$
Bianca Garilli\orcid{0000-0001-7455-8750},$^{15}$
\newauthor
Mauro Giavalisco\orcid{0000-0002-7831-8751},$^{16}$
Carlotta Gruppioni\orcid{0000-0002-5836-4056},$^{17}$
Lucia Guaita\orcid{0000-0002-4902-0075},$^{18}$
Gayathri Gururajan\orcid{0000-0002-7472-7697},$^{4,19}$
\newauthor
Nimish Hathi\orcid{0000-0001-6145-5090},$^{20}$
Daichi Kashino\orcid{0000-0001-9044-1747},$^{21,22}$
Nick Scoville\orcid{0000-0002-0438-3323},$^{23}$
Margherita Talia\orcid{0000-0003-4352-2063},$^{4,19}$
Daniela Vergani\orcid{0000-0003-0898-2216},$^{4}$
\newauthor
Elena Zucca\orcid{0000-0002-5845-8132}$^{4}$
\\
$^{1}$Department of Physics and Astronomy, University of California, Davis, One Shields Ave., Davis, CA 95616, USA\\
$^{2}$Gemini Observatory, NSF’s NOIRLab, 670 N. A’ohoku Place, Hilo, HI 96720, USA\\
$^{3}$INAF Osservatorio di Astrofisica e Scienza dello Spazio di Bologna, Via Piero Gobetti 93/3, 40129 Bologna, Italy\\
$^{4}$University of Hawai’i, Institute for Astronomy, 2680 Woodlawn Drive, Honolulu, HI 96822, USA\\
$^{5}$Department of Physics and Astronomy, Texas A\&M University, College Station, TX 77843-4242 USA\\
$^{6}$George P. and Cynthia Woods Mitchell Institute for Fundamental Physics and Astronomy, Texas A\&M University, College Station, TX 77843-4242 USA\\
$^{7}$Aix Marseille Université, CNRS, LAM (Laboratoire d’Astrophysique de Marseille) UMR 7326, 13388 Marseille, France\\
$^{8}$Max-Planck-Institut für Astronomie, Königstuhl 17, 69117 Heidel- berg, Germany\\
$^{9}$INAF-IASF Milano, Via Alfonso Corti 12, 20159 Milano, Italy\\
$^{10}$Dipartimento di Fisica e Astronomia Galileo Galilei, Università degli Studi di Padova, Vicolo dell’Osservatorio 3, 35122 Padova Italy\\
$^{11}$Academia Sinica Institute of Astronomy and Astrophysics (ASIAA), No. 1, Section 4, Roosevelt Road, Taipei 10617, Taiwan\\
$^{12}$Center for Frontier Science, Chiba University, 1-33 Yayoi-cho, Inage-ku, Chiba 263-8522, Japan\\
$^{13}$Department of Pure and Applied Physics, Graduate School of Advanced Science and Engineering, Faculty of Science and Engineering, \\ \, \, Waseda University, 3-4-1 Okubo, Shinjuku, Tokyo 169-8555,
Japan\\
$^{14}$INAF--IASF, via Bassini 15, I-20133, Milano, Italy\\
$^{15}$Astronomy Department, University of Massachusetts, Amherst, MA, 01003 USA\\
$^{16}$INAF - Osservatorio di Astrofisica e Scienza dello Spazio di Bologna via Gobetti 93/3, 40129, Bologna, Italy\\
$^{17}$Instituto de Astrofisica, Departamento de Ciencias Fisicas, Facultad de Ciencias Exactas, Universidad Andres Bello, Fernandez Concha 700, \\ Las Condes, Santiago RM, Chile\\
$^{18}$University of Bologna – Department of Physics and Astronomy “Augusto Righi” (DIFA), Via Gobetti 93/2, 40129 Bologna, Italy\\
$^{19}$Space Telescope Science Institute, 3700 San Martin Drive, Baltimore, MD 21218, USA\\
$^{20}$Institute for Advanced Research, Nagoya University, Nagoya 464-8601, Japan\\
$^{21}$Department of Physics, Graduate School of Science, Nagoya University, Nagoya 464-8602, Japan\\
$^{22}$California Institute of Technology, MC 249-17, 1200 East California Blvd., Pasadena, CA 91125, USA
}
\date{Accepted XXX. Received YYY; in original form ZZZ}

\pubyear{2023}

\begin{document}
\label{firstpage}
\pagerange{\pageref{firstpage}--\pageref{lastpage}}
\maketitle

\begin{abstract}

Simulations predict that the galaxy populations inhabiting protoclusters may contribute considerably to the total amount of stellar mass growth of galaxies in the early universe. In this study, we test these predictions observationally, focusing on the Taralay protocluster (formerly PCl J1001+0220) at $z \sim 4.57$ in the COSMOS field. Leveraging data from the Charting Cluster Construction with VUDS and ORELSE (C3VO) survey, we spectroscopically confirmed 44 galaxies within the adopted redshift range of the protocluster ($4.48 < z < 4.64$) and incorporate an additional 18 such galaxies from ancillary spectroscopic surveys. Using a density mapping technique, we estimate the total mass of Taralay to be $\sim 1.7 \times 10^{15}$ M$_\odot$, sufficient to form a massive cluster by the present day. By comparing the star formation rate density (SFRD) within the protocluster (SFRD$_\text{pc}$) to that of the coeval field (SFRD$_\text{field}$), we find that SFRD$_\text{pc}$ surpasses the SFRD$_\text{field}$ by \color{black} $\Delta$log(SFRD/$M_\odot$ yr$^{-1}$ Mpc$^{-3}$) = $1.08 \pm 0.32$ (or $\sim$ 12$\times$) \color{black}. The observed contribution fraction of protoclusters to the cosmic SFRD adopting Taralay as a proxy for typical protoclusters is $33.5\%^{+8.0\%}_{-4.3\%}$, a value \color{black} $\sim$2$\sigma$ \color{black} in excess of the predictions from simulations. Taralay contains three peaks that are $5\sigma$ above the average density at these redshifts. Their SFRD is $\sim$0.5 dex higher than the value derived for the overall protocluster. We show that \color{black} 68\% \color{black} of all star formation in the protocluster takes place within these peaks, and that the innermost regions of the peaks encase $\sim 50\%$ of the total star formation in the protocluster. This study strongly suggests that protoclusters drive stellar mass growth in the early universe and that this growth may proceed in an inside-out manner.

\end{abstract}




\color{black} \section{Introduction}

The density of the environment that galaxies live in plays an important role in influencing their evolution. In general, the studies of samples of galaxies across cosmic time have shown that the galaxies in the earlier stages of the universe tend to exhibit heightened levels of star formation activity with respect to their lower-redshift counterparts. This activity peaks around $z \sim 2$ and precipitously drops at higher redshifts (\citealt{2014ARA&A..52..415M} and references therein). However, such studies primarily focus on field galaxies, i.e., those residing in typical environments in the universe. Whether the galaxies in overdense environment follow the same trend is uncertain. 

In the local universe, the galaxies in the overdense environment show suppressed star formation activity compared to their field counterparts. Though the Butcher-Oemler effect \citep{1984ApJ...285..426B} exists, the fraction of optically blue galaxies increasing in overdense environment at higher redshifts, the galaxies in the overdense environment at $z < 1$ still seem deficient in star formation activity compared to their field counterparts (e.g. \citealt{2017ApJ...834...53W}, \citealt{2023MNRAS.521.5400H}).
In the epoch of $1 < z < 2$, some galaxies in high density environments display higher star formation activity (see \citealt{2022Univ....8..554A} and references therein),
though the general trend seems to be that galaxies in high density environments at these redshifts show suppressed star formation activity relative to their field counterparts (e.g., \citealt{2003ApJ...584..210G}, \citealt{2004ApJ...615L.101B}, \citealt{2010MNRAS.404.1231V}, \citealt{2012ApJ...746..188M}, \citealt{2017MNRAS.465L.104N}, \citealt{2019MNRAS.484.4695T}, \citealt{2019MNRAS.490.1231L}, \citealt{2020MNRAS.493.5987O}, \citealt{2020ApJ...890....7C}). Additionally, the presence of an overabundance of massive quiescent galaxies in overdense environments at these redshifts (e.g. \citealt{2016A&A...586A..23D}, \citealt{2017MNRAS.472.3512T}) implies that rapid \color{black} stellar mass (SM) \color{black} growth occurred at the early stages of cluster assembly either through some combination of \emph{in situ} star formation processes and \emph{ex situ} galaxy-galaxy merging activity. It is, therefore, necessary to observe overdensities at higher redshift to place constraints on the assembly history and evolution of low- and intermediate-redshift cluster populations. 

Discovering numerous overdense environments at $z > 2$, though challenging, is the first step towards understanding the behavior of the galaxies that inhabit them. Over the past two decades there has been a considerable advancement in the breadth and depth of observations capable of detecting protoclusters - the progenitors of galaxy clusters - that have, in turn, lent themselves to the discovery of 100s of such structures. These nascent galaxy clusters span large areas in the sky (>10$\arcmin$) (e.g., \citealt{2013ApJ...779..127C}, \citealt{2015MNRAS.452.2528M}, \citealt{2016MNRAS.456.1924C}, \citealt{2022Univ....8..554A} and references therein) and have density contrasts relative to the field approximately an order of magnitude less than mature clusters (e.g. \citealt{2019MNRAS.490.1231L, 2022A&A...662A..33L}, \citealt{2023A&A...670A..58M}). They are typically defined as a structure that will eventually collapse into a virialized galaxy cluster of mass $\geq 10^{14}$ M$_\odot$ at $z \geq 0$ \citep{2016A&ARv..24...14O}, though, in practice, such a definition can be difficult to impose observationally. Detecting these structures requires specialized methods that differ from those used to identify galaxy clusters at lower redshifts. This is because the protoclusters typically cannot be detected by looking for presence of overdensity of redder galaxies and/or a hot medium, methods that are commonly used for finding low- to intermediate-$z$ clusters. 

One popular technique for detecting protoclusters involves using rare, more easily observed galaxy populations and/or active galactic nuclei (AGN) as tracers of massive structures and then probing their surroundings. Tracers include quasars 
(e.g., \citealt{2011MNRAS.416.2041M}, \citealt{2013ApJ...773..178B}, \citealt{2015MNRAS.448.1335A}) \color{black} although sometimes targeting quasars did not find overdense environments (\citealt{2016A&ARv..24...14O} and references therein), \color{black} radio galaxies (e.g., \citealt{2008A&ARv..15...67M}, \citealt{2007A&A...461..823V}, \citealt{2013ApJ...769...79W}, \citealt{2016MNRAS.456.3827O},
\citealt{2021ApJ...912...60S}, \citealt{2022ApJ...941..134H}), sub-mm galaxies (e.g., \citealt{2004ApJ...611..725B}, \citealt{2014A&A...570A..55D}, \citealt{2018ApJ...861...43P},
\citealt{2023arXiv230210323C}), \color{black} ultra-massive galaxies (UMGs; \citealt{2022ApJ...926...37M}, \citealt{2023ApJ...945L...9I}, McConachie et al. in prep) \color{black} strong Ly$\alpha$ emitters ($\lambda1216$Å) (e.g., \citealt{2005ApJ...620L...1O}, \citealt{2019ApJ...879...28H}, \citealt{2020ApJ...896..156F}, \citealt{2021NatAs...5..485H}, \citealt{2022ApJ...930..102Y}), Ly$\alpha$ blobs (e.g., \citealt{2022MNRAS.513.3414L}, \citealt{2023ApJ...951..119R}), and strong H$\alpha$ ($\lambda6563$Å) emitters (e.g., \citealt{2011MNRAS.415.2993H}, \citealt{2014MNRAS.440.3262C}, \citealt{2021MNRAS.503L...1K}) or other strong emitters of rest-frame optical lines (e.g., \citealt{2017ApJ...838L..12F}). All of these tracers are more easily detected than typical star forming galaxies at these redshifts (e.g., \citealt{2003ApJ...588...65S}, \citealt{2019A&A...625A..51L}).

Until recently, the number of known protoclusters \color{black} discovered by field spectroscopic surveys of typical star forming galaxies at $z>2$ was limited \color{black}\citep{2016A&ARv..24...14O, 2019BAAS...51c.180O}, but surveys that target typical star forming galaxies are increasing this number (e.g., \citealt{2005ApJ...626...44S}, \citealt{2012ApJ...750..137T, 2016ApJ...826..114T, 2020ApJ...888...89T}, \citealt{2015A&A...576A..79L}, \citealt{2019ApJ...879....9S, 2020ApJ...899...79S}, \citealt{2022ApJ...935..177S}, \citealt{2022PASJ...74L..27U}, \citealt{2023arXiv230715113F}), 
providing an opportunity to study how stellar mass buildup proceeds in these structures in the early universe. Such wide and deep galaxy surveys that target normal star-forming galaxies using deep rest-frame UV spectra over large cosmic volumes are crucial in order to obtain a more representative sample of protoclusters. 


The on-going discoveries of protoclusters through various surveys are revealing that protoclusters are host to diverse stellar populations. The galaxies in the overdense environment at $z > 2$ seem to display substantial star formation. A recent study by \cite{2022A&A...662A..33L} of $\sim$ 7000 spectroscopically confirmed galaxies found a weak but significant trend of star formation rate (SFR) increasing with denser environment for galaxies at $2 < z < 5$. This trend is a reversal of what is observed at $z < 1.5$ (e.g., \citealt{2019MNRAS.484.4695T}, \citealt{2020MNRAS.493.5987O}) and points to accelerated \emph{in situ} stellar mass growth at higher redshift. Large sub-millimeter surveys are also being used to discover protoclusters whose galaxy populations are undergoing incredible amounts of star formation activity, with aggregate star formation rates in excess of 10000 $M_\odot$ yr$^{-1}$ (see \citealt{2022Univ....8..554A} and references therein) more than equivalent volumes in the coeval field (e.g., \citealt{2018Natur.556..469M, 2018MNRAS.476.3336G}). However, some protoclusters at high redshifts appear to contain an overabundance of redder or quiescent galaxies (e.g., \citealt{2014A&A...572A..41L, 2018A&A...615A..77L}, \citealt{2020ApJ...898..133L}, \citealt{2021ApJ...911...46S}, \citealt{2021ApJ...912...60S}, \citealt{2022ApJ...926...37M}, \citealt{2023ApJ...945L...9I}), which implies that both enhancement and suppression of star formation activity is occurring in high-density environments at these redshifts. This diversity underscores the need for the study of a larger sample of protoclusters at higher redshifts in order to understand the earlier stages of cluster formation and galaxy evolution.

Simulations suggest that there is increased star formation activity within higher density environments at higher redshift. In particular, the fraction of star formation rate per unit volume from protoclusters increases with increasing redshift (\citealt{2017ApJ...844L..23C}, though see also \citealt{2018MNRAS.473.2335M}). According to the finding of \cite{2017ApJ...844L..23C}, protoclusters contribute, e.g., $\sim 20$\% of the cosmic star formation rate density (SFRD) at $z \sim 4.5$, despite occupying only 4\% of the comoving volume of the universe at this epoch. Their contribution to the cosmic SFRD is predicted to increase to 50\% at $z \sim 10$. The higher the redshift of the protocluster, on average, the higher its predicted contribution to the cosmic SFR density. These predictions portray protoclusters as important drivers of stellar mass growth in the early universe and emphasize the need to both probe star formation activity in observational data and understand what drives it. 

In order to deepen our understanding of how the surrounding environment influences the process of star formation in protocluster members, it is essential to pursue two complementary approaches: comprehensive studies of individual systems and analysis of large protocluster samples. 
In this study, we focus on the former approach, and present a detailed study of a massive protocluster located in the COSMOS field at $z\sim4.57$ \citep{2018A&A...615A..77L}, here dubbed "Taralay"\footnote{Taralay is a fusion of words Tara and Aalay that mean a star and a house respectively in Marathi, the native language of the first author. Taralay means house of stars.}, in order to study the star-formation activity of member galaxies in detail. Taralay was the inaugural target of the Charting Cluster Construction with ORELSE and VUDS survey (C3VO, \citealt{2022A&A...662A..33L}). Due to being the highest redshift protocluster\footnote{Though we are designating Taralay a protocluster in this work, its size, extent, complex structure, and mass may indicate that it is, in fact, a proto-supercluster. Additional structure (e.g., \citealt{kakimoto2023massive}) has also been discovered around Taralay on relatively large scales, and future work will be needed to disambiguate its true nature.}
in the C3VO sample Taralay was chosen as the focus of this study. It is the excellent spectroscopic coverage that we obtained for this structure along with the deep and wide panchromatic photometry in the COSMOS field, that allows us to accurately measure the star formation rate density (SFRD) of the galaxies in the protocluster. Here we provide one of the first observational tests of the prediction that the volume averaged star formation activity is more vigorous at higher redshifts in denser environments than in the field. In the future, this analysis will be expanded to an ensemble of structures detected in the C3VO survey (detailed in Section \ref{sec:specdata}).

This paper is organized as follows. In Section \ref{sec:data} we describe the data, Section \ref{sec:method} explains the details of analysis to map out the protocluster, Section \ref{sec:properties} describes the characteristics of the protoclusters and its peaks, Section \ref{sec:sfrd} describes the method to obtain the SFRD for the protocluster and coeval field, Section \ref{sec:results} describes the results, Section \ref{sec:discussion} is a discussion and lastly Section \ref{sec:summary} presents a summary and a discussion of future directions. We adopt a $\Lambda$CDM cosmology with $H_0 = 70$ km s$^{-1}$ Mpc$^{-1}$, $\Omega_{\Lambda}=0.73$ and $\Omega_M = 0.27$. For convenience, star formation rates are represented in units of $h_{70}^{-2}M_*$yr$^{-1}$ throughout the paper where $h_{70} \equiv H_0/70$ km$^{-1}$ s Mpc.

\section{Observations}
\label{sec:data}
In this section we describe the photometry and spectroscopy used in this study specifically in the subsection of the COSMOS field spanning from Right Ascension (RA) and Declination (Dec) range of $150.1\degree < \text{RA} < 150.48\degree$, $2.21\degree < \text{Dec} < 2.5\degree$. This subsection of the COSMOS field was chosen to encompass the entirety of the Taralay protocluster as mapped out by our C3VO observations.

 

\subsection{Photometric Data}

The Cosmic Evolution Survey (COSMOS) field \citep{2007ApJS..172...38S} is a large extragalactic survey that covers an area of about 2 deg$^2$ on the sky. It is one of the largest and most comprehensive multiwavelength surveys ever conducted. Relevant for this study, the survey includes data from a wide range of telescopes and instruments, including the \textit{Hubble Space Telescope}, the \textit{Spitzer Space Telescope}, the \textit{Galaxy Evolution Explorer} and ground-based telescopes such as the Subaru Telescope, the Canada France Hawai'i Telescope, and the Visible and Infrared Survey Telescope for Astronomy (VISTA) \color{black} (see \citealt{2007ApJS..172...99C}, \citealt{2016ApJS..224...24L}, \citealt{2022ApJS..258...11W} and references therein) \color{black}. These observations cover UV, optical, as well as near-infrared band-passes. Additional observations from the Very Large Array (VLA), the Chandra X-Ray Observatory, XMM-Newton, the Herschel Space Observatory cover the radio, X-ray and far-infrared wavelengths. Finally, the COSMOS field will be partially covered by the \textit{James Webb Space Telescope (JWST)} as a part of the COSMOS-Web program (\citealt{2022arXiv221107865C}). This makes this field an extremely valuable resource for studying the properties and evolution of galaxies over a wide range of cosmic epochs.

In this study, we utilize three different public photometric catalogs for the COSMOS field, each of which provide progressively deeper data. These catalogs are Capak+07 \citep{2007ApJS..172...99C} (hereafter C07), COSMOS2015 \citep{2016ApJS..224...24L} (hereafter C15), and COSMOS2020 \citep{2022ApJS..258...11W} (hereafter C20). Our primary source of photometry is the C20 catalog, which offers the deepest data in the COSMOS field \color{black} for over 1.7 million objects. \color{black} For this catalog we adopt the CLASSIC version (see below for more details). The choices used for masking and deblending are different for each catalog, and adopting a combined set of catalogs allows us to mitigate the effect of these choices. Our secondary source of photometry is the C15 catalog \color{black} containing more than half million objects \color{black} followed by C07 \color{black} that contains over 1.7 million objects. \color{black}. The basics of each catalog are summarized in Table \ref{tab:catalogdetails}. Below we briefly explain the properties of photometric data important for our science. 

As mentioned above, we primarily use photometry from C20 as it contains a wealth of data from X-ray to Radio. Of these we use optical/near-IR from Subaru Hyper Suprime-Cam (Subaru-HSC) \citep{2018PASJ...70S...1M} and VISTA InfraRed CAMera (VISTA-VIRCAM) \citep{2015A&A...575A..25S} and \textit{Spitzer} (\citealt{2004ApJS..154....1W}, \citealt{2007ApJS..172...86S}) surveys compared to the older catalogs mentioned above. The multiwavelength data ranges from far-UV from \textit{Galaxy Evolution Explorer (GALEX)} \citep{2005ApJ...619L...1M} to Infrared Array Camera (IRAC) channel 4 from \textit{Spitzer} (near-IR) yielding rest-frame wavelength coverage redward of 4000Å and Balmer breaks at $z \sim 4.57$. Such coverage is crucial for reliably recovering physical properties of galaxies such as stellar mass and star formation rates (e.g. \citealt{2014ApJ...786L...4R}, \citealt{2020ApJS..247...61F}). 

\begin{table}
    \centering
    \begin{threeparttable} 
        \caption{Properties of photometric catalogs used, such as the number of filters, the wavelength range and depth of data.}
        \label{tab:catalogdetails}
        \renewcommand{\arraystretch}{1.5}
        \begin{tabular}{lccr} 
            \hline
            Catalog & No. of Filters & Wavelength Range & Depth of Data\tnote{a} \\
            \hline
            COSMOS2020 & 40 & 1600 - 80000 Å & 25.7 in IRAC1 \\
            COSMOS2015 & 33 & 1600 - 80000 Å & 24.8 in IRAC1 \\
            Capak+2007 & 14 & 1600 - 45000 Å & \color{black} 24.8 in IRAC1 \color{black}\\
            \hline
        \end{tabular}
        \begin{tablenotes}\footnotesize
            \item[a] 3$\sigma$ depth of data in 3" aperture is directly reported for COSMOS2020 from \cite{2022ApJS..258...11W} and COSMOS2015 from \cite{2016ApJS..224...24L}. The Capak+2007 IRAC1 depth \color{black} reported here is 3$\sigma$ in 3" aperture that we converted from the 5$\sigma$ value reported in \cite{2007ApJS..172...99C}. \color{black}
        \end{tablenotes}
    \end{threeparttable}
\end{table}

The C20 photometry is processed in two separate catalogs, CLASSIC and FARMER, which use different methods. As mentioned earlier, we adopt the CLASSIC version for our study as it is consistent with the approach used by other catalogs in the COSMOS field. For this study, we corrected C20 catalog for Milky Way extinction as well as aperture correction using the code provided along with the catalog. We correct the C15 catalog for aperture correction, systematic offsets and Milky Way extinction using formulae 9 and 10 in \cite{2016ApJS..224...24L}. The C07 photometry contains optical and near-IR data in the COSMOS field. 
We exclude the IRAC channel 3 and 4 magnitudes in C07 from our analysis for the reasons described in \cite{2018A&A...615A..77L}. The issue seems to be limited to C07 photometry and we do use the data in these bands from C15 and C20. To correct the C07 photometry for zero point offsets, we subtract the zero point correction given in Table 13 of \cite{2007ApJS..172...99C} from the reported magnitudes in corresponding bands. \color{black} C20, C15 and C07, all three catalogs contain photometric redshifts in the redshift range of $0 < z < 6$. \color{black}

\subsection{Spectroscopic Data}
\label{sec:specdata} 
The spectroscopic data used in this study is taken from a variety of different surveys. We start with the VIMOS Ultra-Deep survey \citep{2015A&A...576A..79L} as it was through this survey that \cite{2018A&A...615A..77L} discovered the target of this work, the protocluster Taralay. Next we describe the Charting Cluster Construction with VUDS and ORELSE (C3VO) survey that re-targeted this structure. Lastly, we describe ancillary spectroscopic data from the DEIMOS 10k Spectroscopic Survey \citep{2018ApJ...858...77H} and the zCOSMOS Spectroscopic Survey that consists of zCOSMOS-Bright survey \citep{2007ApJS..172...70L, 2009ApJS..184..218L} and zCOSMOS-Deep survey (Lilly+ in prep, \citealt{2013ApJ...765..109D, 2015ApJ...802...31D}). 

\subsubsection{The VIMOS Ultra-Deep Survey}
\label{sec:vuds} 
The VIMOS Ultra-Deep Survey (VUDS) \citep{2015A&A...576A..79L} is a large spectroscopic redshift survey designed to study galaxies beyond $z \simeq 2$. With 640h of observing time on VIMOS spectrograph of Very Large Telescope (VLT), this survey targeted $\sim$ 10000  faint ($i_{AB} \sim 25$) galaxies over $2 < z \cong 6$ in COSMOS, the Extended Chandra Deep Field South (ECDFS) and the 02h field of the VIMOS Very Large Telescope Deep Survey 02h field (VVDS-02h), spanning a total of 1 deg$^2$ coverage. The VIMOS spectroscopic observations covered wavelength range of 3650 - 9350 Å spectroscopically confirming Ly$\alpha$ Emitters (LAEs) and Lyman Break Galaxies (LBGs) that do not exhibit Ly$\alpha$ in emission, \color{black} over $2 < z < 6$ \color{black} which resulted in a sample of galaxies roughly representative of the star-forming galaxy population at the epochs over the luminosity range of $0.3L^*_\text{UV} \leq L_\text{UV} \leq 3L^*_\text{UV}$ \citep{2022A&A...662A..33L}. Such a sample made it possible to study a range of overdense environments at different times in the history of the universe. 

Reliability flags were assigned to galaxies with spectroscopic redshifts ($z_\text{spec}$) that describe the confidence level of the assigned spectroscopic redshift being correct \citep{2015A&A...576A..79L}. These flags range from X0 to X4 with higher values of the second digit indicating greater confidence in $z_\text{spec}$. X varies from 0, 1, 2 or 3 where 0 denotes targeted galaxies, 1 denotes a type-1 AGN, 2 denotes a serendipitous detection that is separated in (projected) location from the target and 3 denotes a serendipitous detection that has the same apparent location as the target. Additionally, a flag of X9 is assigned to a galaxy whose spectrum shows a single spectral feature. While the X1 flag represents 41\% probability of the assigned $z_\text{spec}$ being correct, flags X2/X9 and X3/X4 indicate $\sim70\%$ and $\sim99.3\%$ probability of the assigned $z_\text{spec}$ being correct respectively. We refer the reader to \cite{2022A&A...662A..33L} for more details. We adopt flags X2, X3, X4 and X9 from the VUDS observations as secure\footnote{Secure flags are the flags that represent the probability of the assigned $z_\text{spec}$ being correct to higher than 70\%. For the VUDS, DEIMOS10k and zCOSMOS surveys the secure flags are X2, X3, X4, X9. For the C3VO survey, secure flags are 3 and 4 which represent $\geq 95$\% probability of the assigned $z_\text{spec}$ being correct.} flags.

We utilize a total of 709 objects from the VUDS survey over our adopted sky region: $150.1\degree < \text{RA} < 150.48\degree$, $2.21\degree < \text{Dec} < 2.5\degree$ and redshift range of $0 < z < 6$. 453 of these entries have secure flags.

The VUDS survey was monumental for the discovery of $\sim 50$ spectroscopic overdensities (e.g., \citealt{2014A&A...572A..41L, 2018A&A...615A..77L}, \citealt{2014A&A...570A..16C, 2018A&A...619A..49C}, \citealt{2021ApJ...912...60S}, \citealt{2023arXiv230715113F}, \citealt{shah2023identification}). Discovered by \cite{2018A&A...615A..77L}, Taralay protocluster is the highest redshift structure identified through the VUDS survey. 

\subsubsection{The C3VO Survey}
\label{sec:c3vo}

\begin{table*}
    \centering
    \begin{threeparttable}
        \caption{Observing details of the C3VO campaign targeting Taralay protocluster with DEIMOS on Keck. The details include mask details, total integration time, seeing, grating, order blocking filter and central wavelength.}
        \label{tab:observing}
        \renewcommand{\arraystretch}{1.5}
        \begin{tabular}{cccccccc} 
    	\hline
    	Mask & Center of the Mask & PA & Total Integration Time & Seeing\tnote{a} & Grating & Order Blocking Filter & $\lambda_{\rm{c}}$ \\
    	\hline
            dongN1 & 10:01:22:9, 2:21:41.6 & 90.0 & 3h30m & $\sim0.6-0.8"$ & 600 l mm$^{-1}$ & GG400 & 6500 Å\\
            dongS1 & 10:01:02.3, 2:17:20.0 & 90.0 & 4h35m & $\sim0.6-0.8"$ & 600 l mm$^{-1}$ & GG400 & 6500 Å\\
            dongD1 & 10:01:27.4, 2:19:27.0 & 50.0 & 5h20m & $\sim0.7-1.3"$ & 600 l mm$^{-1}$ & GG455 & 7200 Å \\
            dongD2 & 10:01:05.4, 2:22:49.9 & 50.0 & 6h & $\sim0.7-1"$ & 600 l mm$^{-1}$ & GG455 & 7200 Å \\
            dongA1 & 10:01:29.0, 2:21:10.0 & 25.0 & 4h59m & $\sim0.6-0.8"$ & 600 l mm$^{-1}$ & GG455 & 7200 Å\\
            dongA2 & 10:00:44.6, 2:16:08.7 & 78.0 & 4h & $\sim0.45-0.55"$ & 600 l mm$^{-1}$ & GG455 & 7200 Å \\
    	\hline
        \end{tabular}
        \begin{tablenotes}\footnotesize
            \item[a] No meaningful cloud coverage for the duration of observations for all masks.
        \end{tablenotes}
    \end{threeparttable}
\end{table*}

The Charting Cluster Construction with VUDS and ORELSE (C3VO) survey was devised as an extension to higher redshift of the ORELSE (the Observations of Redshift Evolution in Large Scale Environments) survey 
\citep{2009AJ....137.4867L}, aimed at studying a statistical sample of groups and clusters at $z \sim 1$. The C3VO survey is dedicated to mapping out the structures previously discovered through VUDS at $2 < z < 5$ in a manner consistent with that which ORELSE mapped out intermediate redshift structures. The survey was devised to both perform detailed studies of individual protoclusters and their populations and to statistically connect progenitor protoclusters to their descendent clusters. The main objective of C3VO is to better understand the relationship between stellar mass, star-formation, Active Galactic Nuclei (AGN) activity and local environmental density and in turn to probe the evolution of galaxies in large scale structures across cosmic times from $z = 5$ to $z = 0.6$.

C3VO is an ongoing campaign targeting protoclusters with the DEep Imaging Multi-Object Spectrograph (DEIMOS,  \citealt{2003SPIE.4841.1657F}) and the Multi-Object Spectrometer For Infra-Red Exploration (MOSFIRE,  \citealt{2008SPIE.7014E..2ZM}) on the Keck Telescopes, Simultaneous Color Wide-field Infrared Multi-object Spectrograph (SWIMS, \citealt{2012SPIE.8446E..7PK}) and Multi-Object InfraRed Camera and Spectrograph (MOIRCS,  \citealt{2006SPIE.6269E..16I}) on the Subaru Telescope, and Wide Field Camera 3 (WFC3, \citealt{2008SPIE.7010E..1EK}) $F160W$ imaging and $G141$ grism on the \textit{Hubble Space Telescope}. We focus on the Keck observations of the C3VO survey that are designed to further map out the six most prominently detected protocluster environments in VUDS at $2 < z < 5$, including Taralay at $z \sim 4.57$, and to target similar types of galaxies that do not have a spectroscopic redshift from rest frame UV surveys in the field. 

The Taralay protocluster at redshift $z \sim 4.57$ was targeted with 6 masks with DEIMOS in order to acquire restframe UV spectra of prospective member galaxies. The highest priority targets were limited to $i_{AB} \leq 25.3$ in order to obtain continuum redshifts, corresponding to $L^*_\text{FUV}$ at $z \sim 4.5$ where $L^*$ is the characteristic luminosity (\citealt{2007ApJ...670..928B}). We also included fainter objects by extending the limit to $i_{AB} \approx 26.7$ to acquire redshifts from Ly$\alpha$ emission. The observing details for each mask can be found in Table \ref{tab:observing}. 

For this campaign, the targets were selected using photometric redshifts ($z_\text{phot}$) from C15 catalog. The targeting priorities and the selection criteria used for the first two masks are described in \cite{2022A&A...662A..33L}. The same targeting priorities and the selection criteria were largely used for the four remaining masks as well. 
\color{black} In addition to the original targeting, our last two DEIMOS masks, dongA1 and dongA2, included targets from the UV-selected Atacama Large Millimeter Array (ALMA) Large Program to Investigate C$^+$ at Early Times survey (ALPINE-ALMA, \citealt{2020ApJS..247...61F}; \citealt{2020A&A...643A...1L}; \citealt{2020A&A...643A...2B}). The ALMA observations of these targets revealed close companions with [CII] 158$\mu$m emission (see \citealt{2020A&A...643A...1L} and \citealt{2020A&A...643A...7G}). These companions lacked rest-frame UV spectral information and were targeted by DEIMOS in an attempt to recover that information. \color{black}


With the total integration time of approximately 28 hours for all masks (the time per mask is given in Table \ref{tab:observing}), we obtained 204 secure redshifts. 44 of these galaxies are in the fiducial redshift range of Taralay, $4.48 < z < 4.64$ (see Section \ref{sec:pc}), making them potential Taralay protocluster members\footnote{The term potential member is used here due to these galaxies satisfying two out of three criteria adopted in this work to define true membership of the Taralay protocluster. The two criteria are: 1) the RA, Dec range of $150.1\degree < \text{RA} < 150.48\degree$, $2.21\degree < \text{Dec} < 2.5\degree$ and 2) redshift range of $4.48 < z < 4.64$. This potential member sample will be further refined by the third criterion used to define true membership, an overdensity ($\sigma_\delta$) cut (see Section \ref{sec:pc}). The total number of galaxies that satisfy the above two criteria as well as the number of galaxies that ultimately qualify as true protocluster members are given in Table \ref{tab:uniquegals}.}. These galaxies have $\geq 95\%$ probability of the assigned $z_\text{spec}$ being correct. The C3VO-DEIMOS data were reduced following the method described in \cite{2022A&A...662A..33L} using a modified version of spec2D (\citealt{2012ascl.soft03003C}, \citealt{2013ApJS..208....5N}). A modified version of the \textsc{zpsec} tool was used to assign redshifts by two independent users with a flagging code that is similar to that of the DEEP2 redshift survey (\citealt{2003SPIE.4834..161D}, \citealt{2013ApJS..208....5N}). Secure flags -1, 3 and 4 were assigned where the flag -1 denotes a star whereas flags 3 and 4 denote $\geq 95\%$ probability of the assigned $z_\text{spec}$ being correct for a galaxy. For this study, we adopt flags 3 and 4 as being secure extragalactic redshifts.  

Having the secure redshifts for 44 galaxies makes this structure one of the most extensively studied at such high redshifts. Based on the additional redshifts from the C3VO campaign, we adopt the redshift range of $4.48 < z < 4.64$ for the Taralay protocluster, one of the criteria that we use to define this structure. 
Overall, we observed 716 objects with C3VO-DEIMOS for this study over our adopted sky region and in the redshift range of $0 < z < 6$, including serendipitous detections, of which 204 have secure extragalactic redshifts. 

\subsubsection{Ancillary Spectroscopic Data}

Some of the ancillary spectroscopic data for this project comes from the DEIMOS 10k Spectroscopic Survey (\citealt{2018ApJ...858...77H}). For this survey $10,718$ objects were observed with DEIMOS in the COSMOS field over the redshift range $0 < z < 6$. Two or more spectral features were observed for $6617$ objects, while 1798 objects have spectra with a single spectral feature that was consistent with the photometric redshift. The magnitude distribution of objects targeted in this survey peaks at $I_{AB} \sim 23$. We utilized a total of 1161 entries from the DEIMOS10k catalog over our adopted sky region, with redshifts ranging from $0 < z < 6$ for this study. 840 of these entries have secure quality flags. Some of these objects show broad lines in their spectrum. This information is represented by adding 10 to their quality flags, i.e., 11-14, 19.

The majority of the ancillary spectroscopic data is obtained from the zCOSMOS Spectroscopic Survey, a large VLT/VIMOS redshift survey in the COSMOS field, with the vast majority of the galaxies in the redshift range $0 < z < 3$. This survey consists of zCOSMOS-Bright survey \citep{2007ApJS..172...70L, 2009ApJS..184..218L} and zCOSMOS-Deep survey (Lilly+ in prep, \citealt{2013ApJ...765..109D, 2015ApJ...802...31D}). The zCOSMOS-Bright survey spans 1.7 deg$^2$ COSMOS ACS field, targeting 20,000 galaxies over $0.1 \leq z \leq 1.2$ range with $I_{AB} < 22.5$. The zCOSMOS-Deep survey spans central 1 deg$^2$ and targets 10,000 galaxies over $1.4 \leq z \leq 3$ that were selected through color-selection criteria. The flagging system for zCOSMOS survey is effectively the same as that of VUDS survey. We used an updated version of the zCOSMOS catalog provided by one of the authors (DK) that changed the assigned redshift for a small number of entries ($\sim$1\% of the secure spectral redshifts) and revaluated the assigned flags. Some of the entries that were previously assigned flags X2 and lower or X9 received a demoted flag. This catalog was previously used in \cite{2022ApJ...925...82K} and will be described more fully in an upcoming paper. We utilize a total of 3637 entries in our fiducial region of interest, with 2088 of these entries having secure flags (i.e., X2, X3, X4, X9). In addition, we include a small number of galaxies from \cite{2015ApJ...808L..33C} and \cite{2015ApJ...808...37C}.

\begin{figure}
    \centering
    \includegraphics[width = 0.47\textwidth]{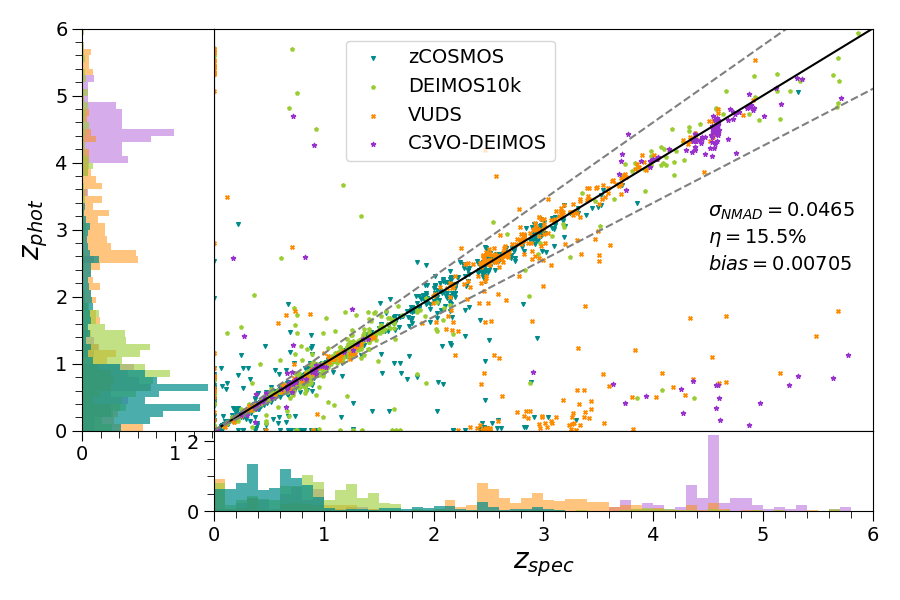}
    \caption{Comparison of photometric redshifts, $z_\text{phot}$ from COSMOS2020, COSMOS2015 catalogs and Capak+2007 photometry with spectroscopic redshifts obtained from C3VO, VUDS, DEIMOS10k, zCOSMOS surveys. The galaxies for which this comparison is performed include serendipitous C3VO detections and have either quality flags of 3, 4 (C3VO) or X2-X4 and X9 (VUDS, DEIMOS10k, zCOSMOS) ensuring high confidence level in the assigned $z_\text{spec}$. See Section \ref{sec:specdata} for the discussion about quality flags. The $z_\text{phot}$ and $z_\text{spec}$ histograms are normalized such that the total area under the histogram sums up to 1; so the height of each bar in the histogram corresponds to the probability density rather than the count of redshifts in that bin. The $\sigma_\text{NMAD}$, catastrophic outlier rate ($\eta =  |z_\text{spec} - z_\text{phot}|/(1+z_\text{spec}$) > 0.15), and bias which is defined as the median of the difference between $z_\text{phot}$ and $z_\text{spec}$ are shown in the main panel.}
    \label{fig:specdata}
\end{figure}

The VUDS and zCOSMOS flagging system has well measured reliabilities and form the basis of our statistical framework. To establish whether the confidence intervals for DEIMOS10k flags are the same as those of VUDS and zCOSMOS surveys, we compare their flagging system. This check is important because in our statistical framework that combines spectroscopic and photometric redshifts (explained in Section \ref{sec:MC}), the spectroscopic redshifts are handled according to their quality flags. 
\color{black} To make this check, we examine a sample of 70 DEIMOS10k targets with flag=X2/X9 which were observed in other surveys and assigned quality flag=X3/X4 therein. \color{black} This process involves assessing the catastrophic outlier rate, which is calculated as $|z_\text{other} - z_\text{DEIMOS10k}|/(1+z_\text{other}) > 5\sigma_\text{NMAD}$ where NMAD is the normalized median absolute deviation of $z_\text{other} - z_\text{DEIMOS10k}/(1+z_\text{other})$. Out of 70 objects, we find that 15 have a catastrophic outlier rate greater than 5$\sigma_\text{NMAD}$, resulting in a reliability of 79.6\%. We repeat this process for 751 DEIMOS10k targets with flag=X3/X4, comparing them with spectroscopic redshifts from another survey also with flag=X3/X4. In this case, we identify 38 objects with a catastrophic outlier rate greater than 5$\sigma_\text{NMAD}$, corresponding to a reliability of 94.9\%. These results broadly align with the flagging system employed by the VUDS and zCOSMOS surveys, indicating that the DEIMOS10k survey shares the same reliability.

\begin{figure*}
    \centering
    \includegraphics[width = 1\linewidth]{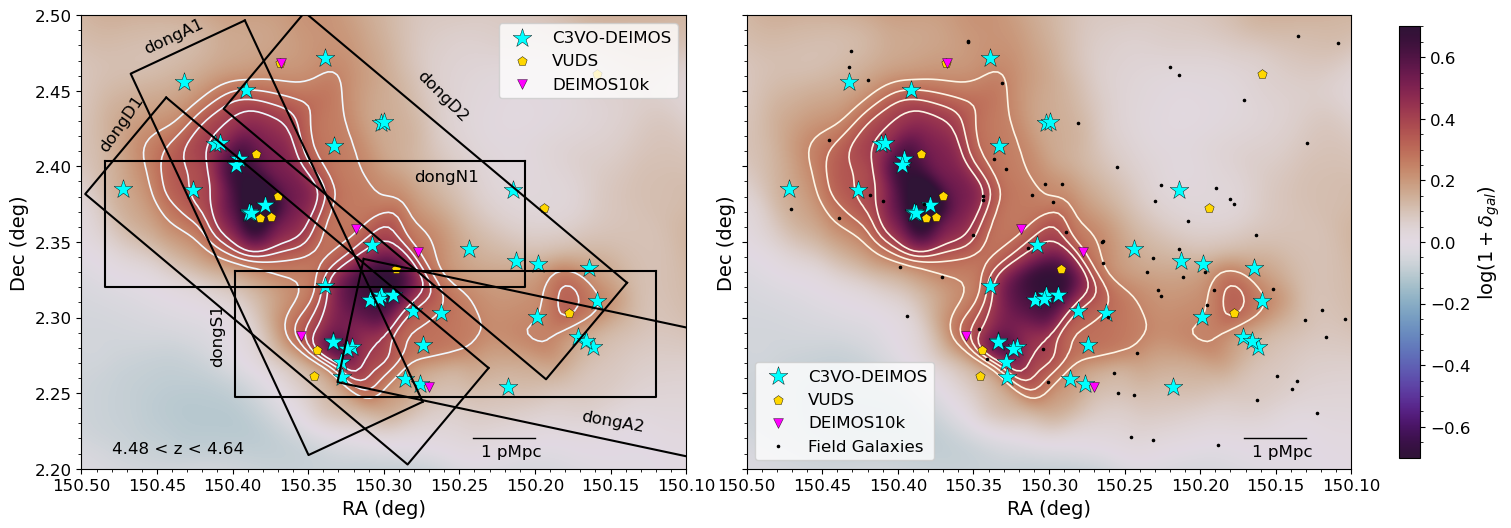}
    \caption{Both the panels of this figure show a sky plot for the Taralay protocluster at $z\sim4.57$. The cyan stars, yellow pentagons and fuchsia inverted triangles are spectroscopically-confirmed potential members (see Section \ref{sec:c3vo} for the definition of potential members) of Taralay observed as a part of C3VO survey, VUDS survey and DEIMOS10k survey, respectively. The fiducial redshift range for the structure, $4.48 < z < 4.64$ is indicated at the bottom left corner. The black bar indicating 1 proper Mpc towards the right corner roughly corresponds to $0.04^\circ$. The color-scale varies with overdensity $\log(1+\delta_\text{gal})$ such that the darker the region, the denser is the environment at those sky coordinates. In the left panel, the black rectangles laid on the protocluster are six masks that we observed this structure with, detailed in Table \ref{tab:observing}. The right panel of this figure shows unique field galaxies from C3VO, VUDS, DEIMOS10k surveys in the redshift range of $4.2 < z < 4.48$ and $4.64 < z < 4.93$. We note that the outermost contour that roughly marks the shape of Taralay in both panels of this figure does not correspond to the $\sigma_\delta \geq 2$ boundary of Taralay in the right panel of Figure \ref{fig:vol} due to the difference in methods with which these two figures are constructed.}
    \label{fig:skyplot}
\end{figure*}

In Figure \ref{fig:specdata}, we show a comparison of the $z_\text{spec}$ from the surveys mentioned above and $z_\text{phot}$ from C20, C15 and C07 that are utilized in this study. The galaxies included for this comparison are brighter in IRAC channel 1 and/or 2 than the completeness limits  given in \protect\cite{2022ApJS..258...11W} \color{black} (i.e., 25.7 and/or 25.6 in the [3.6] and [4.5] bands, respectively) \color{black} in order to mitigate the effects of Malmquist bias \citep{1922MeLuF.100....1M, 1925MeLuF.106....1M}. This cut is also made for our entire spectroscopic and photometric sample. 

When combining data from these surveys, entries need to be resolved. This process of identifying and resolving duplicates is explained in Section \ref{sec:getphot}. The number of spectroscopically confirmed unique galaxies, before making an IRAC channel 1 and/or IRAC channel 2 cut, that fall within the range of the protocluster redshift range, $z_{pc} = 4.48 < z < 4.64$ (see Section \ref{sec:pc}), and the spectroscopically confirmed unique galaxies that fall within the field redshift range $z_\text{field} = 4.2 < z < 4.48, 4.64 < z < 4.93$ (see Section \ref{sec:pc}) are listed in Table \ref{tab:uniquegals}. The unique entries in $z_{pc}$ range are shown in both panels of Figure \ref{fig:skyplot}. In addition, the left panel of this figure also shows the DEIMOS masks given in Table \ref{tab:observing} and used for targeting the Taralay protocluster in the C3VO campaign. The right panel of the figure additionally shows unique entries in $z_\text{field}$ range. 

Applying the conditions outlined in \ref{sec:pc} on these unique entries gives us the final protocluster and the field sample given in Table \ref{tab:uniquegals}. All galaxies, except the ones from C3VO survey, have quality flags of X2, X3, X4, or X9 indicating the high confidence in their redshift measurements. The galaxies included from C3VO survey have quality flags of 3 and 4 indicating more than $95\%$ confidence in the assigned spectroscopic redshift. \color{black} As such,  the flagging systems of C3VO, VUDS, DEIMOS10k, and zCOSMOS are all compatible with each other in terms of the likelihood of a spectroscopic redshift of a given flag to be correct. This homogeneity is crucial when we utilize the spectroscopic information in the various Monte Carlo processes in our analysis (see next section). \color{black} Overall we utilize a \color{black} catalog with total 6629 spectral entries. After undergoing the IRAC cut described earlier ([3.6]$<$25.7 and/or [4.5]$<$25.6), there are 115 and 3169 unique secure spectral redshifts over the redshift ranges of $4.2 < z < 4.93$ and $0 < z < 6$, respectively, drawn from all surveys. \color{black}

\begin{table*}
    \centering
    \begin{threeparttable}
        \caption{
        \color{black} The table shows the total number of secure spectroscopic redshifts ($z_\text{spec}$) from the four surveys that satisfy the various criteria we impose throughout this study. 
        \color{black}}
        \label{tab:uniquegals}
        \renewcommand{\arraystretch}{1.5}
        \begin{tabular}{ccccccc}
    	\hline
    	Survey & Total Entries in $z_{pc}$\tnote{a} & Unique Entries in $z_{pc}$\tnote{b} & PC Members\tnote{c} & Total Entries in $z_\text{field}$\tnote{d} & Unique Entries in $z_\text{field}$\tnote{e} & Field Galaxies\tnote{f}\\
    	\hline
            C3VO-DEIMOS & 44 & 44 & 34 & 46 & 46 & 41\\
            VUDS & 16 & 11 & 7 & 16 & 15 & 13\\
            DEIMOS10k & 15 & 7 & 2 & 24 & 15 & 13\\
            zCOSMOS & 22 & 0 & 0 & 27 & 0 & 0\\
    	\hline
        \end{tabular}
        \begin{tablenotes}
        \color{black}
            \item[a] The total number of secure $z_\text{spec}$ galaxies that fall in the protocluster redshift range, $z_{pc} = 4.48 < z < 4.64$.
            \item[b] The unique secure $z_\text{spec}$ entries from each survey in $z_{pc}$ range.
            \item[c] The number of secure $z_\text{spec}$ galaxies from each survey that satisfy the final criterion for being a protocluster member where the final criterion is residence in a region that is $\sigma_\delta \ge 2$ overdense than the field and also contains a $\sigma_\delta \ge 5$ peak.
            \item[d] The total number of secure $z_\text{spec}$ galaxies that fall in the field redshift range, $z_\text{field} = 4.2 < z < 4.48, 4.64 < z < 4.93$.
            \item[e] The unique secure $z_\text{spec}$ entries from each survey in $z_\text{field}$ range.
            \item[f] The final number of secure $z_\text{spec}$ field galaxies. We exclude any galaxy as a field galaxy if it falls in a protocluster-like structure, i.e, a region that is $\sigma_\delta \ge 2$ overdense than the field and also contains a $\sigma_\delta \ge 5$ peak.
        \end{tablenotes}
    \end{threeparttable}
\end{table*}

\section{Methods}
\label{sec:method}


The primary objective of this study is to examine the rate of in-situ stellar mass growth in 
the Taralay protocluster at $z \sim 4.57$ compared to that of the field. We use the spectral energy distribution (SED) fitting method (explained in Section \ref{sec:sed}) to determine the physical properties such as stellar mass, star formation rate for all galaxies. However, before we can apply the SED fitting, it is necessary to assemble a catalog of high redshift galaxies with $z_\text{spec}$ and find their photometric counterparts. We explain this procedure in Section \ref{sec:getphot}. Though we have extensive spectroscopy, the galaxies with $z_\text{spec}$ may not be representative of the full underlying galaxy population and the galaxies without spectral redshifts far outnumber those with $z_\text{spec}$. To utilize all the data available, we create a a framework, explained in Section \ref{sec:MC}, that statistically combines $z_\text{spec}$ and $z_\text{phot}$. We then use the Voronoi Tessellation Monte Carlo (VMC, \citealt{2017MNRAS.472..419L, 2018A&A...615A..77L, 2022A&A...662A..33L}, \citealt{2017MNRAS.472.3512T}, \citealt{2018A&A...619A..49C}, \citealt{2020MNRAS.491.5524H}) map technique which relies on this statistical framework to reconstruct the density field in Section \ref{sec:densmap}. The reconstruction of the density field allows for an estimation of the local overdensity, map out the structure and get the volume and mass of Taralay. We then use the overdensity information to update the criteria for a galaxy to qualify as the protocluster or field galaxy in Section \ref{sec:pc}. 


\subsection{Catalog Matching Procedure}
\label{sec:getphot} 

To identify the photometric counterparts for the spectroscopic data, an angular separation within which a match can be found is determined based on the sky error\footnote{The sky error represents the potential discrepancy between the measured coordinates of an astronomical source and its true celestial coordinates. The sky error can arise from various factors, including instrumental limitations, atmospheric effects, and the accuracy of the astrometric calibration.}. To choose the matching radius appropriate given the unknown sky error associated with all the catalogs, we start by matching with a radius of 1$\arcsec$ and look at the distribution of the angular separation of the closest match between galaxies with $z_\text{spec}$ and the photometric counterparts they matched to. A local minimum was observed at 0.3$\arcsec$, which strongly indicates that the matches at distance of $< 0.3\arcsec$ are likely genuine and the matches that are at $> 0.3\arcsec$ distances are likely contaminated with impurities. 

Starting with the spectroscopic data from C3VO, VUDS, DEIMOS10k and zCOSMOS surveys within our adopted sky region, we first look for photometric counterparts in C20 which contains the deepest data to date that covers the entirety of the COSMOS field. If there are no photometric counterparts within a circle of $0.3\arcsec$ radius for a $z_\text{spec}$ entry, the search is expanded to include the C15 and C07 catalogs.

Upon conducting this search for \color{black} the 6229 total entries in our spectral catalog 
\color{black} over \color{black} a redshift range of $0 < z < 6$, we found that C20 has \color{black} at least one \color{black} counterparts for 89.4\% \color{black} of the total $z_\text{spec}$  entries, \color{black} with 7.1\% \color{black} of the total $z_\text{spec}$entries \color{black} matching to two photometric sources and approximately 0.4\% of \color{black} entries \color{black} matching to more than two photometric sources. C15 counterparts were found for the 52.93\% of the remaining 614 $z_\text{spec}$ \color{black} entries \color{black} (5.22\% of the total $z_\text{spec}$ \color{black} entries \color{black}), with 2.60\% \color{black} entries \color{black} matching to two photometric sources and 0.16\% of \color{black} entries \color{black} matching to more than two photometric sources. Finally, of the 280 $z_\text{spec}$ \color{black} entries \color{black} that had no counterparts in C20/C15, C07 photometry counterparts were identified for 51.78\% of the remaining $z_\text{spec}$ \color{black} entries \color{black} (2.32\% of the total $z_\text{spec}$ \color{black} entries \color{black}), 0.71\% of \color{black} entries \color{black} matching to two photometric sources. No photometric counterparts were found for 135 (2.17\% of the total $z_\text{spec}$) \color{black} entries \color{black}.

In situations where there are multiple $z_\text{spec}$ \color{black} entries \color{black} that correspond to a single $z_\text{phot}$ counterpart, a set of rules was established to choose the most likely counterpart and avoid duplicates. These rules are as follows: 

\begin{enumerate}
    \item If the $z_\text{spec}$ \color{black} entries \color{black} have different redshift quality flags, the galaxy with the most secure redshift quality flag is given priority.  

    \item If the $z_\text{spec}$ \color{black} entries \color{black} come from different spectroscopic surveys but have the same redshift quality flags, then the priority order is as follows: C3VO-DEIMOS entries, then VUDS entries, followed by DEIMOS10k entries, and lastly zCOSMOS entries. For flags X2, X9, the priority order is VUDS observations, then DEIMOS10k observations, followed by zCOSMOS survey observations, and lastly C3VO-DEIMOS observations, as lower quality flags from C3VO were considered unreliable.

    \item zCOSMOS-Deep was prioritized over zCOSMOS-Bright when there were only zCOSMOS entries and the flags and $z_\text{spec}$ were effectively identical.
\end{enumerate}

We also looked at some cases by eye, where both the matches had identical $z_\text{spec}$ values and very similar quality flags, to confirm that these rules result in a match that is the most sensible in each case. By following these rules, the appropriate match is selected and duplicates are resolved when multiple spectroscopic redshifts are associated with a single photometric counterpart.

We conducted a comparison test between our selection method and an alternative method developed by \cite{2023arXiv230715113F} to verify the accuracy of our choice of a 0.3" radius for identifying the correct $z_\text{phot}$ counterpart for each $z_\text{spec}$ \color{black} entry. \color{black} This alternative method involves finding the nearest match for each $z_\text{spec}$ \color{black} entry \color{black}, updating the coordinates of the $z_\text{spec}$ \color{black} entries \color{black} by calculating the median positional offsets between the spectroscopic survey and photometric catalog, and determining new $z_\text{phot}$ counterparts using the updated coordinates. This is necessary to account for differences in astrometry between the catalogs.

After resolving duplicates, we found that our selection method agreed with the alternative method for 99.3\% of the $z_\text{spec}$ \color{black} entries \color{black}. For the remaining $\sim0.7\%$ of $z_\text{spec}$ \color{black} entries \color{black}, the alternative method provided better matches outside the 0.3$\arcsec$ radius. For those $\sim0.7\%$ $z_\text{spec}$ \color{black} entries \color{black}, the matches resulting from the alternative method were adopted. 

In addition to the spectroscopic data described in Section \ref{sec:specdata}, we take a comprehensive approach of incorporating all of the available photometric data for our study. This includes 36232 entries from the C20 catalog within our adopted sky region that survived the IRAC cut. By incorporating this additional data, we can leverage the extensive photometric information contained in the C20 catalog to complement and enrich our analysis. The completed master catalog contains different types of galaxies, including those with spectroscopic redshifts ($z_\text{spec}$) from various surveys, their photometric counterparts with associated $z_\text{phot}$ values, and galaxies with photometric information. 

\subsection{Statistical Framework with Monte Carlo}
\label{sec:MC}

\color{black} As stated earlier, we utilize all the available data, objects with $z_\text{spec}$ as well as $z_\text{phot}$ for this analysis since the galaxies with $z_\text{spec}$ may not be representative of the full underlying galaxy population. \color{black} 
To establish our framework that statistically combines $z_\text{spec}$ and $z_\text{phot}$, we refer to the statistical model described in Appendix A of \cite{2022A&A...662A..33L} and perform Monte Carlo on redshifts for 100 iterations. We begin with the output catalog from Section \ref{sec:getphot} with combined spectroscopic and photometric data. The statistical framework described below is necessary to map overdense structures using the density mapping technique (Section \ref{sec:densmap}), to perform SED fitting (Setion \ref{sec:sed}) and obtain the SFRD of the Taralay protocluster (Section \ref{sec:sfrtosfrd}).

For each Monte Carlo realization of a galaxy we sample a value, $\xi$, from likelihood probability density function based on the spectral quality flag, with the likelihood representing the chance that the spectroscopic redshift is correct. The likelihood values and their associated uncertainties for quality flags are given in Appendix A of \cite{2022A&A...662A..33L}. Next, we sample a value, $\chi$, from the uniform distribution between 0 and 1. If $\chi < \xi$, $z_\text{spec}$ is retained as the redshift for that galaxy in that iteration. If  $\chi \geq \xi$, then we assign a $z_\text{phot}$ drawn from the redshift probability density function (zPDF) constructed as an asymmetric Gaussian using the median and photometric redshift confidence intervals given in the catalog of the photometric counterpart. The authors of the photometric catalogs have derived the confidence intervals by performing SED fitting using the code LePhare (\citealt{2002MNRAS.329..355A}, \citealt{2006A&A...457..841I}).

For photometric objects, with no measured $z_\text{spec}$, we sample an asymmetric Gaussian distribution based on the zPDF to obtain redshifts instead of relying on the $z_\text{phot}$ assigned to them. The final master catalog encompasses galaxies with $z_\text{spec}$ from different surveys, their photometric counterparts with $z_\text{phot}$, and galaxies with $z_\text{phot}$ information but no spectroscopy, all satisfying $4.2 < z < 4.93$ redshift range as that is our redshift range of interest (see Section \ref{sec:pc}). 


\subsection{Density Mapping and the Size of Taralay}
\label{sec:densmap}

\begin{figure*}
    \centering
    \includegraphics[width = 1\linewidth]{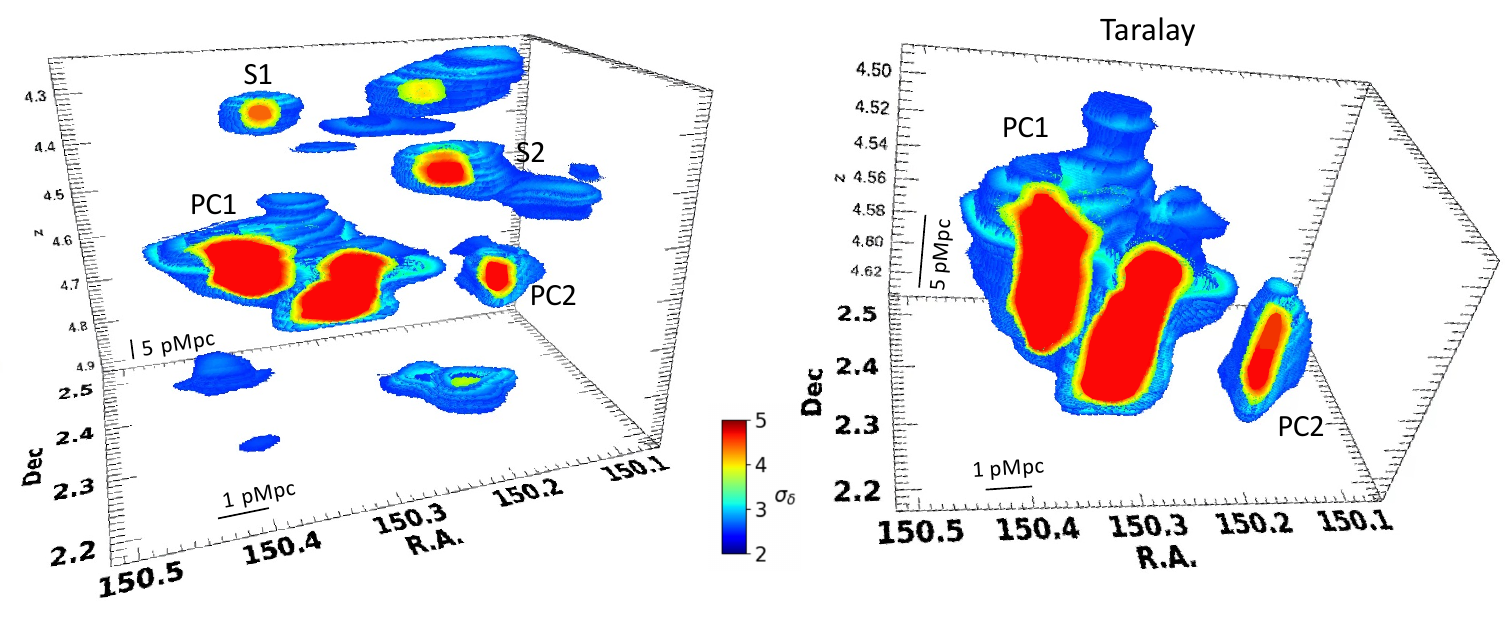}
    \caption{Reconstructed galaxy overdensity using the VMC mapping technique. The panel on the left shows the Taralay protocluster and other overdensities within the volume that contains the coeval field sample. The blue regions are bounded by \color{black}over\color{black}density isopleths of $\sigma_\delta \geq 2$. The right panel of the figure shows the 3D structure of only the Taralay protocluster. \color{black}The red regions have an overdensity of $\sigma_\delta \geq 5$ relative to the field, whereas the blue regions set the $\sigma_\delta \geq 2$ boundary. \color{black} The different colors and their corresponding overdensity values are shown by the color scale which represents the galaxy overdensity. \color{black} Scale bars representing proper distances of varying size are shown in each panel.\color{black}}
    \label{fig:vol}
\end{figure*}

To map out Taralay and measure the underlying density field, we use the Voronoi Tessellation Monte Carlo (VMC) map technique. The VMC method is a statistical approach that employs Voronoi tessellation to estimate the density of galaxies in a given region of the sky, but does so over a large number of Monte Carlo realizations of the input data. Voronoi tessellation divides space into polygons around each galaxy, with each polygon containing all the points in space that are closer to that galaxy than any other. The area of each polygon is inversely proportional to the local density of galaxies around the corresponding galaxy. The VMC technique is explained in great detail in \citet{2017MNRAS.472..419L, 2018A&A...615A..77L, 2019MNRAS.490.1231L, 2022A&A...662A..33L}, \citet{2017MNRAS.472.3512T,  2019MNRAS.484.4695T}, \citet{2017MNRAS.472..998S, 2018MNRAS.477.4707S, 2019MNRAS.484.2433S, 2021ApJ...912...60S}, \cite{2018A&A...619A..49C}, \cite{2019MNRAS.482.3514P}, \cite{2020MNRAS.491.5524H}, Hung et al in prep. We adopt the version outlined in \cite{2022A&A...662A..33L} for this work. As a result of this process we get a 3D cube with a local overdensity at every single voxel which is defined as
\begin{equation}
    \log\left(1 + \delta_{gal}\right) = \log\left(1 + \frac{\left(\Sigma - \Tilde{\Sigma}\right)}{\Tilde{\Sigma}}\right)
\end{equation}
where $\delta_{gal}$ is the local galaxy overdensity. We can then assign $\log\left(1 + \delta_{gal}\right)$ values to galaxies by tethering a galaxy to \color{black} the voxel which contains the RA, Dec, and redshift of the galaxy. \color{black}

We fit the distribution of $\log\left(1 + \delta_{gal}\right)$ with a Gaussian for each redshift slice of depth 7.5 pMpc and obtain its $\mu$ and $\sigma$. The $\sigma$\color{black}(z) \color{black} and $\mu$\color{black}(z) \color{black} are then fitted as a function of redshift with a fifth order polynomial from which we obtain $\mu_\delta$ and $\sigma_\delta$, \color{black} where 
\begin{equation}
    \sigma_\delta = \frac{\log(1+\delta_{gal}) - \mu(z)}{\sigma(z)}
\end{equation}
Both $\mu(z)$ and $\sigma(z)$ are in units of $\log(1+\delta_{gal})$. Over $4.2 < z < 4.93$, $\mu(z)$ is $\sim 0$ and $\sigma(z) \sim 0.1$ . 
\color{black} This methodology is explained in more detail in \cite{2018A&A...619A..49C}. For the rest of this paper, we adopt $\sigma_\delta$ to describe the overdensity.

By plotting the regions above a certain $\sigma_\delta$ we can visualize the density field and map out the protocluster as well as any additional structures that may exist along the line of sight (LoS) as shown in Figure \ref{fig:vol}. The left panel in this figure shows all overdense structures with $\sigma_\delta \geq 2$ along the LoS. The right panel of the figure shows only the protocluster Taralay. The full distribution of the $\sigma_\delta$ values across the redshift range of interest is shown in Figure \ref{fig:zN}. The resulting underlying galaxy density field is converted into a matter density field using a bias factor (see Appendix \ref{sec:overdensappen} for more discussion). For this study, we use a bias factor of $b=3.6$, which is based on previous works (\citealt{2013ApJ...779..127C}, \citealt{2018A&A...612A..42D}). 

\subsection{Definition of Field and Protocluster sample}
\label{sec:pc}

Here we outline the full criteria used to define our protocluster and the coeval field. To ensure a fair and unbiased comparison of the SFRD between the protocluster and the coeval field, we define the boundaries of the region for analysis as a rectangular area around all 6 masks used on DEIMOS to target Taralay as a part of the C3VO campaign. This approach helps us mitigate any unforeseen selection biases that could affect the comparison. These boundaries are the aforementioned sky region of interest: $150.1\degree < \text{RA} < 150.48\degree$ and $2.21\degree < \text{Dec} < 2.5\degree$.

We define the Taralay protocluster as a region with $\sigma_\delta \geq 2$ that also contains a $\sigma_\delta \geq 5$ peak within the redshift range of $4.48 < z < 4.64$. At the time of discovery, \cite{2018A&A...615A..77L} established the redshift range of Taralay to be $4.53 < z < 4.6$. We extend this original redshift range to include all galaxies in the full extent of the $\sigma_{\delta} \ge 2$ voxels associated with Taralay. The left panel of Figure \ref{fig:vol} shows two other structures along the line of sight that also satisfy the conditions for being a protocluster, i.e. $\sigma_\delta \geq 2$ region with a $\sigma_\delta \geq 5$ peak. These are denoted S1 and S2. Both of these structures are outside the redshift range $4.48 < z < 4.64$ and so do not comprise part of Taralay. The properties of these two protocluster candidates are briefly discussed in Section \ref{sec:properties}. 

For the coeval field, we aim to select field galaxies such that the sample has a similar average redshift to the systemic redshift of the protocluster. To achieve this, we establish a redshift range that encompasses a temporal window of $\pm 100$ Myr around the systemic redshift of the protocluster. The reason for this choice is the expectation that, due to the relatively short time span, the properties of galaxies in the field sample will be comparable to those of member galaxies within the protocluster, as their evolution may not have diverged significantly. All galaxies in the redshift range $4.2 < z < 4.48$ and $4.64 < z < 4.93$ are considered field galaxies except for those in S1 and S2. We exclude these two structures from our definition of the coeval field because they resemble protoclusters and may confuse our analysis. The properties of these two potential protoclusters are given in Table \ref{tab:2struc}.

\begin{figure}
    \centering
    \includegraphics[width = 0.45\textwidth]{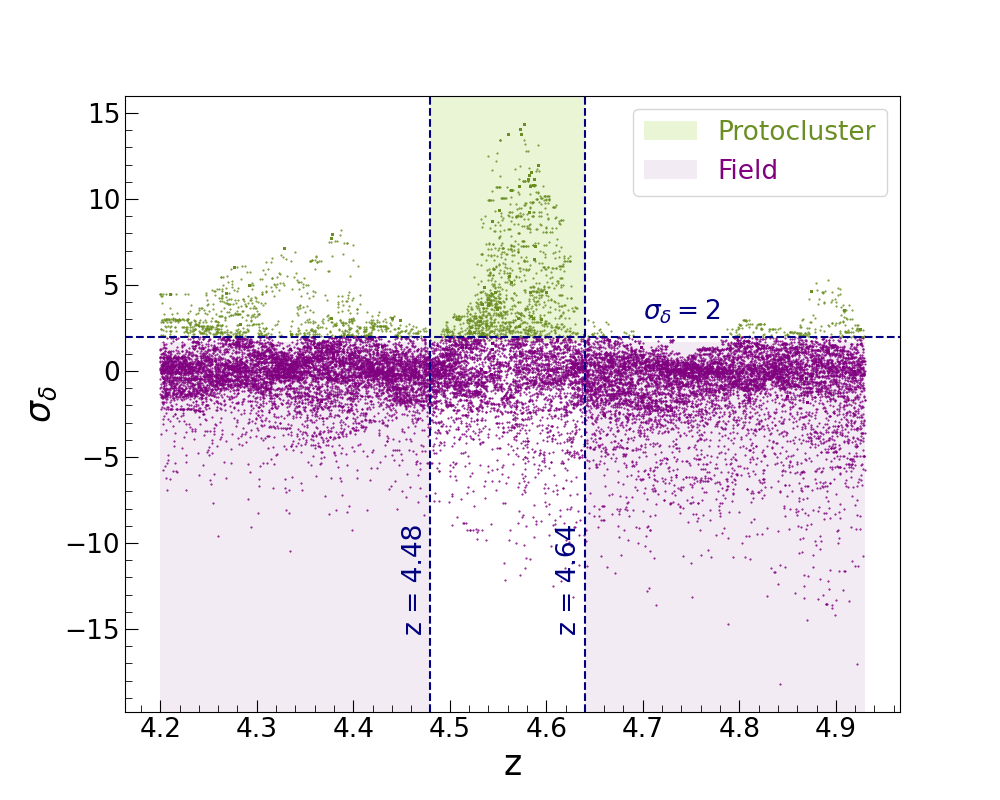}
    \caption{This figure shows the spread of $\sigma_\delta$ \color{black} (overdensity) \color{black} values associated with each galaxy that forms the final master catalog across all MC iterations (see Section \ref{sec:MC}). The green dots represent galaxies with $\sigma_\delta \geq 2$ while the purple dots represent galaxies with $\sigma_\delta \leq 2$. The green region, bounded by $4.48 < z < 4.64$ and $\sigma_\delta \geq 2$, represents galaxies that qualify as protocluster members. The two purple shaded regions show galaxies that qualify as coeval field galaxies. For more discussion, see Section \ref{sec:pc}. 
    }
    \label{fig:zN}
\end{figure}

The galaxies that make up the coeval field sample are shown in Figure \ref{fig:zN}. The galaxies in the purple regions, bounded by $4.2 < z < 4.48$, $4.64 < z < 4.93$ and $\sigma_\delta = 2$ make up most of the field galaxy sample. The field galaxies that are part of the $\sigma_\delta \geq 2$ overdensities as well as the galaxies that are neither field galaxies or protocluster members i.e., the galaxies in S1 and S2, are shown by green points on either side of the green region. We see a clear structure emerging in the green region within $4.48 < z < 4.64$ and $\sigma_\delta \geq 2$. The galaxies in this region are the protocluster galaxies. The galaxies in $4.48 < z < 4.64$ with $\sigma_\delta \leq 2$ are excluded from our field given their close proximity to the protocluster.


\begin{table}
    \centering
    \caption{Location and overdensity conditions satisfied by galaxies in order to quality as protocluster members, protocluster peak members or coeval field members.}
    \label{tab:conditions}
    \renewcommand{\arraystretch}{1.5}
    \begin{tabular}{c@{\hspace{25pt}}c} 
        \hline
	  Region & Condition \\
        \hline
        Protocluster Peak & 150.1\degree < RA < 150.48\degree, \ 2.21\degree < Dec < 2.5\degree \\
         & 4.48 < z < 4.64 \\
         & $\sigma_\delta \ge 5$ \\
         \hline
        Protocluster & 150.1\degree < RA < 150.48\degree, \ 2.21\degree < Dec < 2.5\degree \\
         & 4.48 < z < 4.64 \\
         & $\sigma_\delta \ge 2$ region with a $\sigma_\delta \ge 5$ peak\\
         \hline
         Field & 150.1\degree < RA < 150.48\degree, \ 2.21\degree < Dec < 2.5\degree \\
          & 4.2 < z < 4.48 or 4.64 < z < 4.93 \\
          & All galaxies except the ones in S1 and S2 \\
        \hline
    \end{tabular}
\end{table}

\section{Properties of Taralay}
\label{sec:properties}

In the initial discovery paper \citep{2018A&A...615A..77L}, Taralay had nine members with secure spectroscopic redshifts. With the new data obtained through the C3VO campaign, combined with the VUDS data that led to the discovery and the data from DEIMOS10k and zCOSMOS, we are able to reestablish the morphology, extent, and the internal structure of this protocluster. We found that the Taralay has two substructures, shown in the right panel of Figure \ref{fig:vol}, that do not connect by a density isopleth of $\sigma_\delta \geq 2$. We refer to the bigger structure as PC1, which is very roughly similar in location and extent to Taralay at the time of discovery. However, the smaller structure that we refer to as PC2, was not detected in the original work. PC1 hosts two $\sigma_\delta \geq 5$ overdense peaks PC1\textunderscore P1 and PC1\textunderscore P2 while PC2 has one $\sigma_\delta \geq 5$ overdense peak PC2\textunderscore P.

To investigate the properties of Taralay, the method that we use to characterize the protocluster and its $\sigma_\delta \geq 5$ peaks is identical to that in \cite{2023arXiv230715113F} which is, in turn, identical to the method used in \cite{2018A&A...619A..49C} and \cite{2021ApJ...912...60S}. The mass and volume of the Taralay protocluster along with some other properties are summarized in Table \ref{tab:pcprop}. The properties of $\sigma_\delta \geq 5$ peaks of the protocluster are given in Table \ref{tab:characteristics}. We also list some of the properties of two potential foreground protoclusters S1 and S2 (see Section \ref{sec:pc}) in Table \ref{tab:2struc}. 
 
The apparent comoving volume of the Taralay protocluster obtained by summing the volume of all voxels (see Section \ref{sec:densmap}) within the $\sigma_\delta \geq 2$ envelope is $\sim 33695 \ \text{cMpc}^3$. This apparent volume is artificially increased due to the redshift elongation that originates from the uncertainties in the photometric redshifts and induced motion. To correct the apparent volume, we consider the anisotropic interpretation of the different dimensions. The transverse dimensions are distinct from the LoS dimension, requiring us to factor this discrepancy when calculating the characteristic radii of the structure in each dimension. To correct for this effect, we use the same approach as \cite{2018A&A...619A..49C} defining an effective radius that depends on the density and position of each galaxy as well as the barycenter of the overdensity in question. This effective radius is defined for all three dimensions, i.e. x, y and z. To calculate the elongation, we take a ratio of the effective radius in the z (LoS) direction with the mean of effective radii in x and y directions. The corrected volume, then, is the apparent volume divided by elongation. Table \ref{tab:formulae} lists the formulae used to calculate these quantities, where $\rho_m$ is the comoving matter density, $V$ is the apparent volume and $\delta_m$ is the mass overdensity in the region under consideration. $\text{RA}_\text{peak}$, $\text{Dec}_\text{peak}$, $\text{z}_\text{peak}$ are the barycentric position in RA, Dec and z respectively with $R_x$, $R_y$, $R_z$ being the effective radii in the three dimensions. $E_{z/xy}$ is the elongation correction factor, $V_\text{corr}$ is corrected volume for elongation effect and $<\delta_\text{gal}>_\text{corr}$ is the corrected average galaxy overdensity.

The comoving volume of the Taralay protocluster, corrected for elongation, is $\sim 12620 \ \text{cMpc}^3$, the value we use to calculate the SFRD of the protocluster. We obtain the upper and lower uncertainty in the volume by using density thresholds of $\sigma_\delta \geq 2.2$ and $\sigma_\delta \geq 1.8$ respectively, and calculate the resultant elongation-corrected volume, which results in a final value of $\sim 12620^{+1042}_{-956} \ \text{cMpc}^3$. 

The mass of the Taralay is calculated using a formula given in Table \ref{tab:formulae} for $M_\text{tot}$ with a bias factor of $b=3.6$ (see Section \ref{sec:densmap}). \color{black} Due to our ignorance on the precise value of the bias factor appropriate for our tracer population, we additionally vary the bias factor between 4.5 and 3.12, values which are obtained from \cite{2023MNRAS.523.4693E} and \cite{2021MNRAS.500.3194A}, respectively, and propagate this uncertainty into the mass uncertainty. \color{black} The uncertainty in mass due to the variation in the bias factor \color{black} is added in quadrature with the uncertainty \color{black} in mass coming from varying the density threshold (as described above). We estimate \color{black}  the mass of Taralay to be $1.74^{+1.36}_{-0.77} \times 10^{15}$ M$_\odot$. This value is $\sim$6 times higher ($\sim2\sigma$) than the value reported in \cite{2018A&A...615A..77L} at the time of discovery. \color{black} This difference is likely due to the $\sim$5$\times$ increase in the number of spectroscopic members, the larger adopted redshift extent, the higher spectral redshift fraction overall which decreases the dilution from photometric redshifts (see Hung et al.\ \emph{in prep}), and the discovery of the substructure PC2.  \color{black} 

\subsection{Dynamical versus Overdensity mass}

\begin{figure*}
    \centering
    \includegraphics[width = \textwidth]{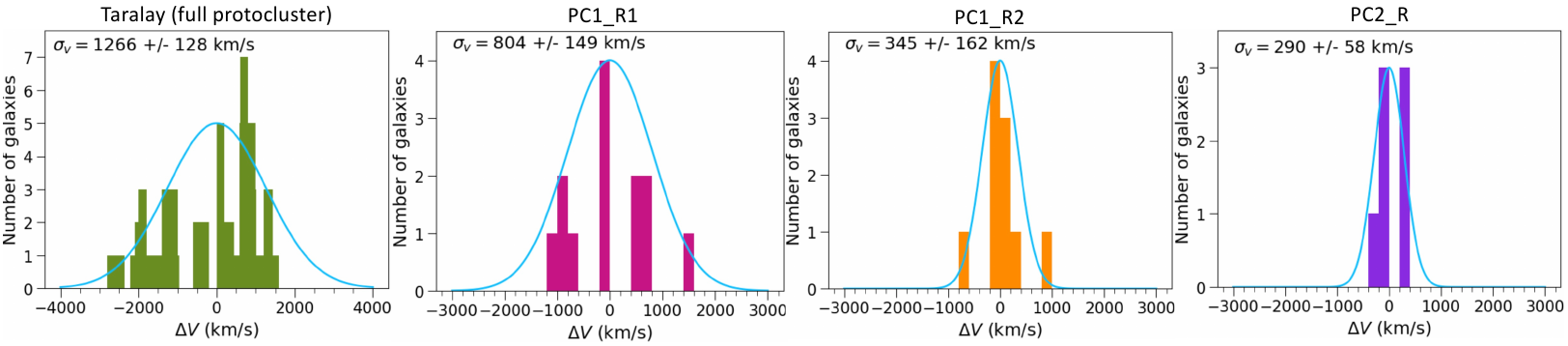}
    \caption{Velocity histograms for Taralay, two $\sigma_\delta \geq 4.5$ regions in PC1 named PC1\textunderscore R1 and PC1\textunderscore R2, and one $\sigma_\delta \geq 2.8$ region in PC2 named PC2\textunderscore R along with the fits to these histograms to estimate velocity dispersion $\sigma_v$.}
    \label{fig:veldisphists}
\end{figure*}

We compare the mass obtained from the overdensity with the mass obtained from the estimated LoS velocity dispersion $\sigma_v$. We do this comparison for the protocluster, two $\sigma_\delta \geq 4.5$ regions in PC1 (we will call them PC1\textunderscore R1 and PC1\textunderscore R2) and one $\sigma_\delta \geq 2.8$ region in PC2 (we will call it PC2\textunderscore R). These values are a compromise between a sufficiently large sample size to measure $\sigma_{v}$ and a region that is sufficiently small such that gravitational interactions between galaxies are reasonably likely. The LoS velocity dispersion $\sigma_v$ is estimated using the gapper method (\citealt{1990AJ....100...32B} and references therein) with jackknifed confidence intervals for PC1\textunderscore R1, PC1\textunderscore R2 and PC2\textunderscore R regions. For the full protocluster we use biweight method also with the jackknifed confidence intervals to estimate $\sigma_v$. PC1\textunderscore R1, PC1\textunderscore R2 and PC2\textunderscore R regions have sample size of $n \sim 10$ and the entire protocluster has a sample size of $n \sim 40$ with secure spectroscopic redshifts. The gapper method is used for the individual peaks as the smaller sample size makes this the preferred method \citep{1990AJ....100...32B}, while the sample size is sufficiently large to adopt the biweight estimator in the case of the full protocluster. Figure \ref{fig:veldisphists} shows the velocity histograms and the fit to these histograms to estimate $\sigma_v$.

The dynamical mass which refers to the mass enclosed in $R_{200}$, the radius within which the density is 200 times the critical density, is calculated from the $\sigma_v$ using the following formula


\begin{equation}
    M_{200} = \frac{3}{G}\sigma_v^2R_{200},
\end{equation}
where
\begin{equation}
    R_{200} = \frac{\sqrt{3}\sigma_v}{10 H(z)}.
    \label{eq:r200}
\end{equation}

\begin{figure}
    \centering
    \includegraphics[width = 0.45\textwidth]{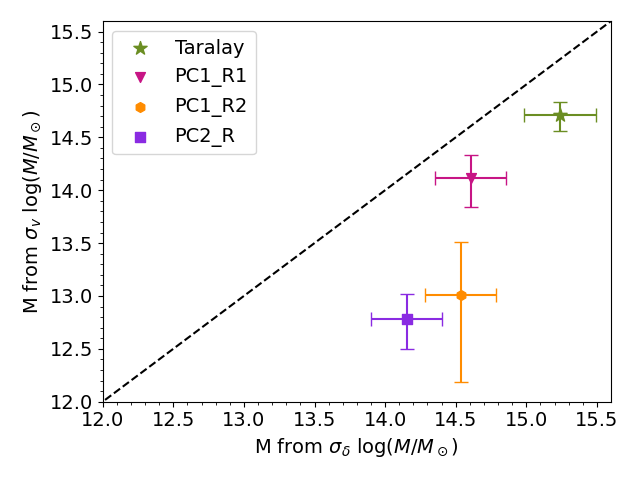}
    \caption{The comparison of mass estimated from the line of sight velocity dispersion to the mass estimated from overdensity for the Taralay protocluster, PC1\textunderscore R1, PC1\textunderscore R2 and PC2\textunderscore R regions (see Section \ref{sec:properties}).}
    \label{fig:veldisp}
\end{figure}

\noindent\citep{1997ApJ...485L..13C}. \color{black} The LoS velocity dispersion of $1266 \pm 128$ km/s for the Taralay protocluster results in the $M_{200}$ of 
\begin{equation*}
    \log(M_{200}/M_\odot)_{z \sim 4.57} = 14.71^{+0.12}_{-0.14}
\end{equation*}
which agrees within the errors with the dynamical mass estimated by \cite{2018A&A...615A..77L}. The $\sigma_v$ and $M_{200}$ for PC1\textunderscore R1, PC1\textunderscore R2 and PC2\textunderscore R are given in Table \ref{tab:los}. We obtain the error bars for $M_{200}$ by taking into account the error on $\sigma_v$.
The mass from the overdensity method is obtained through density mapping method (Section \ref{sec:densmap}) and the error bars for the mass from overdensity are obtained by varying $\sigma_\delta$ and the bias factor. To this we added a systematic uncertainty of 0.25 dex based on masses estimated through overdensity reconstruction based on comparison to simulation (Hung et al. in prep). For the protocluster, $\sigma_\delta$ is varied to 1.8 and 2.2, for PC1\textunderscore R1 and PC1\textunderscore R2 the $\sigma_\delta$ is varied to 4.3 and 4.7, and for PC2\textunderscore R the $\sigma_\delta$ is varied to 2.6 and 3. For all regions, the bias factor is varied from 3.12 to 4.5. \color{black} The comparison between the overdensity masses and the dynamical masses from Figure \ref{fig:veldisp} shows that the dynamical masses have an average deficit of 2.5$\sigma$ (range of 1.5$\sigma$ to 4$\sigma$) compared to the overdensity masses. The source of this consistent deficit of the dynamical masses relative to the overdensity will be explored in simulations in future work. 
\color{black}

\begin{table}
    \centering
    \caption{Formulae used to calculate the total mass, barycentric position, effective radius, elongation correction factor, corrected volume and corrected average galaxy overdensity of each peak.}
    \label{tab:formulae}
    \renewcommand{\arraystretch}{1.5}
    \begin{tabular}{cc} 
        \hline
	Quantity & Formula \\
        \hline
        $M_\text{tot}$ &  $\rho_m V(1+\delta_{gal}/b)$ \\
        $X_\text{peak}$ & $\sum_i(\delta_{\text{gal},X_i}X_i)/\sum_i(\delta_{\text{gal},X_i})$ \\
        $R_X$ & $\sqrt{\sum_i(\delta_{\text{gal},X_i}(X_i - X_\text{peak})^2)/\sum_i(\delta_{\text{gal},X_i})}$\\
        $E_{z/xy}$ &  $2R_z/(R_x + R_y)$ \\
        $V_\text{corr}$ & $V/E_{z/xy}$ \\
        $<\delta_\text{gal}>_\text{corr}$ &  \color{black}$b[M_\text{tot}/(V_\text{corr}\rho_m) - 1]$ \\
        \hline
    \end{tabular}
\end{table}

\begin{table}
    \centering
    \caption{The LoS velocity dispersion ($\sigma_v$), the resulting dynamical masses \color{black} and masses from the overdensity \color{black} for the Taralay protocluster, two $\sigma_\delta \geq 4.5$ regions in PC1 and one $\sigma_\delta \geq 2.8$ region in PC2 along with the number of redshifts (n) that were fitted in each region to obtain $\sigma_v$.}
    \label{tab:los}
    \renewcommand{\arraystretch}{1.5} 
    \begin{tabular}{ccccc} 
	\hline
	Region & n & $\sigma_v$ & $M_{200}$ & $M_\delta$ \\
            &  & (km/s) & ($\log(M_{200}/M_\odot)$) & ($\log(M_\delta/M_\odot)$) \\
        \hline
        Taralay & 43 & $1266 \pm 128$ & $14.71^{+0.12}_{-0.14}$ & $15.24^{+0.25}_{-0.25}$\\
        PC1\textunderscore R1 & 13 & $804 \pm 149$ & $14.11^{+0.22}_{-0.27}$ & $14.61^{+0.25}_{-0.26}$\\
        PC1\textunderscore R2 & 11 & $345 \pm 162$ & $13.01^{+0.5}_{-0.83}$ & $14.54^{+0.23}_{-0.26}$\\
        PC2\textunderscore R & 7 & $290 \pm 58$ & $12.78^{+0.24}_{-0.29}$ & $14.15^{+0.25}_{-0.25}$\\
	\hline
    \end{tabular}
\end{table}

\begin{table*}
    \centering
    \caption{Properties of the Taralay protocluster.}
    \label{tab:pcprop}
    \renewcommand{\arraystretch}{1.5} 
    \begin{tabular}{ccccccccccccc} 
	\hline
	ID & $\text{RA}_\text{peak}$ & $\text{Dec}_\text{peak}$ & $z_\text{peak}$ & $<\delta_\text{gal}>$ & $R_x$ & $R_y$ & $R_z$ & $E_{z/xy}$ & $V$ & $M_\text{tot}$ & $V_\text{corr}$ & $<\delta_\text{gal}>_\text{corr}$ \\ 
         & (deg) & (deg) & & & (cMpc) & (cMpc) & (cMpc) & & (cMpc$^3$) & $(10^{14}$ M$_\odot)$ & (cMpc$^3$) & \\
	\hline
        PC1 & 150.352213 & 2.354023 & 4.567 & 1.53 & 5.85 & 7.19 & 16.40 & 2.51 & 29622 & 15.49 & 11784 & \color{black}9.28 \\
        PC2 & 150.177104 & 2.301292 & 4.592 & 1.07 & 2.05 & 2.77 & 11.93 & 4.94 & 4133 & 1.95 & 836 & \color{black}19.51 \\
	\hline
    \end{tabular}
\end{table*}

\begin{table*}
    \centering
    \caption{Properties of the $5\sigma$ overdense peaks of Taralay protocluster at $z \sim 4.57$.}
    \label{tab:characteristics}
    \renewcommand{\arraystretch}{1.5} 
    \begin{tabular}{ccccccccccccc} 
	\hline
	ID & $\text{RA}_\text{peak}$ & $\text{Dec}_\text{peak}$ & $z_\text{peak}$ & $<\delta_\text{gal}>$ & $R_x$ & $R_y$ & $R_z$ & $E_{z/xy}$ & $V$ & $M_\text{tot}$ & $V_\text{corr}$ & $<\delta_\text{gal}>_\text{corr}$ \\ 
         & (deg) & (deg) & & & (cMpc) & (cMpc) & (cMpc) & & (cMpc$^3$) & $(10^{14}$ M$_\odot)$ & (cMpc$^3$) & \\
	\hline
        PC1\textunderscore P1 & 150.385321 & 2.380454 & 4.567 & 3.37 & 2.27 & 2.86 & 13.24 & 5.16 & 4902 & 3.76 & 950 & \color{black}35.28 \\
        PC1\textunderscore P2 & 150.315664 & 2.300566 & 4.578 & 2.95 & 2.51 & 3.00 & 13.41 & 4.86 & 4631 & 3.32 & 952 & \color{black}30.64 \\
        PC2\textunderscore P & 150.178475 & 2.298961 & 4.593 & 1.98 & 1.04 & 2.05 & 10.34 & 6.69 & 1063 & 0.64 & 159 & \color{black}35.93 \\
	\hline
    \end{tabular}
\end{table*}

\begin{table*}
    \centering
    \caption{Properties of the two potential foreground protoclusters in the area surrounding Taralay.}
    \label{tab:2struc}
    \renewcommand{\arraystretch}{1.5} 
    \begin{tabular}{cccccccccc} 
	\hline
	ID & $\text{RA}_\text{peak}$ & $\text{Dec}_\text{peak}$ & $z_\text{peak}$ & $<\delta_\text{gal}>$ & $E_{z/xy}$ & $V$ & $M_\text{tot}$ & $V_\text{corr}$ & $<\delta_\text{gal}>_\text{corr}$ \\ 
         & (deg) & (deg) & & & & (cMpc$^3$) & $(10^{14}$ M$_\odot)$ & (cMpc$^3$) & \\
	\hline
        S1 & 150.351892 & 2.478142 & 4.327 & 0.99 & 6.02 & 2587 & 1.20 & 429 & \color{black}24.08 \\
        S2 & 150.195099 & 2.306873 & 4.397 & 0.77 & 4.29 & 11003 & 4.89 & 2564 & \color{black}15.16 \\
	\hline
    \end{tabular}
\end{table*}

\section{Galaxy Properties of Taralay}
\label{sec:sfrd}

Now that we have established the morphology and internal structure of the protocluster as well as some of its characteristics, we can investigate the galaxy properties such as the SFR. Understanding the rate at which stars form within galaxies is essential to understand their evolution and behavior. Various indicators can be employed to estimate the star formation rate, all of which involve analyzing the emitted light at different wavelengths. These indicators include UV luminosity, IR luminosity, UV and IR luminosity, as well as the strength of recombination lines or their proxies. While these indicators are generally accessible for samples at lower redshifts, estimating the star formation rate becomes progressively more challenging in the high redshift universe. Acquiring the required recombination line data for a large set of galaxies can be an overwhelming task due to \color{black} a variety of different factors \color{black} and atmospheric transmission issues like absorption and scattering. Moreover, specialized equipment such as the Atacama Large Millimeter/sub-millimeter Array (ALMA) and \textit{JWST} are necessary not only to mitigate transmission issues but also to obtain data far enough in the infrared to allow for a more reliable picture of star forming activity in the early universe.

To navigate these challenges, we use a more accessible method for estimating the star formation rate of high redshift galaxies. Model-based SED fitting provides an alternative approach to estimate the SFR using not just the spectroscopic data but also the available multi-wavelength photometric data. The SED fitting process is performed to determine the SFR of each galaxy in order to estimate the SFR per volume per environment i.e. the SFRD in the protocluster versus the coeval field. We explain this process below. 

\subsection{SED Fitting}
\label{sec:sed} 

SED fitting is a powerful tool in astrophysics that involves modeling the spectral energy distribution of celestial objects ideally across a broad range of wavelengths. A model SED is constructed by combining various components or sources of emission that are expected to contribute to the observed spectrum. Some of the important contributors at high redshift are stellar emission, nebular emission, thermal emission from dust, self absorption and extinction. These components are varied to find the best fit to the observed data points.

To perform our SED fitting, we chose the CIGALE (Code Investigating GALaxy Emission) \citep{2019A&A...622A.103B} software, which uses models that describe the various components of a galaxy, such as the stellar population, dust, and gas. These models are constructed based on physical principles and observations of objects in our local universe. The user can select the models and parameters to be included in the fitting, such as the star formation history, metallicity, dust attenuation, and emission lines. CIGALE compares the observed photometric data (i.e., flux densities measured at various wavelengths) with the model predictions and finds a set of values for a range of parameters that best match the data. 

For our SED fitting using CIGALE, we select a set of models to describe the various components of galaxies. The star formation history (SFH) is  modeled with sfhdelayed which stands for a delayed SFH: $\text{SFR} \propto t \times \text{exp}(-t/\tau)$, where t represents time passed since the onset of star formation and $\tau$ is the time when the SFR peaks. Our choice of this SFH is based on the study of \cite{2017A&A...602A..35T} which found this SFH to be an appropriate choice for large samples of galaxies at high redshifts. 

We model the spectra for composite stellar population with the libraries of single stellar populations (SSPs) from \cite{2003MNRAS.344.1000B} (module bc03). We choose the bc03 module over, e.g., that of \cite{2005MNRAS.362..799M} (module m2005) \color{black} because observations \citep{2008ApJ...674...29V} suggest that the \cite{2003MNRAS.344.1000B} initial mass function (IMF) is more appropriate for higher redshift galaxies than an, e.g., Salpeter \citep{1955ApJ...121..161S} IMF, and the former is supported in the bc03 module but not supported in the m2005 module.\color{black} 

The dust in a galaxy absorbs UV and NIR radiation and re-emits it in the mid- and far-IR. Two curves associated with this process are extinction curve which only depends on the dust grain size and attenuation curve which depends on the dust grain size as well as the geometry, i.e. where the dust grains are relative to the source of radiation and the observer. These curves are taken into account through two different modules, the \cite{2000ApJ...539..718C} based "dustatt\textunderscore modified\textunderscore CF00" module and the \cite{2000ApJ...533..682C} based "dustatt\textunderscore modified\textunderscore starburst module. We choose the dustatt\textunderscore modified\textunderscore starburst" module as it offers more flexibility in terms of slope of the curve and the presence of 217.5 nm bump. This module also includes Small and Large Magellanic Cloud extinction curves of \cite{1992ApJ...395..130P} along with the Milky Way curve of \cite{1989ApJ...345..245C} with \cite{1994ApJ...422..158O} update. More discussion about how various models and the choices of free parameters affect the SED fitting is in Appendix \ref{sec:appendixb}. We compare SFR and SM fitted with CIGALE and LePhare in Appendix \ref{sec:appendixc}.

By selecting these models and adjusting their parameters, we fit the galaxies and model their spectral energy distribution to recover parameters like stellar mass and star formation rates. The parameter values for each of the modules in this fitting are given in Table \ref{tab:parameters} although see more discussion on the choice of these parameters in Appendix \ref{sec:sedappen}. Their detailed description can be found in \cite{2019A&A...622A.103B}. \color{black} We discuss the effect of lack of FIR data on the estimated SFR in Appendix \ref{sec:firabsent}. \color{black}

\begin{table}
    \centering
    \begin{threeparttable}
        \caption{SED fitting parameters of CIGALE modules sfhdelayed, bc03 and dustatt\textunderscore modified\textunderscore starburst}
        \label{tab:parameters}
        \renewcommand{\arraystretch}{1.5}
        \begin{tabular}{lcr} 
            \hline
    	modules & parameter values \\
            \hline
            tau\textunderscore main (Myr)\tnote{a} & 100 to 30000 \\
            age\textunderscore main (Myr)\tnote{b} & 50 to 1400\\
            tau\textunderscore burst (Myr)\tnote{c} & 100, 300 \\
            f\textunderscore burst\tnote{d} & 0, 0.0001, 0.0005, 0.001, 0.005, 0.01 \\
            imf & 1\tnote{e} \\
            metallicity ($Z_\odot$) & 0.008, 0.02 \\
            E\textunderscore BV\textunderscore lines\tnote{f} & 0, 0.05, 0.1, 0.15, 0.2, 0.25, 0.3, 0.35, 0.4, 0.5 \\
            E\textunderscore BV\textunderscore factor\tnote{g} & 1 \\
            uv\textunderscore bump\textunderscore amplitude & 0, 1, 2, 3 \tnote{h}\\
            powerlaw\textunderscore slope\tnote{i} & -0.5, -0.25, 0 \\
            Ex\textunderscore law\textunderscore emission\textunderscore lines & 1\tnote{j}, 3\tnote{k} \\
            \hline
        \end{tabular}
        \begin{tablenotes}\footnotesize
            \item[a] tau\textunderscore main in CIGALE refers to the e-folding time of the main stellar population model.
            \item[b] age\textunderscore main in CIGALE refers to the age of the main stellar population in the galaxy.
            \item[c] tau\textunderscore burst in CIGALE refers to the e-folding time of the late starburst population model.
            \item[d] f\textunderscore burst in CIGALE refers to the mass fraction of the late burst population.
            \item[e] IMF of 1 in CIGALE refers to the Chabrier IMF.
            \item[f] E\textunderscore BV\textunderscore lines in CIGALE refers to the colour excess of the nebular lines light for both the young and old population. 
            \item[g] E\textunderscore BV\textunderscore factor in CIGALE refers to the reduction factor to apply to E\textunderscore BV\textunderscore lines to compute the stellar continuum attenuation.
            \item[h] uv\textunderscore bump\textunderscore amplitude of 3 in CIGALE refers to the Milky Way.
            \item[i] powerlaw\textunderscore slope in CIGALE refers to the slope delta of the power law modifying the attenuation curve.
            \item[j] Ex\textunderscore law\textunderscore emission\textunderscore lines of 1 in CIGALE refers to the Milky Way.
            \item[k] Ex\textunderscore law\textunderscore emission\textunderscore lines of 3 in CIGALE refers to the Small Magellanic Cloud.
        \end{tablenotes}
    \end{threeparttable}
\end{table}

\subsection{From SFR to SFRD}
\label{sec:sfrtosfrd}

After using Monte Carlo to generate 100 realizations of our master spectroscopic plus photometric catalog (see Section \ref{sec:MC}), in each realization, any galaxy may fall into one of three redshift bins: the protocluster, the field or outside the redshift range of interest. The probability of a galaxy falling into one of these categories depends on several factors, including whether the galaxy has a spectroscopic redshift or not, the quality of the spectroscopic redshift and the width of the zPDF.

For each realization, we determine which environmental category each galaxy falls in and calculate the total SFR for all galaxies identified as protocluster galaxies ($\text{SFR}_\text{pc}$) and the total SFR for all galaxies identified as coeval field galaxies ($\text{SFR}_\text{field}$) according to the definitions given in Section \ref{sec:pc}. We then calculate the star formation rate density (SFRD) of the protocluster ($\text{SFRD}_\text{pc}$) by dividing $\text{SFR}_\text{pc}$ by the volume of the protocluster obtained from density mapping but corrected for elongation (given in Table \ref{tab:pcprop}). Similarly, we calculate the SFRD of the field ($\text{SFRD}_\text{field}$) by dividing $\text{SFR}_\text{field}$ by the volume of the field. The volume of the coeval field is obtained by subtracting two quantities from the total volume of our region of interest: \color{black} 1) the volume of the region enclosed in $4.48 < z < 4.64$ (which includes the protocluster) 2) the uncorrected volume of S1 and S2. The resulting coeval field volume is $686583 \ \text{cMpc}^3$ as compared to $\sim 12621 \ \text{cMpc}^3$ for Taralay. However, if we instead subtract the elongation corrected volume for S1 and S2 from the field, the volume of field increases only by 1\%, which has negligible effect on our results. \color{black}

\subsection{Contribution of Lower Luminosity Galaxies}
\label{sec:lowgal} 

The spectroscopic and photometric data that this study uses has limitations in terms of depth. To include the fainter galaxies that our data cannot probe and take into account the contribution of these fainter galaxies to the SFRD, we extrapolate our results for the protocluster as well as field to lower luminosity (see Appendix \ref{sec:dustpropappen} for discussion about the effect of dust properties of faint and bright galaxies on the SED fitting process in order to estimate accurate SFR and the corrections we need to perform in order to extrapolate our result safely). In order to do this, we begin by selecting a sample of objects with $4.2 < z_\text{phot} <4.93$ whose IRAC channel 1 magnitudes fall within 25.3 and 25.5. We look at the distribution of Far-UV magnitude of this sample and take the 80\% completeness limit to probe the depth of our UV and optical data in a given window of IRAC channel 1 (see Appendix B in \citealt{2018A&A...615A..77L} for the basic idea). The range of $25.3 < M_\text{IRAC1} < 25.5$ is chosen because it is brighter than the IRAC channel 1 completeness limit \color{black} (25.7 in [3.6] band) \color{black} stated in \cite{2022ApJS..258...11W} making it likely that all objects at this brightness are detected.
For the sample in this window, we sort the $M_\text{FUV}$ magnitudes and remove the faintest 20\% objects of the sample. This is done to obtain 80\% completeness limit. The resulting $M_\text{FUV}$ distribution is then corrected by the average difference between the range of IRAC channel 1 window we choose and the completeness limit \color{black} (25.7) \color{black} in order to get the $M_\text{FUV}$ completeness of our sample. For example, the average difference between the window $25.3 < M_\text{IRAC1} < 25.5$ and the completeness limit \color{black} of IRAC channel 1 (25.7) \color{black} is 0.3. 

\color{black} The above calculation is based on an assumption that the change in $M_\text{IRAC1}$ corresponds exactly to the change in $M_\text{FUV}$ for the galaxy population considered here. This is not necessarily the case. In order to check this, we repeated this exercise with a different IRAC channel 1 window, $24.8 < M_\text{IRAC1} < 25.0$, and found that the median $M_\text{FUV}$ is offset by 0.36 mags between the windows $25.3 < M_\text{IRAC1} < 25.5$ and $24.8 < M_\text{IRAC1} < 25.0$ as compared to the change in $M_\text{IRAC1}$ of 0.5 mags between these two windows.\color{black} To account for this difference, we correct the measured $M_\text{FUV}$ distribution in our original window not by the average difference between the median IRAC1 magnitude in our chosen window and the corresponding completeness limit \color{black} but by the expected corresponding change in $M_{FUV}$ coming from the above exercise. Ultimately, this results in a very small correction to the $M_{FUV}$ distribution ($\sim$0.2 mags). 
This exercise results in the $M_\text{FUV}$ completeness limit of our sample to be approximately -19.3, a value that is not strongly dependent on the various windows chosen in this exercise. \color{black}

With the depth of our data established at $M_\text{FUV} = -19.3$, we extrapolate our results for SFRD to $M_\text{FUV} = -17$ in order to include the contribution of the fainter UV galaxies not detected in the data used in this study. This also lets us compare our results with studies that report the SFRD values corrected to include the contribution of the fainter galaxies. We use the Schechter function \citep{1976ApJ...203..297S} to extrapolate $\text{SFRD}_\text{pc}$ and $\text{SFRD}_\text{field}$ down to $M_\text{FUV} = -17$. \color{black} The faint limit of $M_\text{FUV} = -17$ is chosen because the behaviour of the galaxy luminosity function is not well known beyond $M_\text{FUV} = -17$ and may deviate from a simple Schechter function (e.g. \citealt{2015ApJ...807L..12O}, \citealt{2019A&A...628A...3D}, \citealt{2019MNRAS.483.2983Y}). We use the following equation to determine the correction factor:
\begin{equation}
    \begin{split}
        CF = \int_{-\infty}^{-17}\Big[dM(0.4\times \log 10)\phi^*[10^{0.4(M^* - M)}]^{\alpha + 1}\\
        \times\exp [-10^{0.4(M^* - M)}]^{\alpha + 1}\Big] \Bigg/ \\
        \int_{-\infty}^{-19.3}\Big[dM(0.4\times \log 10)\phi^*[10^{0.4(M^* - M)}]^{\alpha + 1}\\
        \times\exp [-10^{0.4(M^* - M)}]^{\alpha + 1}\Big]
    \end{split}
\end{equation}

The faint end slope $\alpha$ utilized in our study is determined by combining values obtained from multiple studies (e.g., \citealt{2004ApJ...611..660O}, \citealt{2005NewAR..49..440G}, \citealt{2006ApJ...653..988Y}, \citealt{2006ApJ...642..653S}, \citealt{2007ApJ...670..928B, 2015ApJ...803...34B}). To obtain a representative value and its associated uncertainty, we construct a joint PDF from the reported values and associated uncertainties in these studies. The mean of this joint PDF serves as the final estimate of $\alpha$ with 16th and 84th percentiles serving as corresponding errors in our study. The value for $\alpha$ is $\alpha = -1.77^{+0.22}_{-0.20}$. The values for $M^*_\text{UV}$, $\phi^*$
\begin{equation*}
\begin{split}
    M^*_\text{UV} = (-20.95 \pm 0.10) + (0.01 \pm 0.06)(z - 6), \\[4pt]
    \phi^* = \big(0.47^{+0.11}_{-0.10}\big)10^{(-0.27\pm 0.05)(z-6)}10^{-3} \text{Mpc}^{-3}
\end{split}
\end{equation*}
are taken from \cite{2015ApJ...803...34B} where we substituted $z = 4.57$ for this study. The correction factor is log(CF) = $0.96 \pm 0.17$ dex, which implies that faint galaxies are contributing significantly to the overall SFRD. Changing the completeness limit by 10\% changes the log correction factor by $\sim$0.1 dex.

\section{Results}
\label{sec:results}

In this section we report the SFRD of the Taralay protocluster at $z \sim 4.57$, SFRD of its three $\sigma_\delta \geq 5$ peaks, SFRD of the coeval field as well as the contribution of the protoclusters at $z \sim 4.57$ to the cosmic SFRD using Taralay as a proxy of all the protoclusters at this redshift. We also report on the SFR-$\sigma_{\delta}$ relation for all galaxies in the protocluster and coeval field. 




\subsection{SFRD of the Field Surrounding Taralay}
\label{sec:sfrdfield}

We find that the SFRD of the galaxies in the coeval field surrounding the Taralay protocluster is \color{black} log(SFRD/$M_\odot$ yr$^{-1}$ Mpc$^{-3}$) = $-0.82^{+0.19}_{-0.29}$ \color{black}. The uncertainty on the $\text{SFRD}_\text{field}$ is a result of the combined uncertainty on the $\text{SFR}_\text{field}$ from performing Monte Carlo on redshifts (Section \ref{sec:MC}), the SED fitting (Section \ref{sec:sed}), the change in the SFR of the galaxies in the field due to varying the boundary of the protocluster (since $\sigma_\delta$ cut dictates which galaxies qualify as field galaxies or protocluster members, see Section \ref{sec:properties}), the uncertainty on the Schechter parameters (Section \ref{sec:lowgal}) and the uncertainty on the volume of the field from changing the boundary of the protocluster (see Section \ref{sec:sfrtosfrd} for calculating the field volume). We compare our $\text{SFRD}_\text{field}$ with various studies in Figure \ref{fig:results}.

A particularly comparable study to our own is that of \cite{2021A&A...649A.152K} where the SFRD value is measured using a spectroscopic sample (from VUDS and DEIMOS) with corrections based on an adopted Far-UV luminosity function and galaxy stellar mass function. This study uses \color{black} rest-frame far-infrared continuum observations with ALMA in order to derive dust-obscured SFR. Using \color{black} a somewhat similar framework to ours, the authors of this work also performed a similar faint-galaxy correction to their SFRD results. 
We find that our result for $\text{SFRD}_\text{field}$ is statistically indistinguishable from the SFRD value estimated by \cite{2021A&A...649A.152K} giving us confidence in our $\text{SFRD}_\text{field}$ value. \color{black} This agreement indicates that the assumptions on the dust attenuation curves that went into our SED fitting in order to derive SFRD values are well accounted for.\color{black}


Although the $\text{SFRD}_\text{field}$ value we report here is higher than what is predicted by \cite{2014ARA&A..52..415M} \color{black} at $z\sim4.57$, the values at these redshifts from \cite{2014ARA&A..52..415M} may be underestimated due the paucity of data at those redshifts a decade ago. Indeed, many of the more contemporary studies reported in Figure \ref{fig:results} recover values in excess of the \cite{2014ARA&A..52..415M} best fit at these redshifts. 

More specifically, values in excess of the \cite{2014ARA&A..52..415M} fit at these redshifts \color{black} is supported by the findings of \cite{2009ApJ...705L.104K}, in which the SFRD values are measured based on gamma-ray bursts, the SFRD values from measurements based on \textit{Herschel} data from \cite{2016MNRAS.461.1100R}, the SFRD value measured using Far-UV luminosity function and the galaxy stellar mass function from \cite{2021A&A...649A.152K}, the SFRD measurement based on \color{black} Far-IR from \cite{2020A&A...643A...8G}, and the SFRD measurement at $z > 3$ derived from radio luminosities and translated to Far-IR luminosities using $q_{TIR}$ \citep{2017A&A...602A...5N}. 
Beyond \cite{2021A&A...649A.152K}, other recent works from the ALPINE survey using Far-IR measurements also recover higher values for the SFRD at $4<z<6$ than the \cite{2014ARA&A..52..415M} relation \citep{2020A&A...643A...8G, 2021A&A...646A..76L}. \color{black} Such results are similarly in tension with the SFRD measured from other studies that use dust-corrected FUV data (e.g., \citealt{2012A&A...539A..31C}, \citealt{2012ApJ...754...83B}, \citealt{2015ApJ...803...34B}, \citealt{2013ApJ...768..196S}, \citealt{2018A&A...620A..51P}, \citealt{2020A&A...634A..97K}). 
\color{black} The reason that likely accounts for the difference between SFRD values derived from IR data versus the SFRD values derived from dust-corrected FUV data is the uncertainty that comes from the IRX-$\beta$ relation that is used for dust-correction (e.g. \citealt{2020ARA&A..58..529S} and references therein). \color{black} In the future, we will investigate further the $\sim 2.5\sigma$ tension with the best fit in \cite{2014ARA&A..52..415M} by probing the field surrounding other structures at other redshifts in the C3VO survey.



\subsection{SFRD of the Taralay Protocluster}
\label{sec:sfrdpc}

\begin{figure*}
    \centering
    \captionsetup{justification=raggedleft}
    \includegraphics[width = \textwidth]{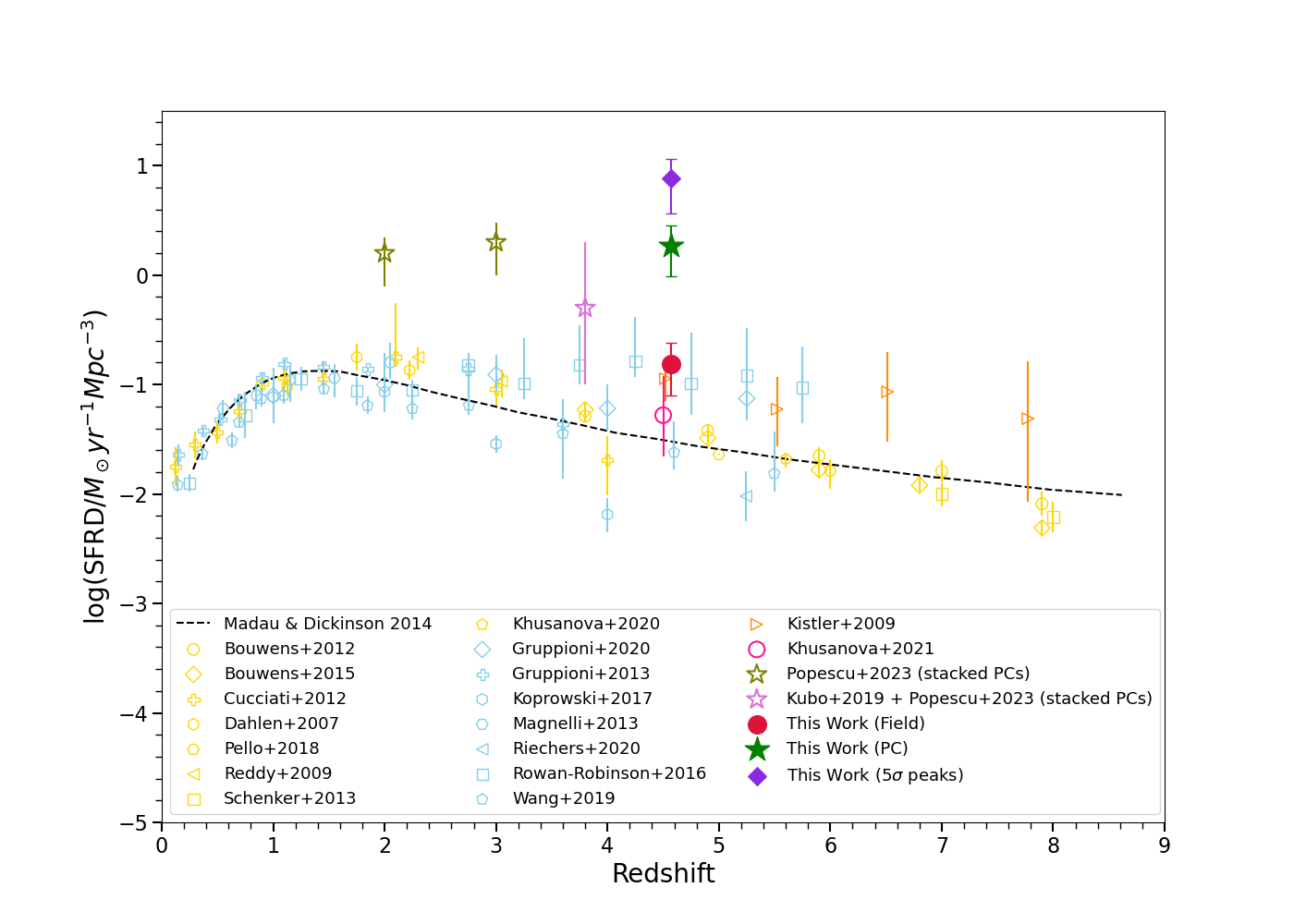}
    \caption{Comparison of protocluster SFRD with the coeval field SFRD. The green star, purple diamond and circle cross show the SFRD values calculated in this work for the Taralay protocluster, its $\sigma_\delta \geq 5$ peaks and the surrounding coeval field respectively. The rest of the points show evolution of SFRD with redshift from various studies. The yellow points are dust-corrected SFRD obtained with FUV data from \protect\cite{2007ApJ...654..172D}, \protect\cite{2009ApJ...692..778R}, \protect\cite{2012A&A...539A..31C}, \protect\cite{2012ApJ...754...83B, 2015ApJ...803...34B}, \protect\cite{2013ApJ...768..196S}, \protect\cite{2018A&A...620A..51P}, \protect\cite{2020A&A...634A..97K}. The skyblue symbols are the SFRD measured from IR due to the re-radiation of dust emission from forming stars. These data points are taken from \protect\cite{2013A&A...553A.132M}, \protect\cite{2013MNRAS.432...23G, 2020A&A...643A...8G}, \protect\cite{2016MNRAS.461.1100R}, \protect\cite{2017MNRAS.471.4155K}, \protect\cite{2019A&A...624A..98W}, \protect\cite{2020ApJ...895...81R}. The orange data points are SFRD from \protect\cite{2009ApJ...705L.104K} calculated using the number of $\gamma$-ray bursts. The pink circle is the extrapolated \color{black} field \color{black} SFRD value \color{black} that is an average of the SFRD values leveraged on  an adopted Far-UV luminosity function and those leveraged on an adopted galaxy stellar mass function in \protect\cite{2021A&A...649A.152K} \color{black}. The olive open stars are the SFRD values calculated by \protect\cite{2023arXiv230800745P} for protoclusters at $2 < z < 4$ using stacked WISE and Herschel/SPIRE images. The orchid open star is the SFRD value of protoclusters at $z \sim 3.8$ where the SFR value is estimated by \protect\cite{2019ApJ...887..214K} and the SFRD value is calculated by \protect\cite{2023arXiv230800745P}.}
    \label{fig:results}
\end{figure*}

The SFRD of the Taralay members is \color{black} log(SFRD/$M_\odot$ yr$^{-1}$ Mpc$^{-3}$) = $0.26^{+0.18}_{-0.28}$ \color{black} (shown in Figure \ref{fig:results}) in excess of $\text{SFRD}_\text{field}$ by $1.08 \pm 0.32$ dex ($\sim 12 \times$). This value is $\sim 6\sigma$ in excess than the best fit in \cite{2014ARA&A..52..415M} indicating that the protocluster galaxies are well outpacing the field. 
The excess with respect to the field \color{black} may mean \color{black} that members of Taralay are rapidly building up their collective stellar mass through star forming processes, well in excess of such growth in the field. \color{black} Later in this section we will show that this is indeed the case. 
\color{black}

We compare our results for SFRD$_\text{pc}$ with the SFRD values from \cite{2023arXiv230800745P} and \cite{2019ApJ...887..214K}, shown by olive and pink open stars in Figure \ref{fig:results} respectively. \cite{2023arXiv230800745P} stacked \textit{Wide-field Infrared Survey Explorer} (\textit{WISE}, \citealt{2010AJ....140.1868W}) and \textit{Herschel/SPIRE} (Spectral and Photometric Imaging REceiver, \citealt{2010A&A...518L...3G}) images for 211 protocluster candidates at $2 < z < 4$ that they selected as Planck cold sources from \cite{2015A&A...582A..30P}. They define sources with redder color as cold sources that peak between 353 and 857 GHz. The redder color corresponds to a cold dust temperature or a high redshift. The SFR of the protocluster candidates was derived through SED fitting method using CIGALE. This SFR was converted into SFRD \color{black} for each candidate protocluster \color{black} by using a volume approximated by a sphere of radius 10.5 comoving Mpc (cMpc) at $z = 2$. \cite{2023arXiv230800745P} also converted the SFR derived in \cite{2019ApJ...887..214K}, a study that stacked \textit{Planck}, \textit{AKARI} \citep{2007PASJ...59S.369M}, \textit{Infrared Astronomical Satellite} \citep{1984ApJ...278L...1N}, \textit{WISE} and \textit{Herschel} images for 179 candidate protoclusters at $z \sim 4$, selected from the Hyper Suprime-Cam Subaru Strategic Program, with the combined IR emission in the observed 12-850 $\mu$m wavelength range. The SFR of LBG-selected protocluster candidates from this study was converted into SFRD by using a spherical volume with a radius of 10.2 cMpc. Albeit at different redshifts, we find a good agreement between our results and the results of these studies (see Section \ref{sec:discussion} for more discussion).

\begin{figure}
    \centering
    \includegraphics[width = 0.45\textwidth]{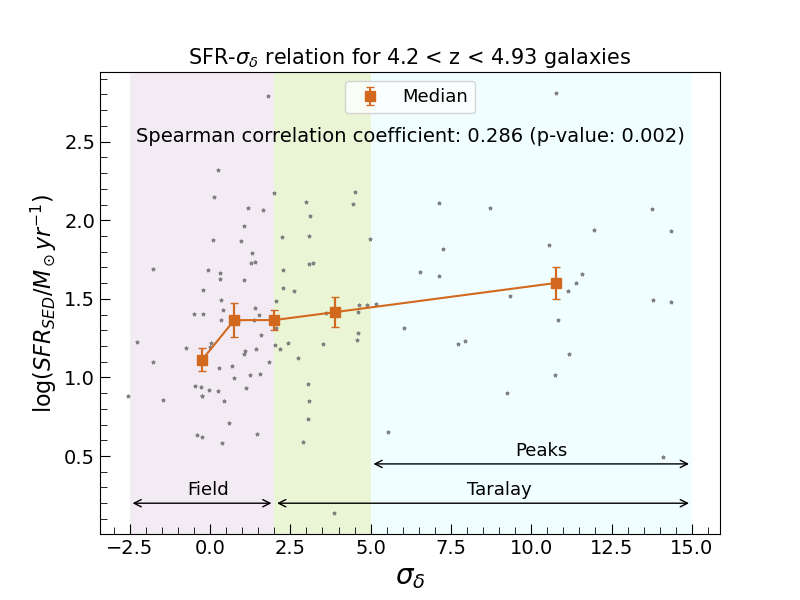}
    \caption{SFR - $\sigma_\delta$ \color{black} (overdensity) \color{black} relation for all spectroscopically-confirmed galaxies that fall in the redshift range of interest $4.2 < z < 4.93$. The SFR for these galaxies are estimated with CIGALE, an SED fitting code (see \ref{sec:sed}). The orange squares are the median SFR plotted at the median overdensity for each bin such that each bin contains approximately equal number of galaxies. The spread on the median is calculated using $\sigma_\text{NMAD}$. The Spearman correlation coefficient is positive indicating a weak correlation between local galaxy density and the SFR.}
    \label{fig:sfroverdens}
\end{figure}

It is tempting to attribute the elevated SFRD of the Taralay protocluster compared to the field solely to the fact that the protocluster hosts a great number of star forming galaxies in a relatively small volume. Here we focus on the SFR-$\sigma_\delta$ relation in order to investigate whether the high SFRD$_\text{pc}$ comes simply from having a large number of galaxies in the protocluster or if it is also a product of the protocluster galaxy members genuinely having an elevated SFR relative to their counterparts in the field. The SFR - $\sigma_{\delta}$ relation shown in Figure \ref{fig:sfroverdens} for galaxies in this sample reveals a positive correlation between the SFR and overdensity. \color{black}
A Spearman test results in a correlation coefficient of 0.286 and a p-value of 0.002. \color{black}An identical exercise is performed with respect to stellar mass later in this section. \color{black} \cite{2022A&A...662A..33L} found a weak but significant trend for SFR-overdensity for the full VUDS+ sample of 6730 star-forming galaxies over the redshift range $2 \leq z \leq 5$. The strength of the correlation seen in our sample is $> 2$ times higher than that measured in \cite{2022A&A...662A..33L} ($\rho = 0.29$ versus $\rho = 0.13$) indicating that members of the Taralay protocluster \color{black} are even more likely to have an increase in the SFR as in denser environments than the overall star-forming galaxy population at $2 \le z \le 5$. \color{black} 

To disentangle the fraction of SFRD that results from the protocluster having a higher number of galaxies versus the fraction that results from the protocluster galaxies having higher SFR on average, we investigated a scenario where we assumed that the average SFR of the protocluster galaxies is the same as the average SFR of the coeval field galaxies. In Figure \ref{fig:sfroverdens}, the average \color{black} log(SFR/$M_\odot$ yr$^{-1}$) \color{black} increases 0.22 for galaxies in the protocluster relative to those in the field -\color{black} log(SFRD/$M_\odot$ yr$^{-1}$ Mpc$^{-3}$) = 1.25 vs. 1.47 \color{black}. We then reduce the $\text{SFRD}_\text{pc}$ by the ratio between the average SFR of protocluster members and that of field galaxies, which results in a \color{black} log($\text{SFRD}_\text{pc}$/$M_\odot$ yr$^{-1}$ Mpc$^{-3}$) = 0.04 \color{black} as compared to the derived value of 0.26. This means that even if protocluster members had the same average SFR as field galaxies, the $\text{SFRD}_\text{pc}$ would still be  higher than the $\text{SFRD}_\text{field}$ by 7.3 times instead of 12. 
In other words, 43\% of the difference in the $\text{SFRD}_\text{pc}$ and the $\text{SFRD}_\text{field}$ is as a result of the elevated SFR of the protocluster members relative to that of field galaxies, with the remaining 57\% resulting from the higher galaxy number density.

In principle, it is possible that this increase in the SFR is due to an increase in the SM (e.g. \citealt{2015ApJ...799..183S}, \citealt{2016ApJ...817..118T}, \citealt{2008MNRAS.385..147D}, \citealt{2014MNRAS.437.3516S}) as the environment gets denser. However, performing a comparable calculation with respect to SM results in no significant evidence of correlation between the SM and overdensity. The Spearman correlation coefficient is weaker (0.155) with a p value of 0.09. We recast these results in the next section. 

\subsection{SFRD of the \texorpdfstring{$\sigma_\delta \geq 5$}{TEXT} Peaks of Taralay}
\label{sec:sfrdpeak}

We calculate the SFRD of the peaks of Taralay to be \color{black} log(SFRD$_\text{peak}$/$M_\odot$ yr$^{-1}$ Mpc$^{-3}$) = $0.87^{+0.18}_{-0.32}$. \color{black} Similar to the full Taralay protocluster, the $\sigma_\delta \geq 5$ peaks of this protocluster also show SFRD well in excess of the coeval field value. 

\begin{figure*}
    \centering
    \includegraphics[width = \textwidth]{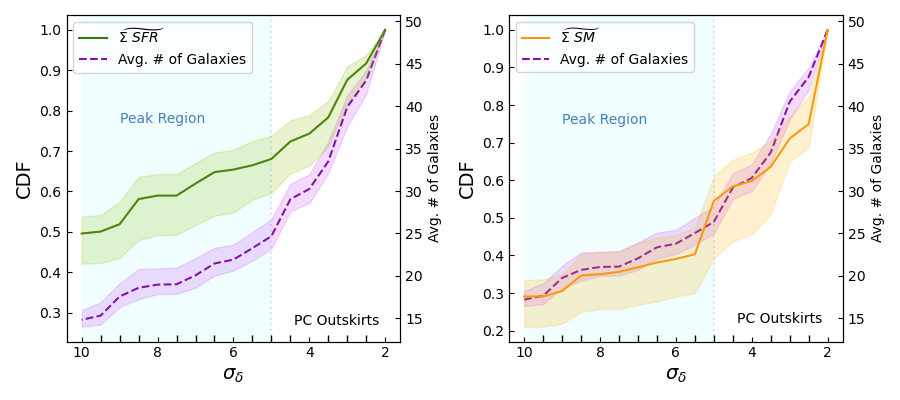}
    \caption{The cumulative distribution function (CDF) of the average total SFR, the average total SM and the average number of galaxies as a function of $\sigma_\delta$. The average for all quantities is taken from 100 iterations of Monte Carlo. The distribution of the average number of galaxies shown on the right hand Y axis for both panels is slightly imprecise due to the statistical nature of the Monte Carlo process. The shaded blue region indicates the peak region of the Taralay protocluster and the $2 \leq \sigma_\delta \leq 5$ area shows the outskirts of the protocluster for both panels.} 
    \label{fig:smsfrncdf}
\end{figure*}

Figure \ref{fig:smsfrncdf} 
shows the cumulative distribution function (CDF) of the SFR, SM and the number of galaxies as a function of $\sigma_\delta$.  \color{black} It is obvious on inspection that the CDF of SFR in the protocluster is skewed toward higher values in the central regions than that of the average number of galaxies, while the CDF of stellar mass tracks the average number of galaxies throughout the entire protocluster. For example, about 68\% \color{black} of the total star formation rate in the protocluster ($\text{SFR}_\text{pc}$) takes place in the $\sigma_\delta \geq 5$ peaks while the inner most regions of the peak ($\sigma_\delta \geq 10$) contains 50\% of the $\text{SFR}_\text{pc}$. This points to the galaxies in the peaks having accelerated evolution and is highly suggestive of inside-out growth. These galaxies might become the more quiescent galaxies that are seen $\sim$850 Myr later at $z \sim 3$ (e.g., \citealt{2020ApJ...890L...1F, 2020ApJ...903...47F}, \citealt{2020A&A...643A..30F}, \citealt{2021ApJ...912...60S}, \citealt{2022ApJ...926...37M}, \citealt{2023ApJ...945L...9I}).

The high SFR in the peaks cannot be attributed only to the large number of galaxies in these regions as less than 30\% protocluster member galaxies are in the inner most regions of the peaks with the entire $\sigma_\delta \geq 5$ peaks hosting less than 50\% of the total protocluster member galaxies. The SM in the peaks is $< 50\%$ of the total SM encased in the protocluster, which largely rules out higher SFR of the galaxies in the peak being a result of those galaxies having higher SMs than counterparts in more rarefied regions. 
The segregation observed in SFR between the densest regions of the protocluster, the protocluster outskirts, and the field, and the lack of stellar mass segregation strongly indicate that the protocluster members, especially those in densest regions, are just beginning to ramp up their star formation activity. If such activity was sustained for even a relatively short time early in the formation history of the protocluster, stellar mass segregation would almost certainly also be observed. 


\subsection{Fractional Contribution of the Protoclusters at \texorpdfstring{$z \sim 4.57$}{TEXT} to the Cosmic SFRD}
\label{sec:fraccontri}
We also estimate the fractional contribution from protoclusters to the cosmic SFRD at $z \sim 4.57$ using Taralay as a representative of a sample of protoclusters at these redshifts (although see Appendix \ref{sec:proxy} for discussion about using such a massive protocluster as a proxy). We estimate this fraction using the following formula:
\begin{equation}
    \begin{split}
        \text{SFRD}_\text{pc} \ \text{fraction} = \frac{\text{SFRD}_\text{pc} \times \text{vol}_\text{pc}}{\text{SFRD}_\text{pc} \times \text{vol}_\text{pc} + \text{SFRD}_\text{field} \times (1 - \text{vol}_\text{pc})}
    \end{split}
\end{equation}
where $\text{vol}_\text{pc}$ refers to the volume occupied by protoclusters. At $z \sim 4.57$, the volume filling factor for protoclusters calculated by \cite{2017ApJ...844L..23C} is 0.04\footnote{The volume filling factor in our data is 0.018 when only Taralay is considered and increases to 0.022 if S1 and S2 are included. The volume filling factor is obtained by dividing the volume of the overdense structure(s) by the volume of the coeval field reported in Section \ref{sec:properties}.}. This makes the fractional contribution to SFRD from protoclusters at $z \sim 4.57$ to be $33.5\%^{+8.0\%}_{-4.3\%}$\footnote{\color{black} Adopting the volume filling factor of our data, 0.022,  rather than that of simulations, 0.04, recovered a fractional contribution to SFRD from protoclusters at $z \sim 4.57$ of $\sim 21$\%. This value is also consistent with the equivalent value from simulations. However, the filling factor estimated from our data is subject to cosmic variance to a much higher level than that of the simulation.}. The uncertainty on the $\text{SFRD}_\text{pc}$ fraction comes from the uncertainty on the SFR from performing Monte Carlo on redshifts (Section \ref{sec:MC}), and change in the SFR and volume of the protocluster due to varying the boundary of the protocluster (Section \ref{sec:densmap}). 

We compare this result with the predictions from \cite{2017ApJ...844L..23C}. The estimated $\text{SFRD}_\text{pc}$ fraction in this study is \color{black} $2.6 \sigma$ and $1.7 \sigma$ in excess of the predicted $\sim 22\%$ and $\sim 26\%$ from \citep{2013MNRAS.428.1351G} and \citep{2015MNRAS.451.2663H} simulations, respectively. \color{black} Though our $\text{SFRD}_\text{pc}$ fraction shows moderate tension with the predictions from simulations, we would like to note that there are several caveats when comparing observational data and simulations. We discuss how the definition for a protocluster that we adopt in this study varies from the definition adopted in \cite{2017ApJ...844L..23C} in Appendix \ref{sec:sim}. We also remind the reader that semi-analytic models (SAMs) assign the SFR to galaxies using a prescription that differs from SED fitting results. However, regardless of these considerations, the simulations, other observational studies, and our measurements of Taralay all indicate that protoclusters at $z \ge 2$ contribute significantly to the stellar mass growth of the universe. 

\begin{figure}
    \centering
    \includegraphics[width = \linewidth]{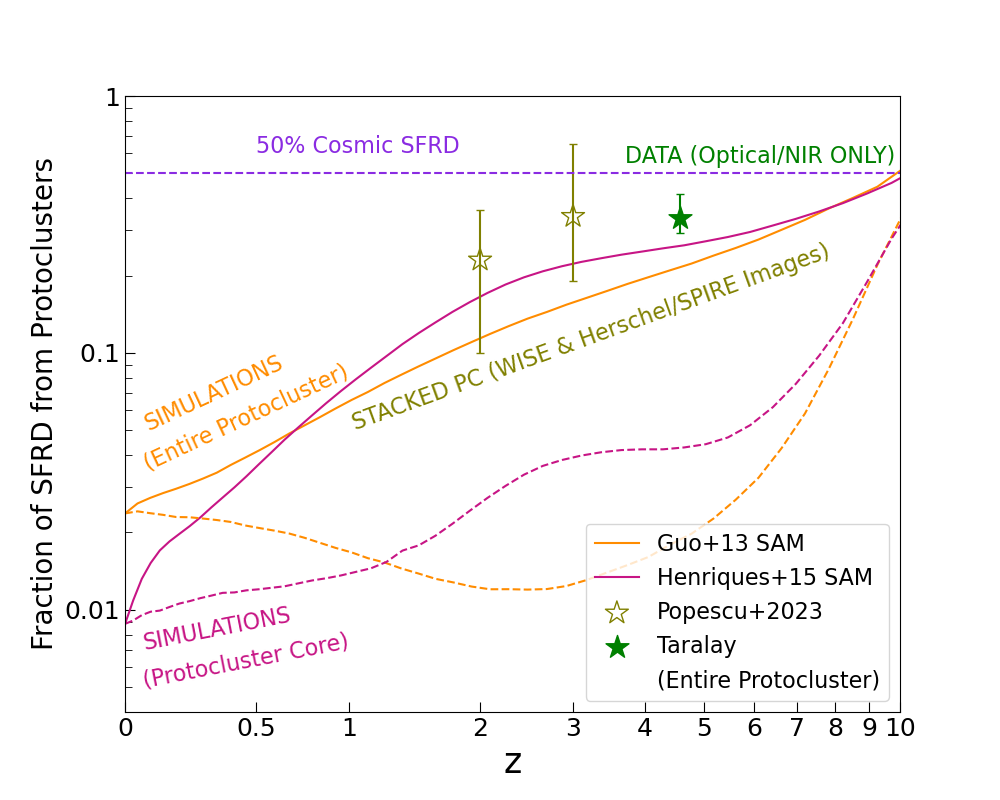}
    \caption{Comparison of the fraction of SFRD from protoclusters at $z \sim 4.57$  estimated using Taralay protocluster as a proxy of all protoclusters at this redshift with the fraction of protocluster SFRD predicted through \protect\cite{2017ApJ...844L..23C} simulation. The olive stars show the fraction of SFRD from protoclusters at $z = 2$ and $z = 3$ obtained from \protect\cite{2023arXiv230800745P}. The solid lines is the fraction predicted for \color{black}all protoclusters in the simulations 
    \color{black} with two different semi-analytic models (SAMs) from \protect\cite{2015MNRAS.451.2663H} and \protect\cite{2013MNRAS.428.1351G}. The dotted lines show the fraction predicted for only the protocluster cores with SAMs mentioned above.}
    \label{fig:sfrdfrac}
\end{figure}

We also compare our results with the SFRD$_\text{pc}$ fractions based on the SFRD values of \cite{2023arXiv230800745P} in Figure \ref{fig:sfrdfrac}. To obtain the SFRD$_\text{pc}$ fractions corresponding to the SFRD values listed in this study at $z = 2$ and $z = 3$, we first measure SFRD$_\text{field}$ at $z = 2$ and $z = 3$ based on the SFRD values in the $1.8 < z < 2.2$ redshift range from \cite{2007ApJ...654..172D}, \cite{2013MNRAS.432...23G, 2020A&A...643A...8G}, \cite{2013A&A...553A.132M}, \cite{2017MNRAS.471.4155K}, \cite{2019A&A...624A..98W} and SFRD values in the $2.8 < z < 3.2$ redshift range from \cite{2013MNRAS.432...23G, 2020A&A...643A...8G}, \cite{2017MNRAS.471.4155K} respectively. The median, 16th and 84th percentile of these samples give us the SFRD$_\text{field}$ and errors at $z = 2$ and $z = 3$ respectively. We find a good agreement between the SFRD$_\text{pc}$ fraction of this study and the SFRD$_\text{pc}$ fractions we calculate based on the SFRD values of \cite{2023arXiv230800745P}. 

We further discuss these results with respect to additional literature in Section \ref{sec:discussion}. 
 
We also calculate the fractional contribution from the $\sigma_\delta \geq 5$ peaks of protoclusters to the cosmic SFRD using equation
\begin{equation}
    \begin{split}
        \text{SFRD}_\text{peak} \ \text{fraction} = \frac{\text{SFRD}_\text{peak} \times \text{vol}_\text{peak}}{\text{SFRD}_\text{pc} \times \text{vol}_\text{pc} + \text{SFRD}_\text{field} \times (1 - \text{vol}_\text{pc})}
    \end{split}
\end{equation}

where $\text{vol}_\text{peak}$ is calculated by as the ratio of volume of peaks estimated with density mapping ($\sim 2060$ cMpc$^3$) divided by the volume of the entire protocluster ($\sim 12620$ cMpc$^3$) and then multiplying this ratio by the volume filling factor for protoclusters, 0.04. The estimated $\text{SFRD}_\text{peak} \ \text{fraction}$ is $22.2\%^{+ 5.4\%}_{-7.3\%}$, a significant portion of the $\text{SFRD}_\text{pc} \ \text{fraction}$, which is indicative of the inside-out growth of protoclusters as discussed in \cite{2017ApJ...844L..23C}.  Note that we do not use the volume filling factor for the protocluster cores as outlined in \cite{2017ApJ...844L..23C} as their definition of core differs from our definition of peaks, discussed in Section \ref{sec:sim}.



\section{Discussion}
\label{sec:discussion}  

The $\text{SFRD}_\text{pc}$ we report in this study is based on optical/Near-IR data. To obtain the total SFR for each galaxy, we perform SED fitting that includes a dust correction. We find that our value is in agreement with the value reported in \cite{2023arXiv230800745P}, a study that focuses on the stacked Far-IR data for 211 protoclusters at $2<z<3$ to estimate the SFR of galaxies using SED fitting (see Section \ref{sec:sfrdpc} for a brief description of analysis employed by \cite{2023arXiv230800745P} to obtain the SFRD values). Because this study concentrates on the Far-IR emission of protocluster galaxies, they are only sensitive to obscured star formation activity. \color{black}At such redshifts the emission in the FIR is, perhaps, a good proxy of the total SFR at $2 < z < 3$, \color{black} as obscured star formation activity is predicted and measured to be an order of magnitude higher than the unobscured SFRD at these redshifts \color{black} on average \color{black} (see, e.g., \citealt{2023MNRAS.518.6142A} and references therein). \color{black} 
By contrast, at $z \sim 4.5$, the unobscured SFRD is thought to be \color{black} in excess or comparable to \color{black} the obscured SFRD (e.g., \citealt{2021A&A...649A.152K, 2023MNRAS.518.6142A}), which makes it a reasonable approximation of the total SFRD when corrections for extinction are applied. 
Note, however, that there are some clear exceptions in protocluster environments, which we discuss in the next paragraph. 
Although the relative contribution of the unobscured and obscured SFRD \color{black} in the field and \color{black} in protoclusters has yet to be studied in great detail, a comparison of our results with those of 
\cite{2023arXiv230800745P} shows a good agreement between the obscured SFRD of protocluster galaxies at redshifts $z \sim 2.5$ with the extinction-corrected unobscured SFRD derived for Taralay at $z \sim 4.5$.

There exist some rare systems \color{black} that contain overdensities of sub-mm galaxies that exhibit extreme star formation activity, \color{black} such as the SPT2349-56 protocluster, which was discovered in the South Pole Telescope (SPT)’s extragalactic mm-wave point-source catalogue (\citealt{2010ApJ...719..763V}, \citealt{2013ApJ...779...61M}, \citealt{2020ApJ...900...55E}) and \color{black} followed up \color{black} with the ALMA telescope \citep{2018Natur.556..469M}. \color{magenta} 
\color{black} This protocluster has over 30 submillimeter-bright galaxies along with LAEs and LBGs and SFRD of over 10$^5$ M$_\odot$yr$^{-1}$Mpc$^{-3}$ at $z \sim 4$ \citep{2022MNRAS.512.4352H}. 
Such rare systems with high star formation activities also exist at lower redshifts, e.g, a system from \cite{2016ApJ...828...56W} at $z \sim 2.5$ that has 9 starburst galaxies in the center whose SFR amounts to 3400 M$_\odot$ yr$^{-1}$ within an 80 kpc region, and four enormous Ly$\alpha$ nebulae from \cite{2022A&A...658A..77N} at $ 2 < z < 3$ with an SFRD of $1200 \pm 300$ M$_\odot$ yr$^{-1}$ Mpc$^{-3}$; however, the SPT2349-56 protocluster is one of the only few systems yet discovered at a comparable redshift to Taralay (see \citealt{2022Univ....8..554A} and references therein) that shows an extremely high star formation activity. The SFRD of such systems, while high in the region of the universe that they exist, averaged over the entire sky is probably much less than typical optical/Near-IR selected systems as such types of systems are extremely rare (e.g., \citealt{2017MNRAS.470.2253N}, \citealt{2022MNRAS.514.5004L}). In this way, Taralay might be a better representative of the underlying population of massive protoclusters at this epoch. More work is needed in the future to compare SFRD of Taralay with an ensemble of protoclusters at these redshifts.

Although the optical/Near-IR selection may lead to a sample of protoclusters that are more representative of an underlying galaxy populations, one of the disadvantages of this approach is that the optical/Near-IR diminished/dark galaxies get left out of the sample. If such galaxies, sometimes called HST-dark galaxies, i.e. the galaxies that are undetected in the current \textit{HST} surveys due to being effectively invisible in the rest-frame ultrablue to the typical depths of \emph{HST} survey observations (e.g. \citealt{2018A&A...620A.152F}, \citealt{2019Natur.572..211W}), exist in the region that is targeted in this study, \cite{2023MNRAS.522..449B} shows that their presence will contribute approximately an order of magnitude less than the rest-frame UV/optically selected galaxies to the total SFRD at these redshifts. 
Due to the volume of the protocluster being a lot smaller than the field, the contribution of the HST-dark galaxies may affect the SFRD of the protocluster more than the field widening the gap between $\text{SFRD}_\text{pc}$ and $\text{SFRD}_\text{field}$. 
The effect of these galaxies on the gap between $\text{SFRD}_\text{pc}$ and $\text{SFRD}_\text{field}$ will depend on the location preferred by the galaxies. 


Studies such as \cite{2004ApJ...611..725B}, \cite{2011MNRAS.417.2057A}, \cite{2017A&A...597A...4S}, \cite{2020A&A...642A.155Z} claim association of sub-millimeter galaxies with overdensities, though it is not clear if these sources are fractionally over represented in protoclusters relative to the field. There is some indication from the ALPINE-ALMA survey that there might be a higher number of extremely dusty star-forming galaxies at elevated redshifts in denser environments compared to the field, including around $z \sim 4.57$ (e.g. \citealt{2020MNRAS.496..875R}, \citealt{2021A&A...646A..76L}, \citealt{2023arXiv230301658F}). If such galaxies are more prevalent in rich environments relative to the field, the $\text{SFRD}_\text{pc}$ we calculate here will be a lower limit 
as will its fractional contribution to the overall SFRD at this redshift. 


The observational SFRD$_\text{pc}$ fraction we found in this study is nonetheless quite large, even in excess of the predictions from simulations and indicates that the protoclusters are a significant contributors to the cosmic SFRD at high redshifts. \cite{2020ApJ...899....5I} also found protoclusters as a driver of stellar mass growth in the early universe. They calculated rest-frame ultrablue luminosity function of g-dropout galaxies in 177 protocluster candidates at $z \sim 4$ selected in the Hyper Suprime-Cam Subaru Strategic Program \citep{2018PASJ...70S...4A}; though their SFRD fraction, 6\%-20\%, is not as high as the finding of this study.

At $z \sim 4.57$, the number density of quiescent galaxies is small (e.g., \citealt{2014ApJ...791L..25S}, \citealt{2015A&A...581A..54T}, \citealt{2017A&A...605A..70D},
\citealt{2023AJ....165..248G}). 
All such galaxies may not be detected in our combined spectroscopic and photometric sample. However, their impact on our results are minimal, as their contribution to the SFR is expected to be very low. For example, a massive quiescent galaxy in an overdense environment at $z = 4.53$ in the COSMOS field that was observed with MOSFIRE on Keck has SFR an order of magnitude less than the SFR of main-sequence galaxies at $z = 4.5$ \citep{kakimoto2023massive}. This galaxy is $\sim 6$ proper Mpc away from the center of the PC1 of Taralay and may or may not be associated with Taralay.

\section{Summary and Conclusion}
\label{sec:summary} 

This study was performed in order to conduct one of the first observational tests of whether the protocluster regions in the high redshift universe contribute significantly to the overall mass assembly of the universe as predicted by simulations. We chose the Taralay protocluster at $z \sim 4.57$ in the COSMOS field as the target of a large Keck/DEIMOS campaign as part of the C3VO \citep{2022A&A...662A..33L} survey.  With the total integration time of $\sim 28$ hours, we obtained 44 new secure spectral redshifts in the redshift range of the protocluster as compared to the 9 secure spectral redshifts from VUDS at the time of the discovery. We also reported on 46 new secure spectral redshifts obtained in the redshift range of the coeval field. We combined this spectroscopic data with spectral redshifts from other spectral surveys and a variety of photometric catalogs in the COSMOS field. Using this wealth of data we measured the SFRD of the Taralay protocluster and the surrounding field. Following are the main conclusions of this paper: 

\begin{itemize}
 \setlength\itemsep{1em}
    \item Using the density mapping technique, we mapped out the Taralay protocluster at $z \sim 4.57$, established its internal structure and characterized its properties. Taralay protocluster displays two sub structures PC1 and PC2 for a density isopleth of $\sigma_\delta \geq 2$ that we use as a boundary to define the outline of this protocluster. While PC1 hosts two $\sigma_\delta \geq 5$ overdense peaks, PC2, the smaller substructure only hosts one $\sigma_\delta \geq 5$ overdense peak. \color{black}
    
    \item The mass of Taralay protocluster is estimated to be $1.74^{+1.36}_{-0.77} \times 10^{15}$ M$_\odot$ which makes it exceptionally massive at these redshifts. This protocluster occupies a comoving volume $12620^{+1042}_{-956} \ \text{cMpc}^3$.

    \item We measured the star formation rate density (SFRD) of the field surrounding the Taralay protocluster to be \color{black} log(SFRD/$M_\odot$ yr$^{-1}$ Mpc$^{-3}$) = $-0.82^{+0.19}_{-0.29}$ \color{black} and found it to be consistent with the most comparable study at these redshifts \citep{2021A&A...649A.152K} but in moderate tension with that of \cite{2014ARA&A..52..415M}.

    \item We compare the masses estimated from overdensities calculated with the density mapping technique to the dynamical masses estimated from the line of sight velocity dispersion for the protocluster, two $\sigma_\delta \geq 4.5$ regions and one $\sigma_\delta \geq 2.8$ region. \color{black} We find that the masses estimated from the line of sight velocity dispersion show a deficit in the range of $1.5\sigma$ to $4\sigma$ with an average deficit of $2.5\sigma$. \color{black} 

    \item The SFRD of the $z \sim 4.57$ Taralay galaxy members is $\sim 12$ times higher or \color{black} log(SFRD/$M_\odot$ yr$^{-1}$ Mpc$^{-3}$) = $1.08 \pm 0.32$ \color{black} in excess of the SFRD of the coeval field galaxies signifying that the environment does play a crucial role in driving the SFRD. Protoclusters like Taralay are clearly drivers of stellar mass growth in the early universe. 
    

    \item We provide one of the first observational tests of simulation predictions \color{black} that protoclusters contribute significantly to the fraction of cosmic star formation rate density in the early universe. \color{black} Our findings indicate that protoclusters drive the stellar mass growth in the early universe contributing $33.5\%^{+8.0\%}_{-4.3\%}$ to the cosmic SFRD at $z \sim 4.57$, in $2.67\sigma$ excess over the $\sim 22\%$ value predicted from simulations.

    \item We find that the contribution to the cosmic SFRD from the $\sigma_\delta \geq 5$ peaks of the Taralay protocluster is $22.2\%^{+ 5.4\%}_{-7.3\%}$, a significant portion of the total SFRD of the protocluster and indicative of the inside-out growth pattern as predicted by simulations.

    \item We find that the $\sigma_\delta \geq 5$ peaks of the Taralay protocluster encase 68\% of the SFR while hosting less than 50\% of the galaxies. The SFRD of $\sigma_\delta \geq 5$ peaks is \color{black} log(SFRD/$M_\odot$ yr$^{-1}$ Mpc$^{-3}$) = $0.87^{+0.18}_{-0.32}$. \color{black}

    \item We find a moderately strong, significant positive correlation between SFR and overdensity for galaxies in and around the Taralay protocluster.

\end{itemize}

In the future, we will be expanding this work to ensembles of protoclusters. With an ensemble of protoclusters, it is possible to divide the protoclusters by mass, dynamical state, redshift, etc. in order to better understand the underlying mechanisms which drive and quench the rapid stellar mass growth. We also plan on including sub-millimeter observations to attempt to characterize the role of highly dusty star-forming galaxies in protocluster environments. It appears that such sources may prefer overdense environments (e.g. \citealt{2020MNRAS.496..875R}, \citealt{2021A&A...646A..76L}, \citealt{2023arXiv230301658F}). The inclusion of these highly dusty star-forming galaxies has the potential to significantly increase the estimated SFRD of this protocluster, and thus its contribution to the overall SFRD of the universe at these redshifts.
The results we have presented in Taralay are tantalizing and, if Taralay is indeed an exemplar of massive protoclusters at these redshifts, our results indicate that protoclusters play a key role in driving stellar mass growth in the early universe. 


\section*{Acknowledgements}
\color{black} We thank the anonymous referee for their valuable feedback. \color{black} This work was supported by the National Science Foundation under Grant No. 1908422. \color{black} YF acknowledges support from JSPS KAKENHI Grant Number JP23K13149. \color{black} We thank the late Olivier Le Fèvre for his pioneering work with galaxy redshift surveys, including the VIMOS Ultra Deep Survey, without which this work would have likely not been possible. \color{black} PS thanks Matthew Staab for providing crucial programming support that made the analysis of this paper possible. \color{black} We thank the teams that did the hard work of compiling the various photometric catalogs in the COSMOS field for making these catalogs and their associated quantities public. 
\color{black} A part of this work is based on the Galaxy Evolution Explorer (GALEX) satellite, Canada-France-Hawaii Telescope (CFHT), Hubble Space Telescope (HST), Subaru telescope, Visible and Infrared Survey Telescope for Astronomy (VISTA), Spitzer Space Telescope. \color{black} Some of the data presented herein were obtained at the W. M. Keck Observatory, which is operated as a scientific partnership among the California Institute of Technology, the University of California and the National Aeronautics and Space Administration. The Observatory was made possible by the generous financial support of the W. M. Keck Foundation. The authors wish to recognize and acknowledge the very significant cultural role and reverence that the summit of Maunakea has always had within the indigenous Hawaiian community. We are most fortunate to have the opportunity to conduct observations from this mountain. 

\section*{Data Availability}
The data used for this study will be shared upon a reasonable request.


\bibliographystyle{mnras}
\bibliography{references} 




\appendix
\section{Caveats}
\label{sec:appendixa}
Here we discuss some of the key caveats that should be considered when interpreting the results of our study.
 
\subsection{Estimating the overdensity}
\label{sec:overdensappen}
In this study, we assume that galaxies can serve as tracers of the underlying matter density field and measure the overdensity in terms of galaxies (e.g. \citealt{2014A&A...570A..16C}). Different types of galaxies, such as quiescent or star-forming, trace the matter density field differently. Sometimes matter density is not traced very well by UV-selected galaxies, as in the case of \cite{2022Natur.606..475N}, which finds unexpectedly low galaxy overdensity where large-scale Ly$\alpha$ absorption is strongest indicating high matter density.
Moreover, observations might not trace the true underlying galaxy population as low luminosity galaxies are harder to detect, especially at high redshifts. By considering the mass of galaxies and their star formation rates, we determine an appropriate bias factor from simulations that scales the galaxy overdensity to the matter overdensity. For this study, we use a bias factor of $b=3.6$, which is based on previous research (e.g. \citealt{2013ApJ...779..127C}, \citealt{2018A&A...612A..42D}) and the upper and lower limit for the bias factor are 4.5 and 3.12 respectively obtained from \cite{2023MNRAS.523.4693E} and \cite{2021MNRAS.500.3194A}.



\subsection{SED Fitting}
\label{sec:sedappen}

The SFR derived from SED fitting method correlates well with the SFR derived from independent measures such as [CII] lines at intermediate and high redshift (e.g., \citealt{2020A&A...643A...3S}). \color{black} We checked if this correlation holds for galaxies that we obtained from Y. Fudamoto in \emph{private communication} based on serendipitously detected [CII] lines with ALMA. \color{black} These three galaxies have log($L_\text{CII}/L_\odot)$ of 8.99, 9.01 and 9.10 respectively. When we compared the [CII] derived SFR with the SFR values obtained for these three galaxies through SED fitting using the parameter set given in Section \ref{sec:sed}, we found that the SED-fit SFR values are consistently lower than the [CII] derived SFR. If we force the fits such that the best model chooses the high end of the E(B-V) value range, the difference between the [CII]-derived SFR and the SED-fit value decreases. However, such fits are clearly disfavored by comparing their reduced $\chi^2$. Note that recent theoretical and observational studies suggest that [CII] is more tightly connected to the molecular gas mass in a complicated way through Kennicut-Schmidt relation (e.g., \citealt{2018MNRAS.481.1976Z}; \citealt{2020A&A...643A.141M}; \citealt{2022ApJ...929...92V}) than to SFR. \color{black}


\color{black} \subsection{Absence of FIR data}
\label{sec:firabsent}

We tested the reliability of SFR derived from SED fitting with optical/NIR data (used in this study). We selected 12 galaxies from ALPINE survey that fall in the redshift range of $4.2 < z < 4.93$ with significant detection in the continuum around restframe $155 \mu m$. We performed the SED fitting on these galaxies with the same parameters as listed in Table \ref{tab:parameters} along with \citep{2012ApJ...761..140C} dust template. We found that, relative to the fits that include the FIR data, the optical/NIR-only fits show higher SFRs by $\sim 0.2$ dex. For these galaxies, the extinction correction is generally overestimated by the optical/NIR-only fits. Because these 12 galaxies may or may not be representative of the true galaxy population, as these galaxies are likely the more massive/more dusty extreme of the ALPINE population, we decided to not make any changes to our analysis. Programs probing sub-mm galaxies in protoclusters are needed in order to get a fuller perspective on the star formation in protoclusters. To check how much the result of our study would change if, indeed, what we find were a generally applicable result, we reduced the SFR of each galaxy in our analysis, SFRs that are derived by optical/NIR-only fitting, by 0.2 dex and performed the analysis again. The result was that the SFRD of Taralay and the field were reduced, but well within the uncertainties. The fractional contribution of Taralay-like systems to overall SFRD at these redshifts remained the same. 

\color{black}

\subsection{Dust properties of bright and faint galaxies}
\label{sec:dustpropappen}

Here we test our assumption that the bright and faint galaxies experience the same level of dust extinction which let us extrapolate the SFRD results to include the lower luminosity galaxies in Section \ref{sec:lowgal}. If this assumption is wrong, we are either underestimating or overestimating the SFR of the fainter galaxies. This is because the different dust extinction, through the process of SED fitting, will cause the SFR to be different for the fainter galaxies compared to the SFR of the brighter galaxies. To test this assumption, we analyse the SED models for nearly 3000 galaxies at $2.5 < z < 3.5$ in the COSMOS field. These galaxies make up our sample because at this redshift range we can probe deeper than $M_\text{FUV} = -19.3$, the depth of our data, at $4.2 < z < 4.93$. For this exercise, a galaxy is considered bright if its FUV absolute magnitude is less than -19, the delineation point between bright and faint. 

We perform a KS test on the $E(B-V)$ distribution of both bright and faint galaxies at $2.5 < z < 3.5$, and find that they prefer the same range of values for the color excess of the nebular line light, for both young and old populations. The range for E\textunderscore BV\textunderscore lines is presented in Table \ref{tab:parameters}. This finding shows that at $ 2.5 < z < 3.5$, the SFR traced by FUV photons is similarly affected by the dust properties of both faint and bright galaxies. We assume that this behaviour is consistent for galaxies at $ 4.2 < z < 4.93 $ and make no further changes to our FUV magnitude correction factor.




\subsection{Comparison to Simulations}
\label{sec:sim}



Our study diverges from the definition and characterization of protoclusters as outlined in \cite{2017ApJ...844L..23C} in several key aspects. In their work, a protocluster encompasses all the dark matter and baryonic matter that will eventually merge into a cluster by $z=0$, with a mass exceeding $10^{14} M_{\odot}$ within $R_{200}$. The protocluster's volume encompasses all the matter that will contribute to the formation of the cluster by $z=0$. 
In contrast, our study defines a protocluster as a structure contained within the $2\sigma$ density isopleth.

Furthermore, while \cite{2017ApJ...844L..23C} designates the most massive halo within the protocluster as its core at any given epoch, having a size of $\sim 0.4$ cMpc at $z \sim 4.57$, our study identifies the protocluster core as the region bounded inside a $5\sigma$ density isopleth with a size $\sim 2$ cMpc. To put our peaks on the same footing as the core defined in \cite{2017ApJ...844L..23C}, our peak size will have to reduce beyond $\sigma_\delta > 10$ where we have spectroscopic data only for $\sim 3$ galaxies. Hence we adopt our definition for peaks. It is crucial to acknowledge that these differing definitions of protoclusters may contribute to the measured difference between the observed and simulated data.

\subsection{Using a Massive Protocluster as a Proxy}
\label{sec:proxy}
Using the density maps, the estimated mass of Taralay protocluster at $z \sim 4.57$ is $1.74\times10^{15}$ M$_\odot$ making this structure exceptionally massive at these redshifts. 

Using such a massive protocluster, which has highly star forming galaxies, as a representative sample of all the protoclusters at $z \sim 4.57$ to calculate the fraction of SFRD from protoclusters (see Figure \ref{fig:sfrdfrac}) may lead to an overestimated value for the derived contribution of protoclusters to the SFRD. Nevertheless, there is a possibility that environmental quenching is affecting the galaxies within this protocluster \citep{2018A&A...615A..77L} which could result in a decrease in the SFRD. To draw a definitive conclusion, a large collection of protoclusters is required for comprehensively testing their contribution to the SFRD at high redshift. 


\section{Effect of parameter choices on the model spectrum}
\label{sec:appendixb}
The SFR values that our results are based on come from the SED fitting process. It is important, therefore, to discuss how the different parameters that are used in the SED fitting process impact the spectrum of a galaxy, affecting the estimated SFR. We discuss this impact here using two different approaches. First, we look at the scatter on the SFR of a large sample of galaxies which sheds light on how the SFR itself varies with different parameters used for SED fitting. Then we look at a model SED of a single galaxy to understand how the SED is impacted with the changes in one parameter while keeping the rest of the parameter inputs constant.

\begin{figure*}
    \begin{tabular}{cc}
    \includegraphics[width=75mm]{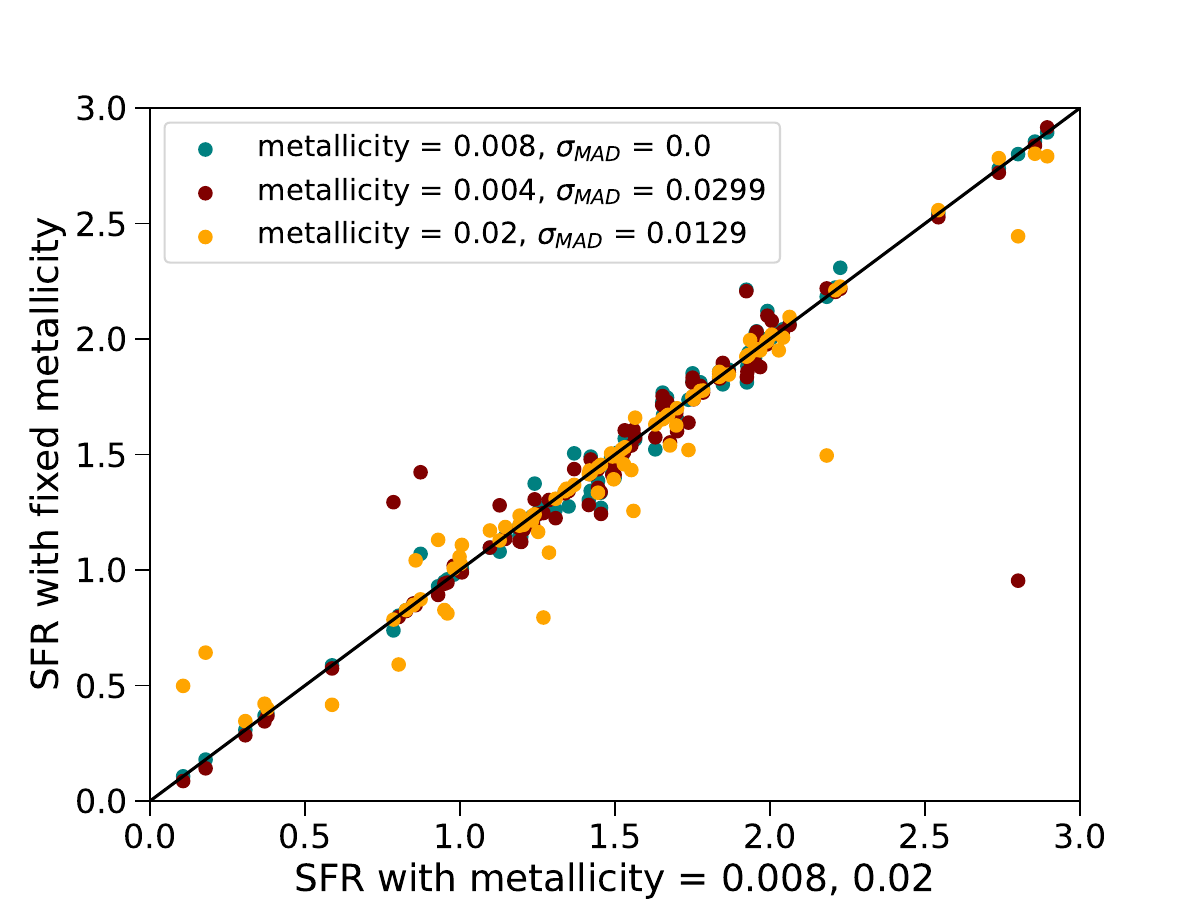} &   
    \includegraphics[width=75mm]{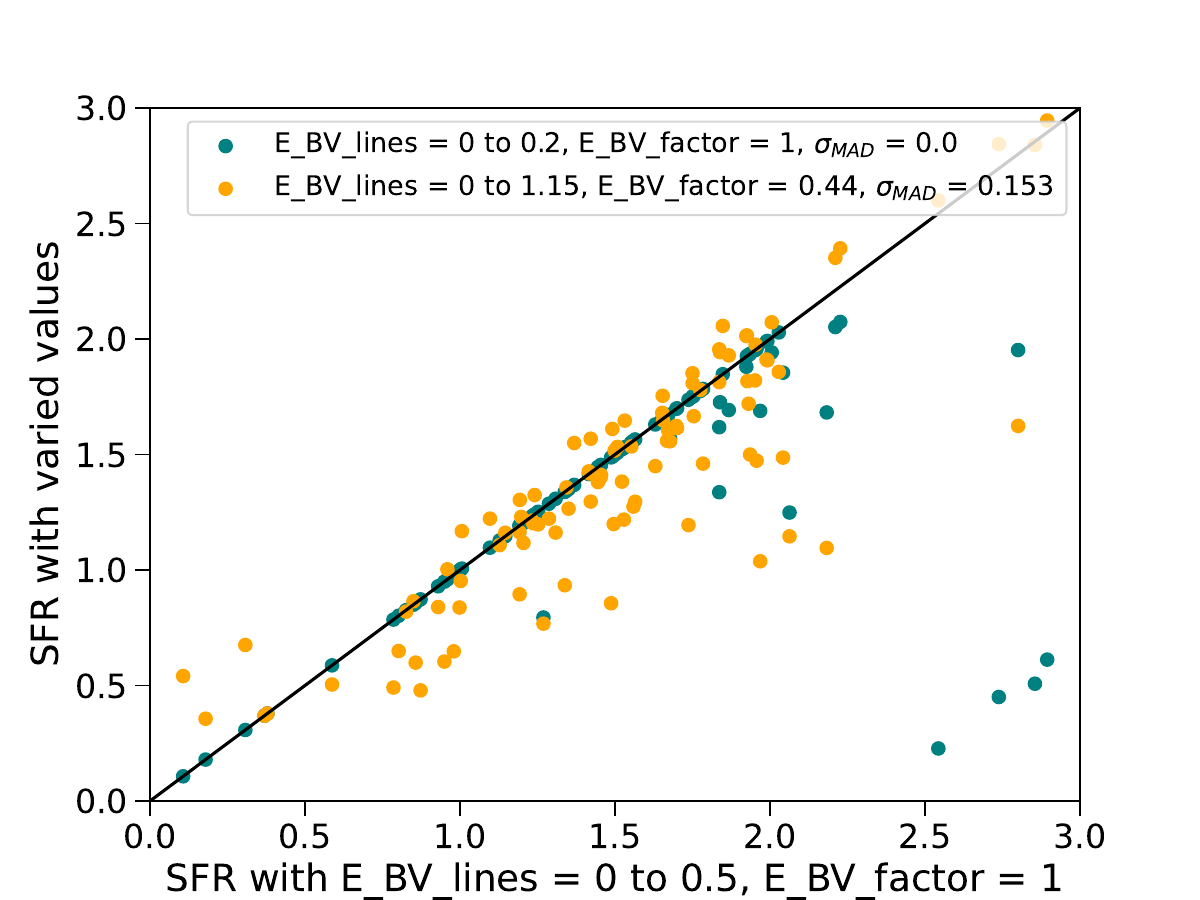}\\
    \multicolumn{2}{c}{\includegraphics[width=75mm]{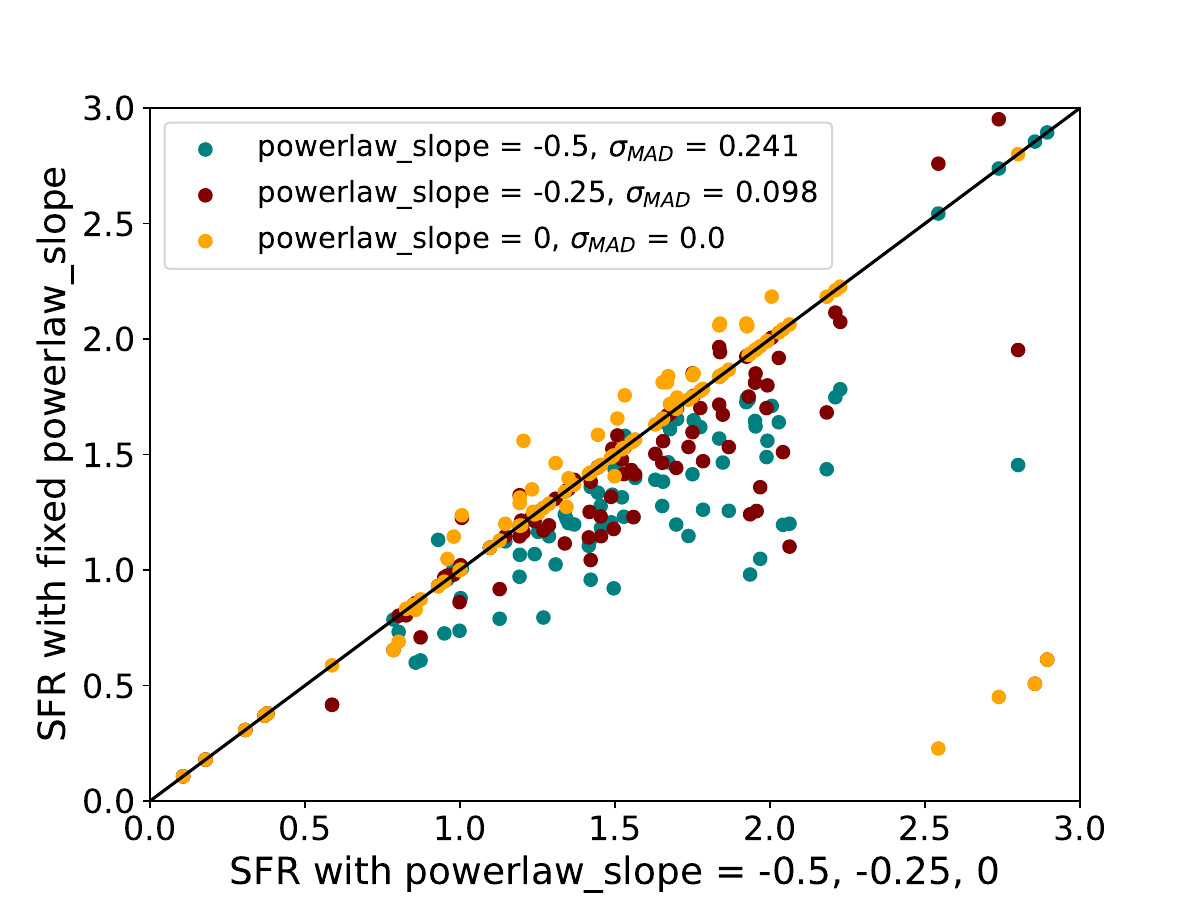} }
    \end{tabular}
    \caption{Comparing SFR values estimated using the ranges of parameters given in Table \ref{tab:parameters} with the SFR values estimated by forcing metallicity to take on a fix value in the top left panel, changing the range of E(B-V) lines in the top right panel and forcing the powerlaw slope to a fix value in the bottom panel. For each plot, the x axis represents SFR obtained from SED fitting the galaxies using a range of a particular parameter. These ranges for metallicity, E\textunderscore BV\textunderscore lines and powerlaw\textunderscore slope are shown on the x axis of each plot. The y axis represents SFR obtained from SED fitting galaxies using a fixed parameter. For both axes for each plot, the SFR values are in $\log(\text{SFR}/M_{\odot}\text{yr}^{-1})$. The $\sigma_{MAD}$ for a distribution in the each figure shows the scatter on the SFR value. The $\sigma_{MAD} = 0$ comes from having a very few outliers with the majority of the galaxies having one to one correlation. }
    \label{fig:fixed params}
 \end{figure*}

In Figure \ref{fig:fixed params} we investigate the change in SFR values by fixing metallicity and powerlaw\textunderscore slope values and by varying E\textunderscore BV lines range. The sample chosen to perform this exercise is one hundred galaxies and the same modules are chosen to estimate their SFR with SED fitting as used in the analysis of this study, i.e. sfhdelayed, bc03, nebular, dustatt\textunderscore modified\textunderscore starburst, redshifting (See Section \ref{sec:sed} for the ranges of these parameters). We find that the scatter on the SFR resulting from changing metallicity is fairly small indicating that a specific value of metallicity is not a strong driver of change in SFR. The scatter on SFR from varying the powerlaw\textunderscore slope and E\textunderscore BV lines is larger indicating these parameters may have an impact on the accuracy of the estimated SFR. The largest systematic uncertainty of $\sim 0.2$ dex coming from the powerlaw\textunderscore slope value of -0.5 affecting the SFR of both the protocluster and the coeval field galaxies equally. Assuming that factors like IMF, E(B-V), powerlaw\textunderscore slope do not dramatically change for the two populations, this uncertainty does not change our results since we compare the SFR of the two populations.

\begin{figure*}
\begin{tabular}{cc}
  \includegraphics[width=75mm]{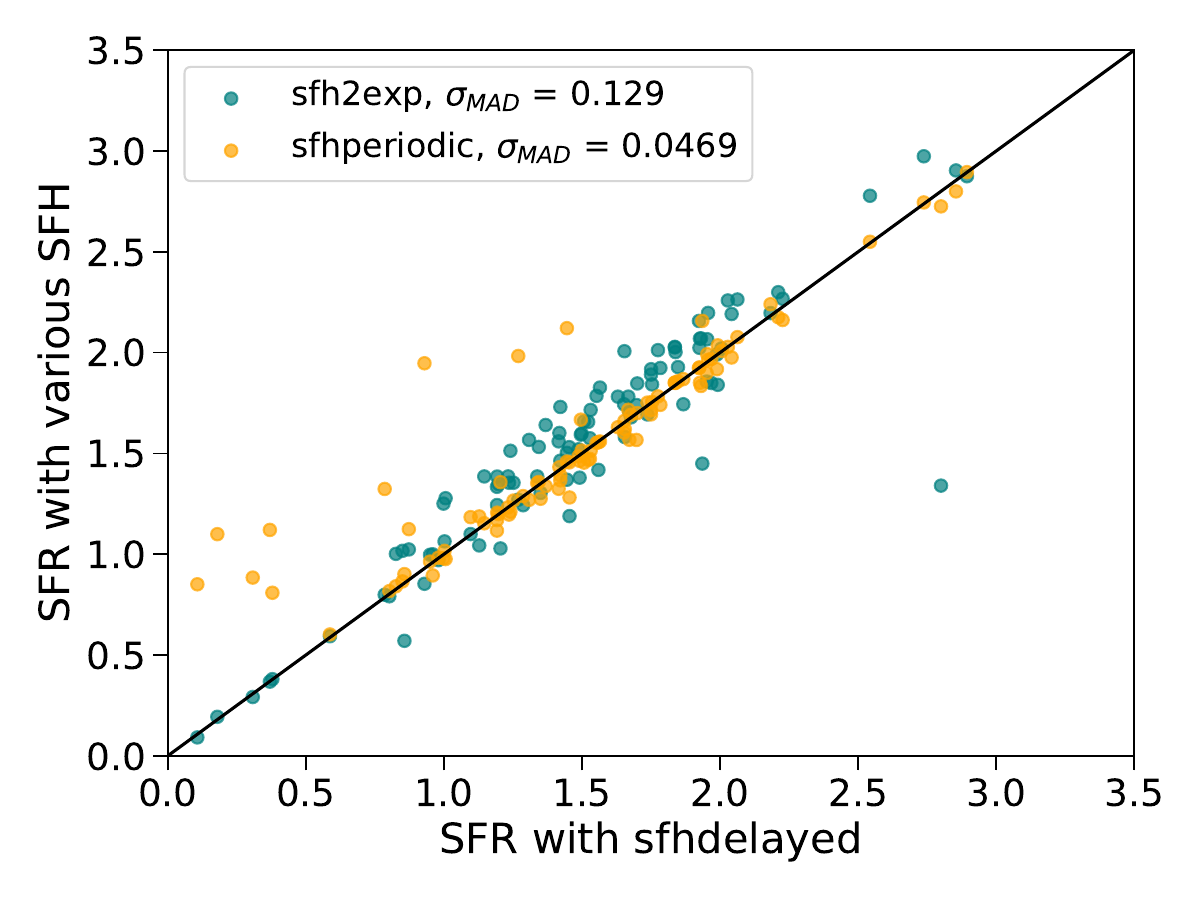} &   
  \includegraphics[width=75mm]{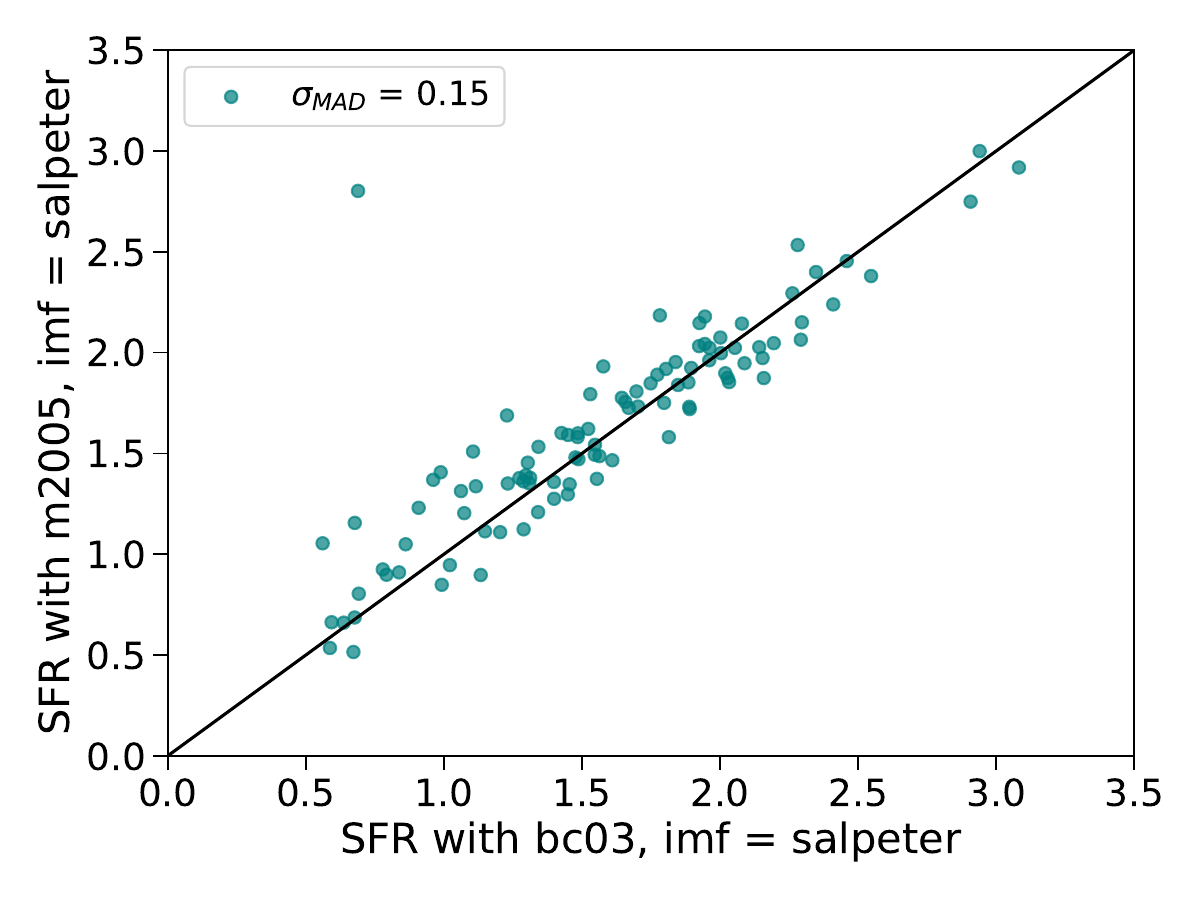}\\
\end{tabular}
\caption{Comparing SFR value obtained from SED fitting using different modules of SFH and SSP. The x axis of the first plot represents SFR obtained from doing SED fitting with sfhdelayed module while the y axis represents SFR obtained from doing SED fitting with sfh2exp and sfhperiodic modules. For the second plot, x axis represents SFR obtained from doing SED fitting with bc03 module with Salpeter IMF while the y axis represents SFR obtained from doing SED fitting with m2005 module with Salpeter IMF. For both axes for each plot, the SFR values are in $\log(\text{SFR}/M_{\odot}\text{yr}^{-1})$. The $\sigma_{MAD}$ for a distribution in the each figure shows the scatter on the SFR value.}
\label{fig:sfhssp}
\end{figure*}

In the left panel of Figure \ref{fig:sfhssp} we compare the SFR estimated with sfhdelayed, the SFH module chosen for perform this study, with two of the other SFH modules available from CIGALE to see how the change in our choice of module can impact our results. We find that the scatter on the SFR is small and a particular choice of SFH is not likely to impact the estimated SFR significantly. In the right panel of this figure, we change the library of SSP to investigate the impact of using bc03 module instead of m2005 module. Because m2005 module is not compatible with the Chabrier IMF originally used in the SED fitting of this study, for this exercise, we use the Salpeter IMF with both bc03 and m2005 module. We do not see a significant difference in the estimated SFR by choosing different SSP modules.

\begin{figure*}
    \begin{tabular}{cc}
     \includegraphics[width=72mm]{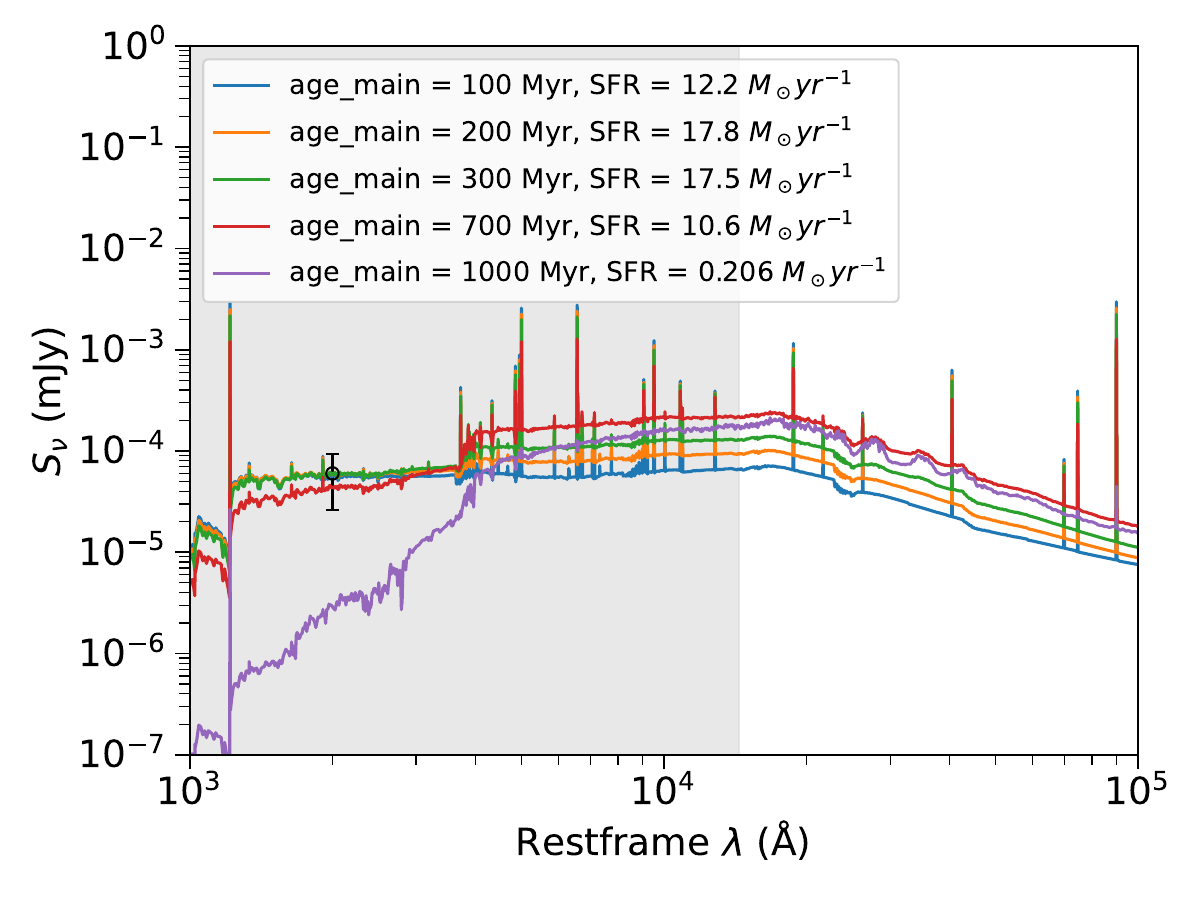} &   
    \includegraphics[width=72mm]{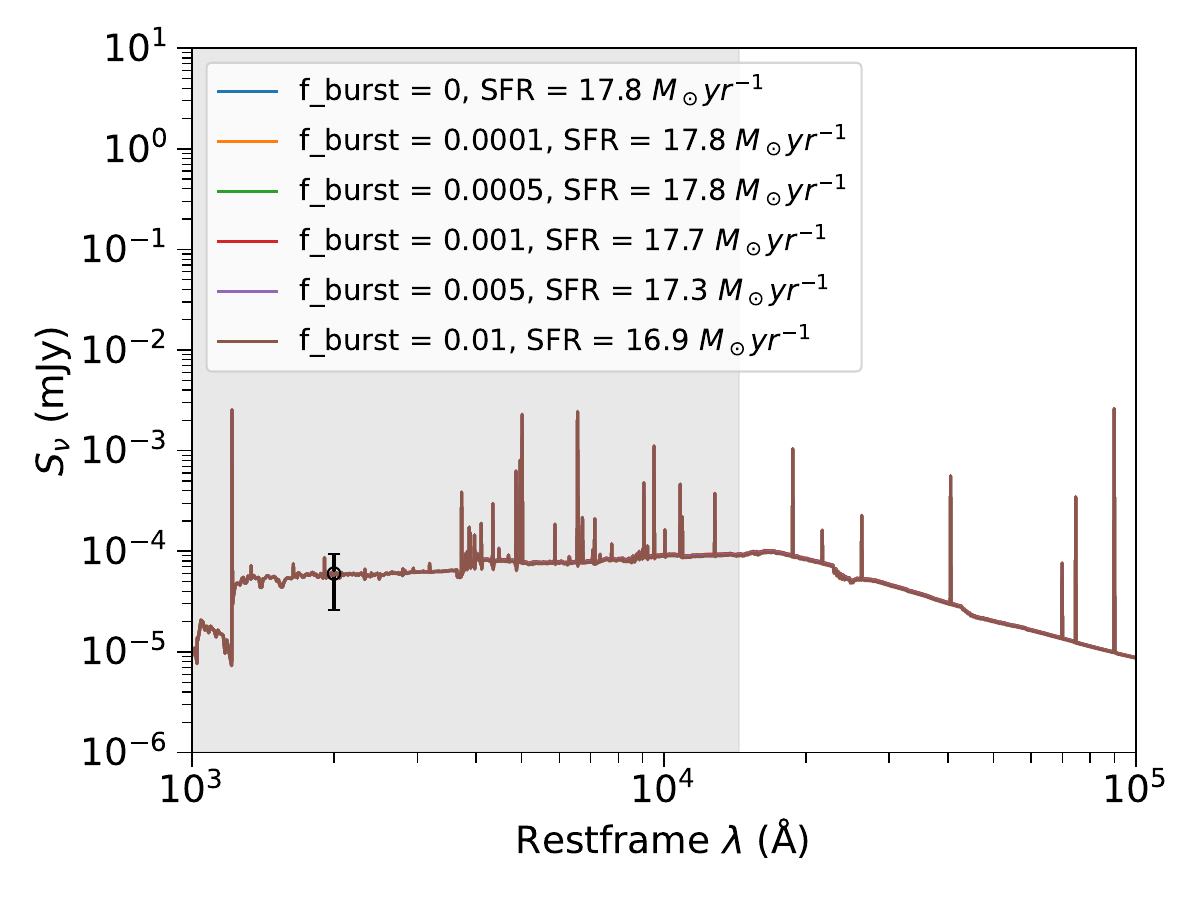} \\
    \includegraphics[width=72mm]{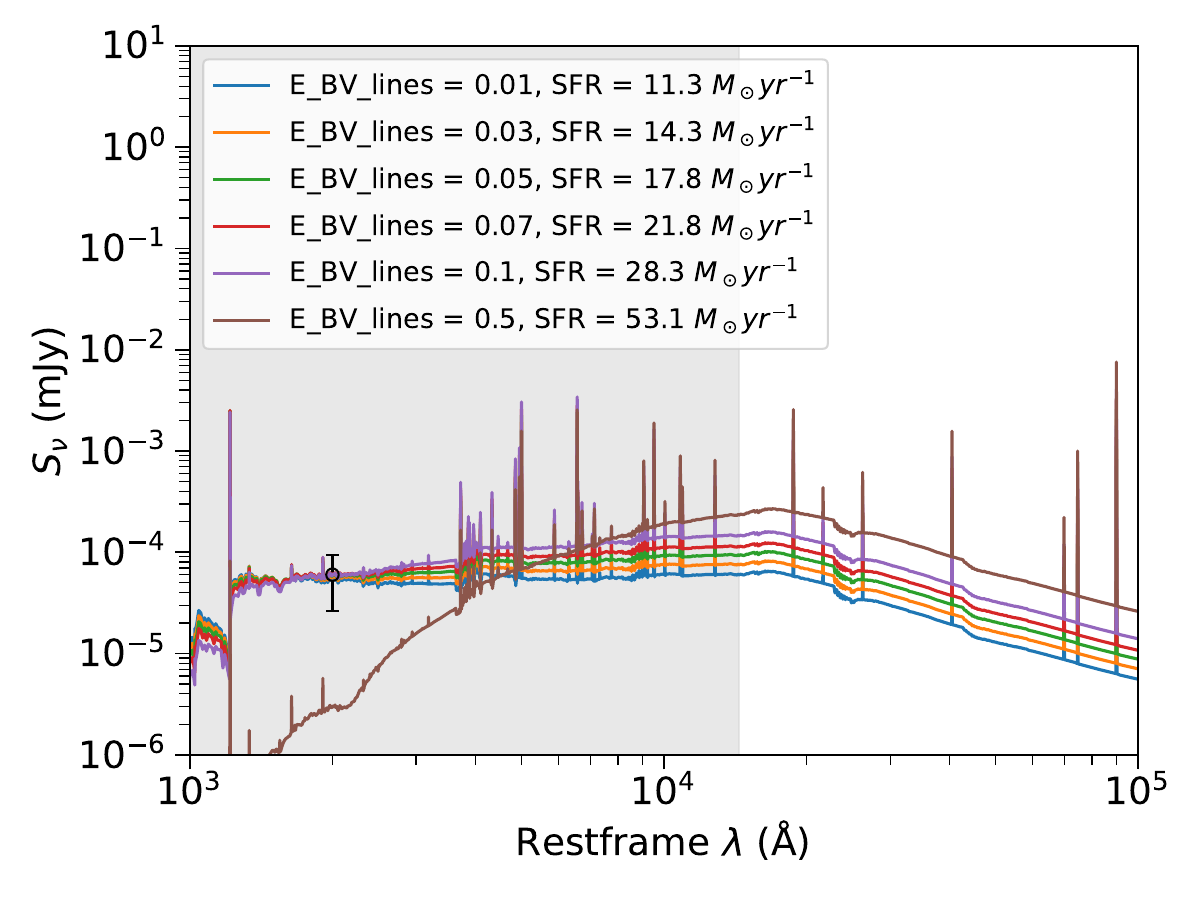} &   
    \includegraphics[width=72mm]{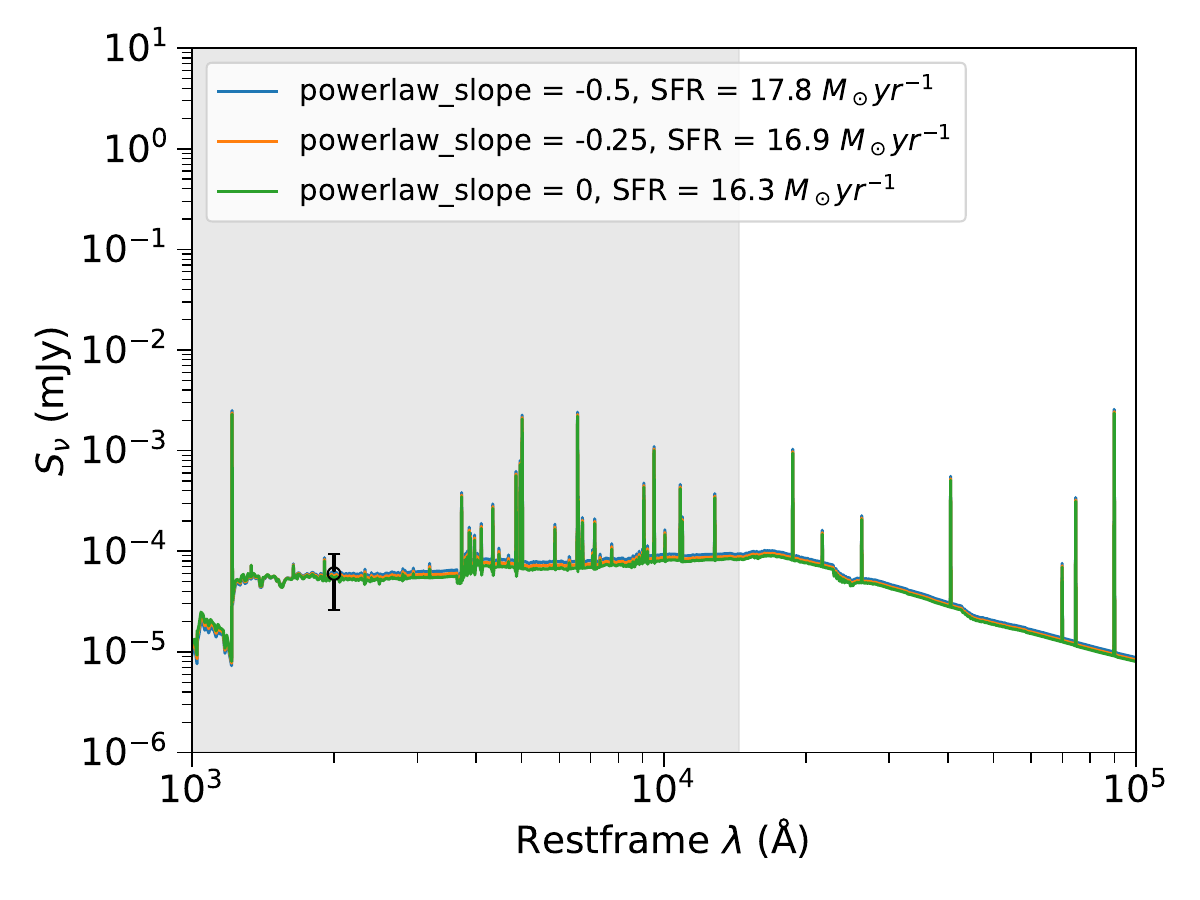} \\
    \includegraphics[width=72mm]{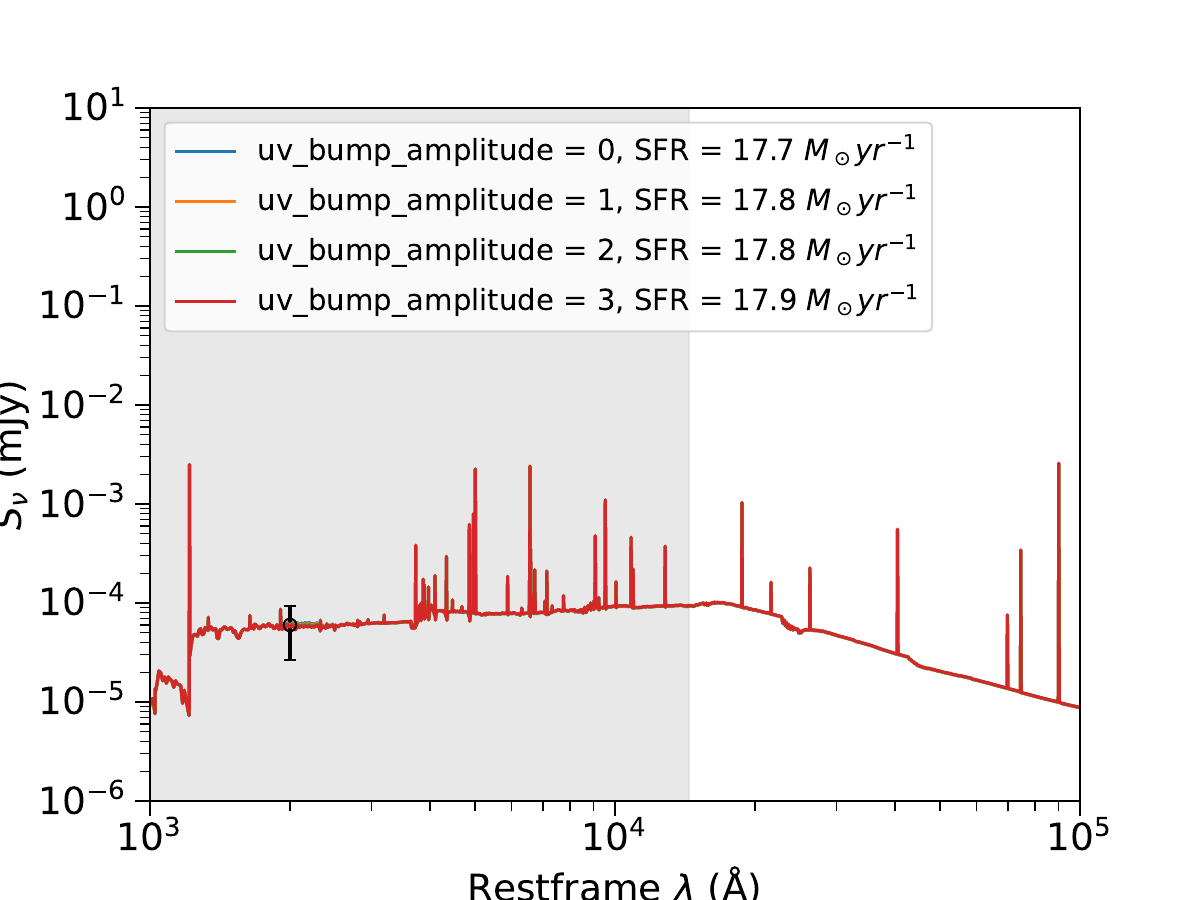} &   \includegraphics[width=72mm]{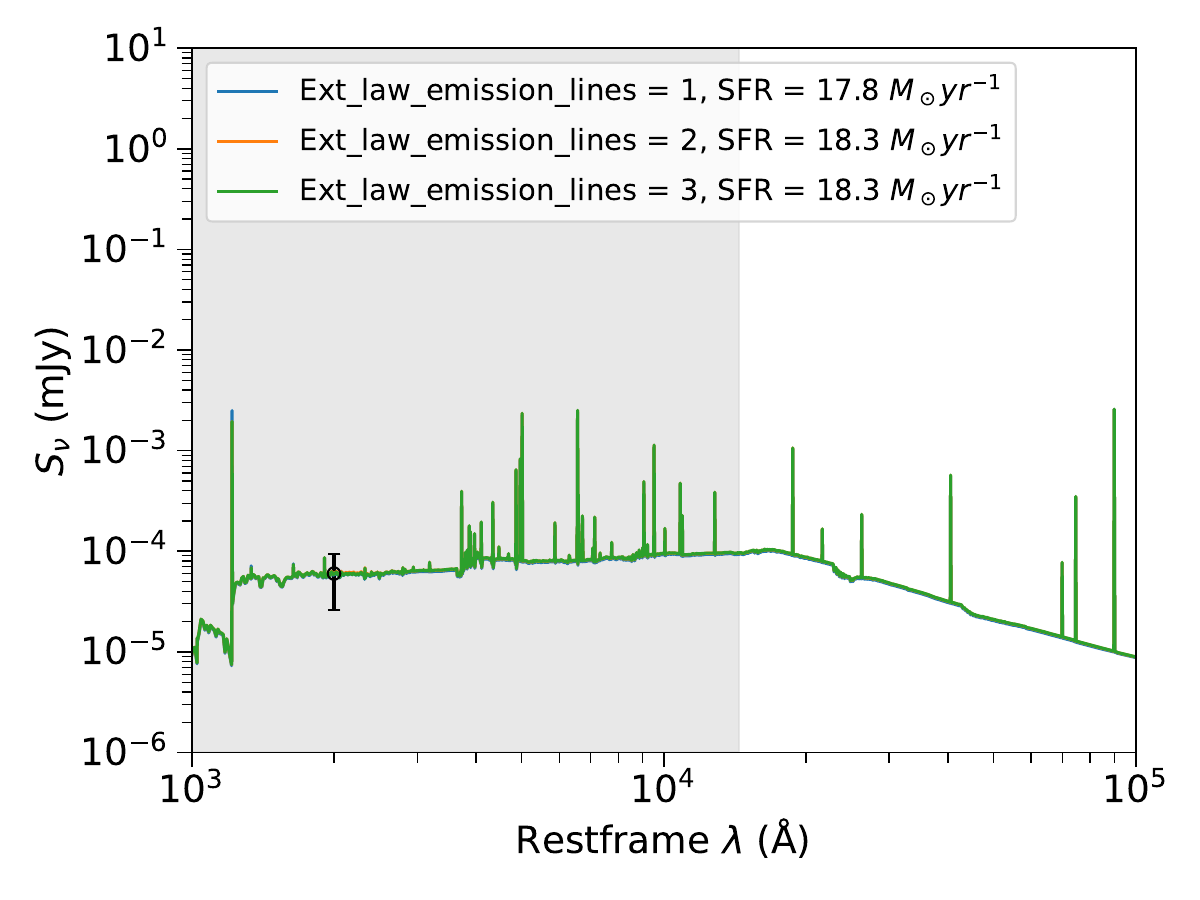} \\
    \multicolumn{2}{c}{\includegraphics[width=72mm]{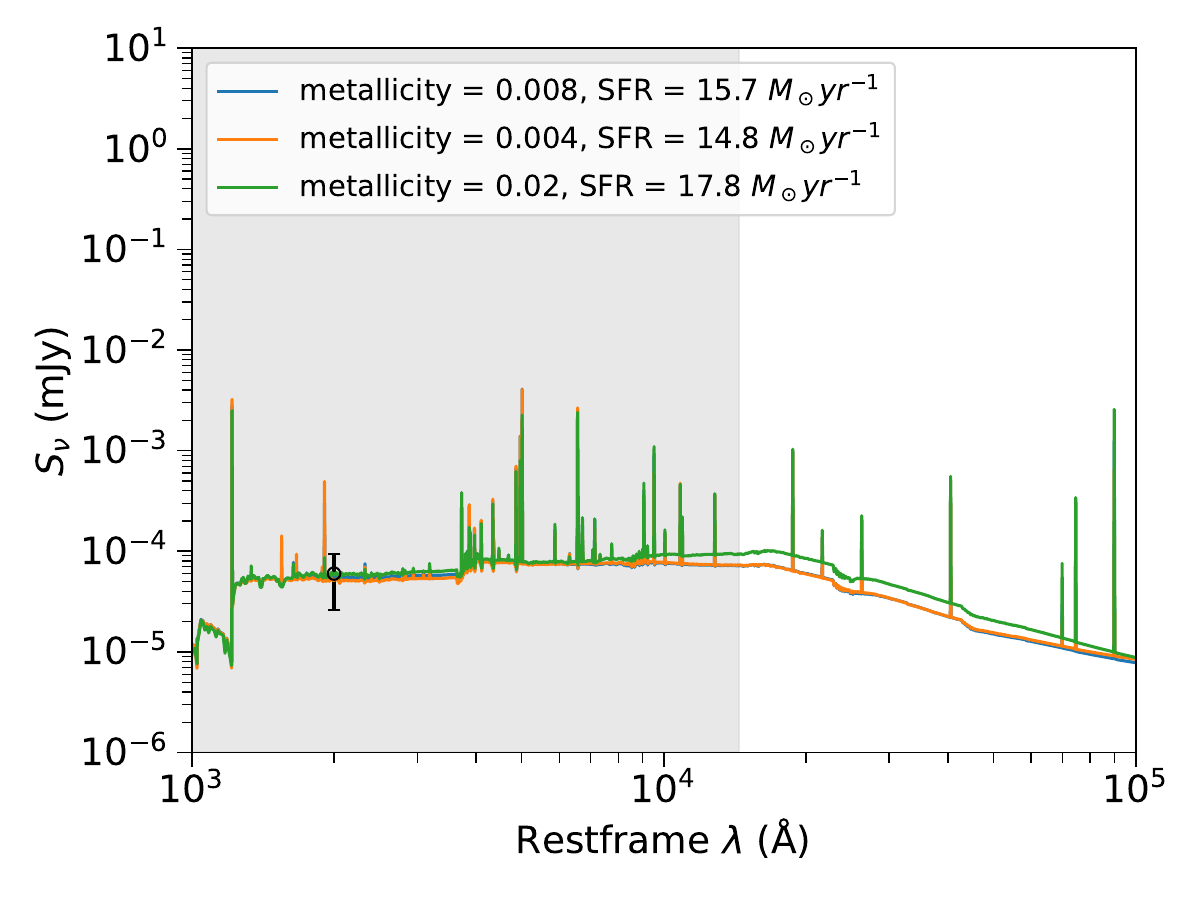} }\\
    \end{tabular}
    \caption{\color{black} Rest-frame SED for a single galaxy with different parameters (see legends of each panels). The SEDs from each panel are such that they have nearly the same SFRs while changing the parameters indicated in the legends in the panels. Data points at $\lambda_\text{rest}$ = 2000 Å in each panels indicate typical errors of the observations. The differences in the SED of the galaxy are within the uncertainty except for the SED obtained with age\textunderscore main of 1000 Myr and E\textunderscore BV\textunderscore lines of 0.5.}
    \label{fig:lonelygal}
\end{figure*}

Next, we investigate the impact of using different parameter ranges on the spectrum of a single galaxy. The SED of the example galaxy in Figure \ref{fig:lonelygal} changes drastically in the restframe UV (shaded region) covered by our data for age\textunderscore main of 1000 Myr and E\textunderscore BV\textunderscore lines of 0.5 showing that it is mainly these two factors that can affect the spectrum of a galaxy. However, our data cannot differentiate between the models with age\textunderscore main differences of $\sim 100$ Myr and E\textunderscore BV\textunderscore lines differences of $\sim 0.05$ as the variations in the models are within the uncertainty of our data (typically of order $\sim 1e^{-5}$ mJy for this example).

\section{SED fitting with CIGALE vs LePhare}
\label{sec:appendixc}

We compare various physical parameters obtained from SED fitting with CIGALE using the parameter set given in Section \ref{sec:sed} to the best-fit parameters given in the COSMOS2020 catalog \citep{2022ApJS..258...11W} and find no significant difference that can be attributed to the choice of using a particular SED fitting software. Here we show this comparison for stellar mass and star formation rate. The sample used for comparison consists of 106 galaxies in the redshift range of $4.2 < z < 4.93$ that have secure spectroscopic redshifts. This sample has also undergone an IRAC channel 1 and/or IRAC channel 2 cut such that the 106 galaxies have magnitudes brighter than the IRAC channel 1 and/or IRAC channel 2 completeness limit listed in \citep{2022ApJS..258...11W}.

\begin{figure*}
    \centering
    \includegraphics[width = \textwidth]{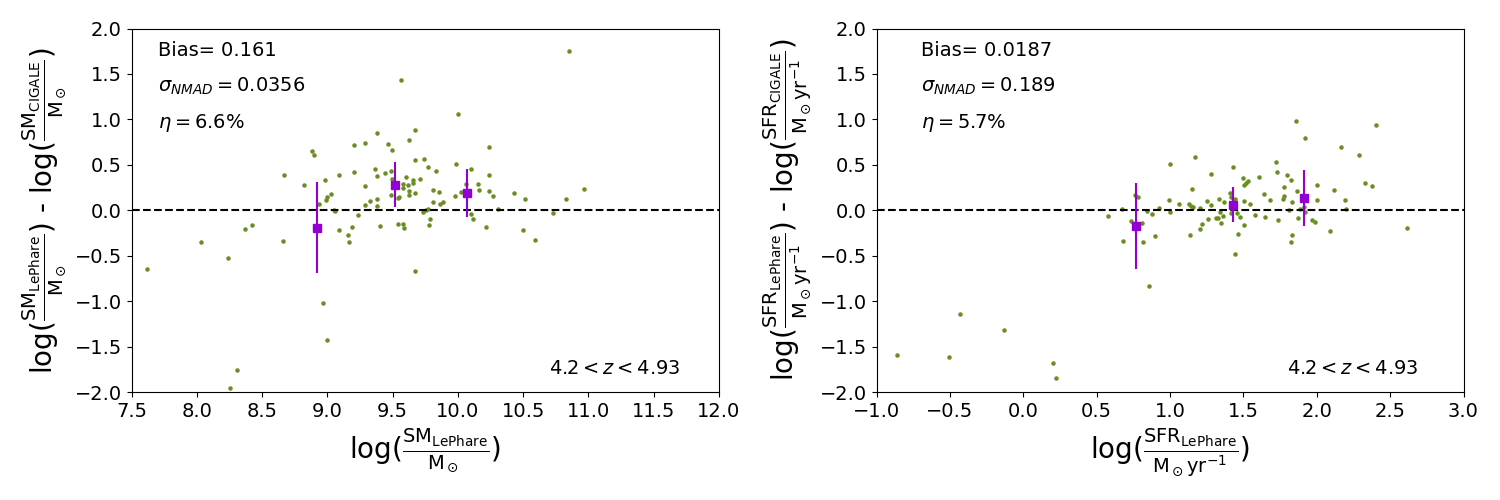}
    \caption{Difference between stellar mass and star formation rate estimated with CIGALE and LePhare, two SED fitting software. The $\sigma_\text{NMAD}$, catastrophic outlier rate ($\eta =  |\log(\text{X}_\text{LePhare}) - \log(\text{X}_\text{CIGALE})|/(1+\log(\text{X}_\text{LePhare}$) > $3 \sigma_\text{NMAD}$), and bias which is defined as the median of the difference between $\log(\text{X}_ \text{LePhare})$ and $\log(\text{X}_\text{CIGALE})$ are shown for both panels where X represents SM/M$_\odot$ for the left panel and SFR/M$_\odot$ yr$^{-1}$ for the right panel.}
    \label{fig:smsfrdiff}
\end{figure*}




\bsp	
\label{lastpage}
\end{document}